\documentclass[11pt,a4papert]{article}
\pdfoutput=1
\usepackage{jinstpub}
\usepackage{subfig}
\usepackage{multirow}

\title{The Cesium Source Calibration and Monitoring System of the ATLAS Tile Calorimeter: Design, Construction and Results.}

\author[a]{G.~Blanchot,}
\author[a]{M.~Bosman,}
\author[b]{J.~Budagov,}
\author[a]{M.~Cavalli-Sforza,}
\author[c]{I.~Efthymiopoulos,}
\author[d]{A.~Isaev,}
\author[a]{Y.~Ivanyushenkov,}
\author[d]{A.~Karyukhin,}
\author[d]{S.~Kopikov,}
\author[c]{M.~Nessi,}
\author[d]{V.~Senko,}
\author[d]{N.~Shalanda,}
\author[d]{M.~Soldatov,}
\author[d]{A.~Solodkov,}
\author[d,1]{O.~Solovyanov,\note{Corresponding author.}}
\author[d]{E.~Starchenko,}
\author[b]{V.~Tsoupko-Sitnikov,}
\author[e,\dag]{I.~Vichou,\note[\dag]{Deceased.}}
\author[d]{and A.~Zaitsev}

\affiliation[a]{Institut de F\'isica d'Altes Energies (IFAE), Barcelona Institute of Science and Technology, Barcelona, Spain}
\affiliation[b]{Joint Institute for Nuclear Research, Dubna, Russia}
\affiliation[c]{CERN, Geneva, Switzerland}
\affiliation[d]{State Research Center Institute for High Energy Physics, Protvino, Russia}
\affiliation[e]{University of Illinois at Urbana-Champaign} 
 
\emailAdd{Oleg.Solovyanov@cern.ch} 

\newcommand{\Cs}{$^{137}$Cs }

\abstract{
This article describes the design, construction and use of a calibration and monitoring system, based on movable \Cs $\gamma$-ray sources, for the ATLAS Tile Calorimeter (TileCal). The sources, propelled by a water-based liquid through tubes that traverse all the calorimeter's cells, produce signals that precisely characterise the response of each tile, thereby providing very granular and accurate data on the response of TileCal to particles. The system has been used to guide and control the quality of the optical instrumentation of all TileCal modules, to set and equalise the dynamic range of the response to physics data, and to set the energy scale of the readout system. In the ATLAS cavern, periodic measurements of the whole detector's response to \Cs sources allow monitoring the uniformity and stability of all the calorimeter's cells as well as maintaining precise knowledge of its energy calibration.
The design of the source hydraulic drive system's hardware and software, the data acquisition system and the data processing algorithms are described. Finally, the results of this two-decade program are shown.
}


\begin{document}

\maketitle
\flushbottom

\section{Introduction}

\begin{figure}[!htbp]
  \centering
  \subfloat[]{\includegraphics[width=0.65\textwidth]{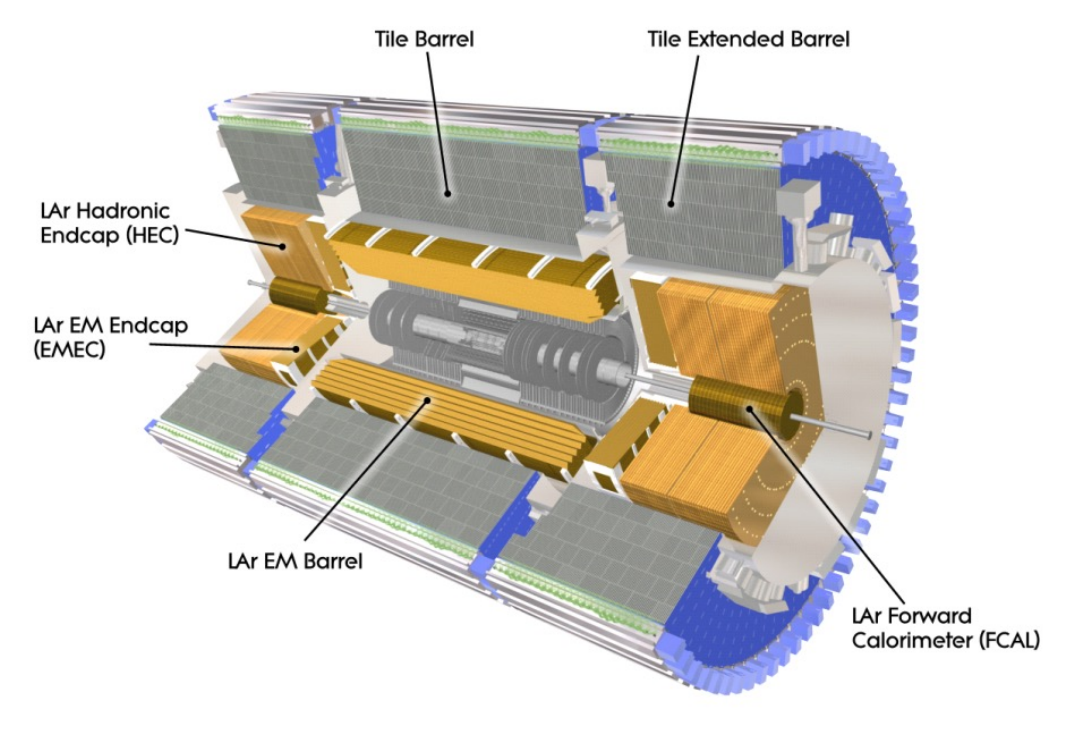}}
  \subfloat[]{\includegraphics[width=0.35\textwidth]{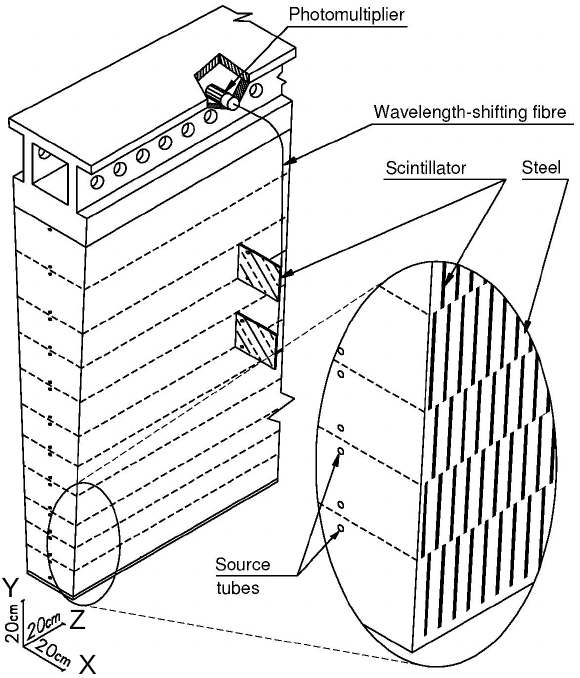}}
  \caption{(a) ATLAS calorimeter system. (b) Tile calorimeter module internal structure~\cite{TileReady}.}
  \label{fig:calorimeter}
\end{figure}

The hadronic Tile Calorimeter (TileCal) is an essential part of the ATLAS experiment~\cite{ATLAS} at the CERN Large Hadron Collider~\cite{LHC}. Together with the Liquid Argon (LAr) electromagnetic and hadronic calorimeters, it provides measurements of the energy of particles and jets produced in LHC collisions as well as information on the missing transverse energy $E^\mathrm{miss}_{\mathrm{T}}$. The ATLAS calorimeter system is shown in figure~\ref{fig:calorimeter}a.
	
TileCal is a sampling calorimeter composed of steel plates as the absorber/radiator and organic scintillating tiles made from polystyrene base as the active material. The design, general features and expected performance of the calorimeter are described in Ref.~\cite{TileCal}, while its operating parameters and performance can be found in Ref.~\cite{TileReady} and Ref.~\cite{TileRun1}.  

A brief description of the salient dimensions of TileCal follows, while the detailed information about the mechanical structure is available in Ref.~\cite{TileMechanics}.

TileCal consists of three sections, known as a Long Barrel (LB) and two Extended Barrels (EB), called EBA and EBC. These three sections have a cylindrical shape with inner and outer radii of 2280 and 4230~mm respectively. The LB is 5640~mm long along the beam axis (Z), while EBA and EBC are 2910~mm long. The mechanical supports for the LAr calorimeter cryostats are located inside the Extended Barrels. 

Each of the TileCal cylinders is subdivided in azimuth into 64 independent modules. Three mm thick scintillating tiles lie within modules in the $r-\phi$ plane at a distance of 18.2~mm along the Z-axis, separated by 14.0~mm steel plates, thereby creating the sensitive material matrix with the periodic structure along the beam line, as shown in figure~\ref{fig:calorimeter}b.

The scintillating tiles are organised along the radius from the beam pipe in 11 tile rows of different sizes, numbered from 1 to 11 starting from the smallest radius. The scintillation light is collected at the exposed edges of each tile by wavelength shifting (WLS) fibres, arranged in pre-shaped opaque plastic ``profiles'' attached to both sides of the modules and running radially. To accommodate calibration tubes and fixing rods, all steel plates and tiles have two \O9.0~mm holes.

\begin{figure}[!htbp]
  \centering
  \includegraphics[width=0.9\textwidth]{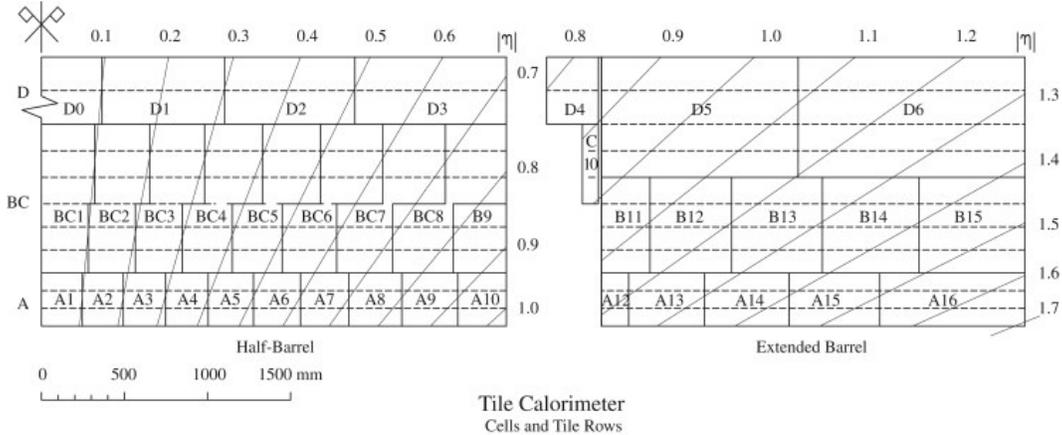}
  \caption{Tile calorimeter segmentation in depth (radius) and pseudo-rapidity ($\eta$)~\cite{TileReady}.}
  \label{fig:tilecells}
\end{figure}

Within a module, the readout cells, shown in figure~\ref{fig:tilecells}, are defined by grouping together fibres which are then bundled and coupled to the photomultiplier tubes (PMTs) that read out each TileCal cell, as described below. Each fibre bundle brings to a PMT the light from one side of a contiguous group of tiles. The light from every cell is read out by two PMTs, which measure the light from two cell sides, thus improving the uniformity of cell response across $\phi$ and the reliability of light collection. The cells span pseudo-rapidity ($\eta$) intervals of 0.1 in layers A, B and C and 0.2 in layer D of the Long Barrel. In the angular range covered by the Extended Barrels, it is not possible to define pseudo-rapidity intervals with similar accuracy, therefore the readout cells span larger pseudo-rapidity intervals. There are 45 cells in each LB module and 14 cells in each EB module.  The total number of TileCal PMTs is~9852. The details of the optics instrumentation can be found in Ref.~\cite{Instrumentation}.

\begin{figure}[!htbp]
  \centering
  \subfloat[]{\includegraphics[width=0.45\textwidth]{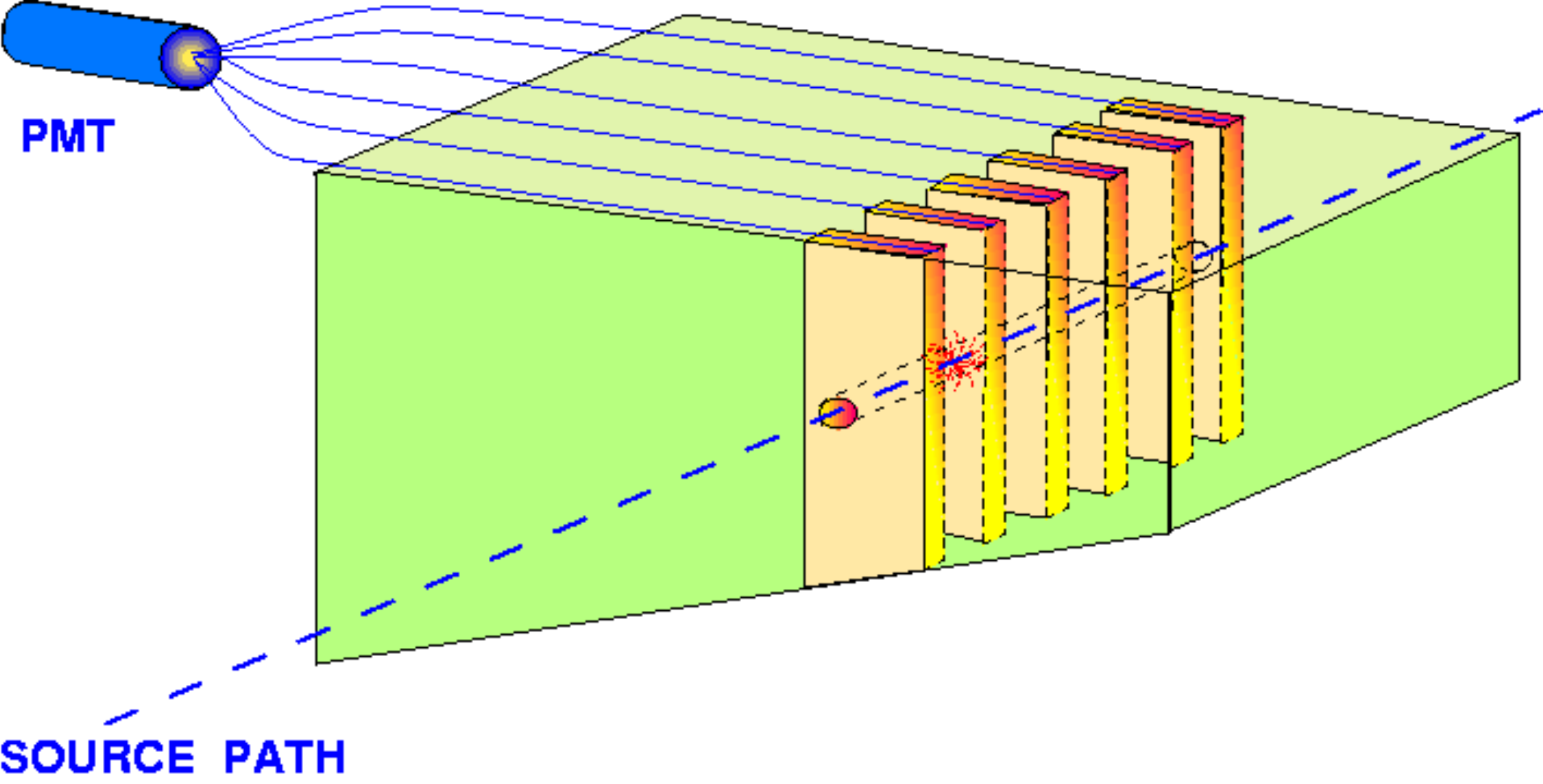}}
  \subfloat[]{\includegraphics[width=0.55\textwidth]{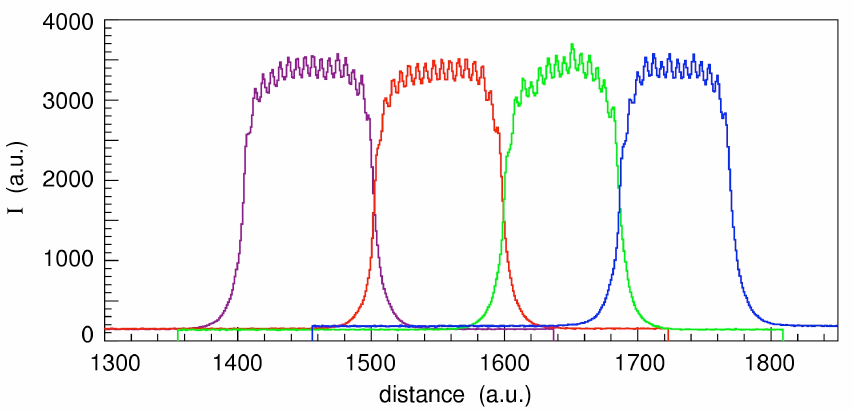}}
  \caption{(a) The concept of the \Cs source calibration system. (b) Typical PMT response from the four neighbouring cells as a function of the source position along its path~\cite{Cs137}.}
  \label{fig:principle}
\end{figure}

To calibrate and monitor TileCal it was decided to use \Cs $\gamma$-ray sources, propelled by a hydraulic system, which traverse all modules and deposit in them the energy of the $\gamma$-ray~\cite{Cs137,Cs137H}. With this design, the source signal, averaged over all the tiles of a cell, will represent the relative response to particles detected in that cell of TileCal. The design and configuration of the system allow to check and to monitor the whole optical path from the scintillating tiles to the PMT with better than 1\% precision. The geometrical concept of the approach is shown in figure~\ref{fig:principle}a. 

When a scintillating tile is traversed by the \Cs source, the emitted 0.662~MeV gamma rays induce light in the scintillator. The light is transported to a PMT by the WLS fibres. The mean-free-path of the gamma rays in the calorimeter structure matches the 18.2~mm calorimeter sampling period, therefore the fine detail of individual tile responses is clearly visible, and any defect can be diagnosed.

The observed PMT signal averaged over a readout cell indicates the tile and fibre optical quality for the entire cell, while peak values can be associated with individual tile responses (figure~\ref{fig:principle}b). The PMT signal read-out electronics is based on a low-noise operational amplifier used as a charge integrator, described in Ref.~\cite{Integrator}.

The relatively long 30.2~year half-life of the \Cs isotope (corresponding to a $\sim$2.3\% intensity loss per year) makes it possible to use the same sources to monitor the calorimeter response stability over a time comparable to the lifetime of the ATLAS experiment.

Two more systems, together with in-situ beam calibration, are used to monitor the TileCal response: a Laser monitoring system~\cite{Laser} to check the PMTs, and a Charge Injection System (CIS)~\cite{CIS} to test and to calibrate the fast front-end electronics. However, these two systems do not measure the properties of the entire optical path (scintillator, fibres, their optical coupling, etc.), whereas the radioactive source system can fill this need. Other options sensitive to the properties of the entire optical path, such as monitoring the calorimeter's operation with real events (with ``minimum bias'' or cosmic-ray triggers), depend on the available statistics of the recorded runs and are not as precise and flexible as the radioactive source system. This has made calibration by a Cs source very important for TileCal, during the module instrumentation and test beam phases and during the physics runs.

An elaborate source drive system, appropriate controls and an advanced online data analysis framework are required to precisely and reliably measure the response of $\sim$463000 tiles, passing through the 192 TileCal modules. The design of the system, the layout of the calibration tubes and the additional equipment had to respect the TileCal detector geometry envelopes and tolerances. 

A system that fulfils these requirements, using a hydraulic drive to drive radioactive sources through the entire volume of TileCal, was designed, constructed and commissioned~\cite{MonSys}. A water-based liquid propels the radioactive sources through a large tubing circuit at a steady velocity of about 35~cm/s, which is adequate to scan the whole TileCal volume in a few hours while providing high-granularity outputs on the optical response of every single tile. An early system was used to calibrate the prototype and production modules of TileCal; the full-blown system was installed in the ATLAS experimental cavern from 2002 to 2008 and has been used since 2009 to monitor the response of TileCal to cosmic rays and to particles produced in LHC collisions.

In this article, the design, construction, installation, and results of the radioactive source system of the ATLAS Tile Calorimeter are described, with an emphasis on the overall aspects of this facility. 

\section{System description}
\subsection{General design}

\begin{figure}[!htbp]
  \centering
  \subfloat[]{\includegraphics[width=0.5\textwidth]{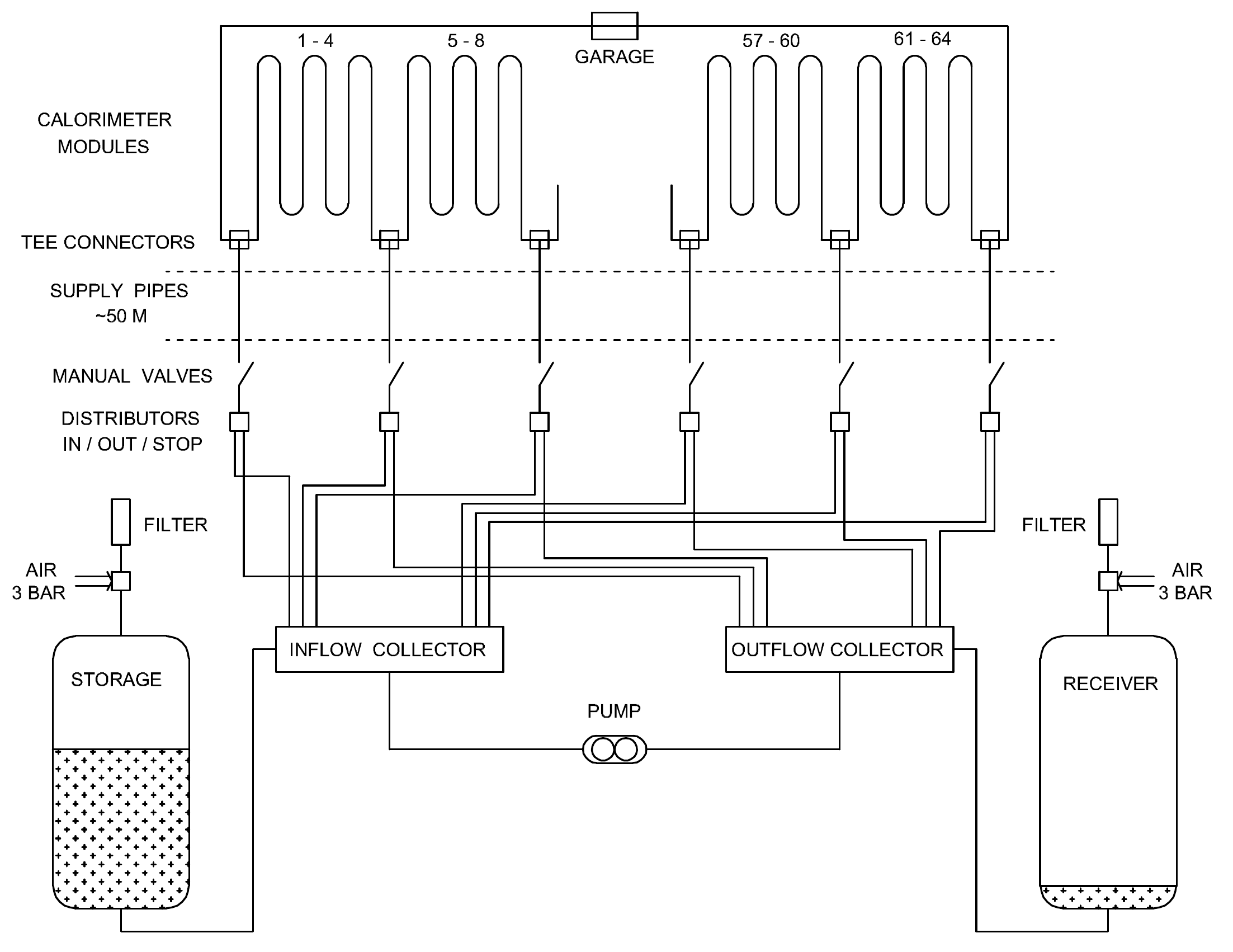}}
  \subfloat[]{\includegraphics[width=0.5\textwidth]{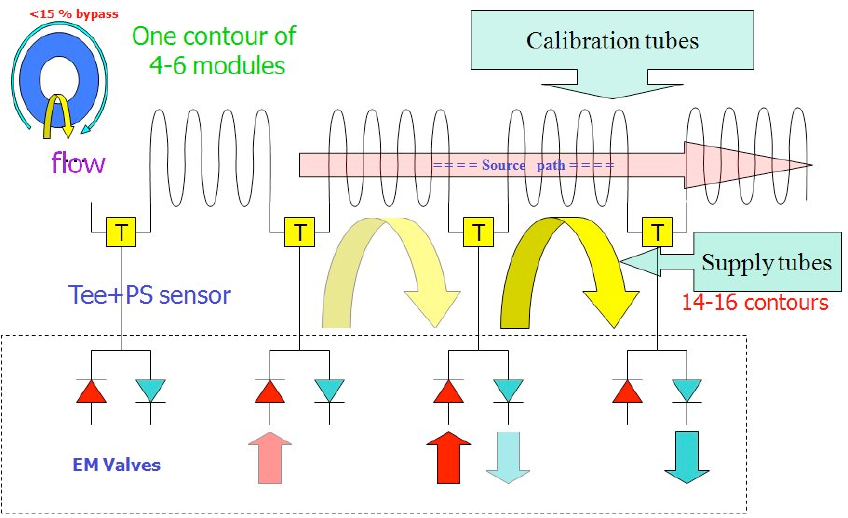}}
  \caption{ (a) Schematics of the Cs system hydraulics. (b) The hydraulic contour switching concept.}
  \label{fig:hydraulics}
\end{figure}

The TileCal Cs source system consists of three separate but functionally identical subsystems, serving the Long Barrel (LB) and the two Extended barrels, EBA and EBC. The layout of a subsystem is schematically shown in figure~\ref{fig:hydraulics}. It consists of the following components:  

\begin{itemize}
\item A circuit of calibration tubes within each of the three calorimeter barrels, that defines the source path, including the source storage devices (``garages''), where the source is kept between scans;
\item The circuit is divided into a number of segments (also known as ``contours'') consisting of 4 to 8 modules each; special tee-joints separate the source path segments and connect to supply pipes that lead the liquid into and out of the calibration circuit's common volume;
\item The supply pipes themselves, that provide the pressure that propels the source capsule within the appropriate circuit segment;
\item Pumping and distributing devices, that provide the liquid flow to each circuit segment, as well as propelling the source capsule in the desired direction within a segment and through the tee-joint to the next segment;
\item Storage vessels, to hold the liquid while the system is idle.
\end{itemize}

The tee-joints set the entry and exit points of the working liquid and define circuit segments with a volume smaller than that of the entire calibration circuit and allow to apply pressure to one circuit segment at a time with enough force to push the capsule in the desired direction. This way, pressures in each section and the overall liquid circuit are kept at appropriately low levels. 

The liquid in other parts of the circuit stays almost at rest. There is a small flow in the opposite direction, estimated to be about 15\% of the contour flows. 

The source capsule moves in the desired direction with the steady flow of the liquid. The stability of the movement depends on the performance of the pumping equipment, the inertia due to the mass of the fluid, the dynamic frictional properties of the source capsule on the surface of the calibration tubes and lubricity and viscosity of the working liquid. The source capsule passes from a segment to the next one through a tee-joint when the liquid flow is switched from one to the next.

\begin{figure}[!htbp]
  \centering
  \subfloat[][Long barrel]{\includegraphics[width=0.42\textwidth]{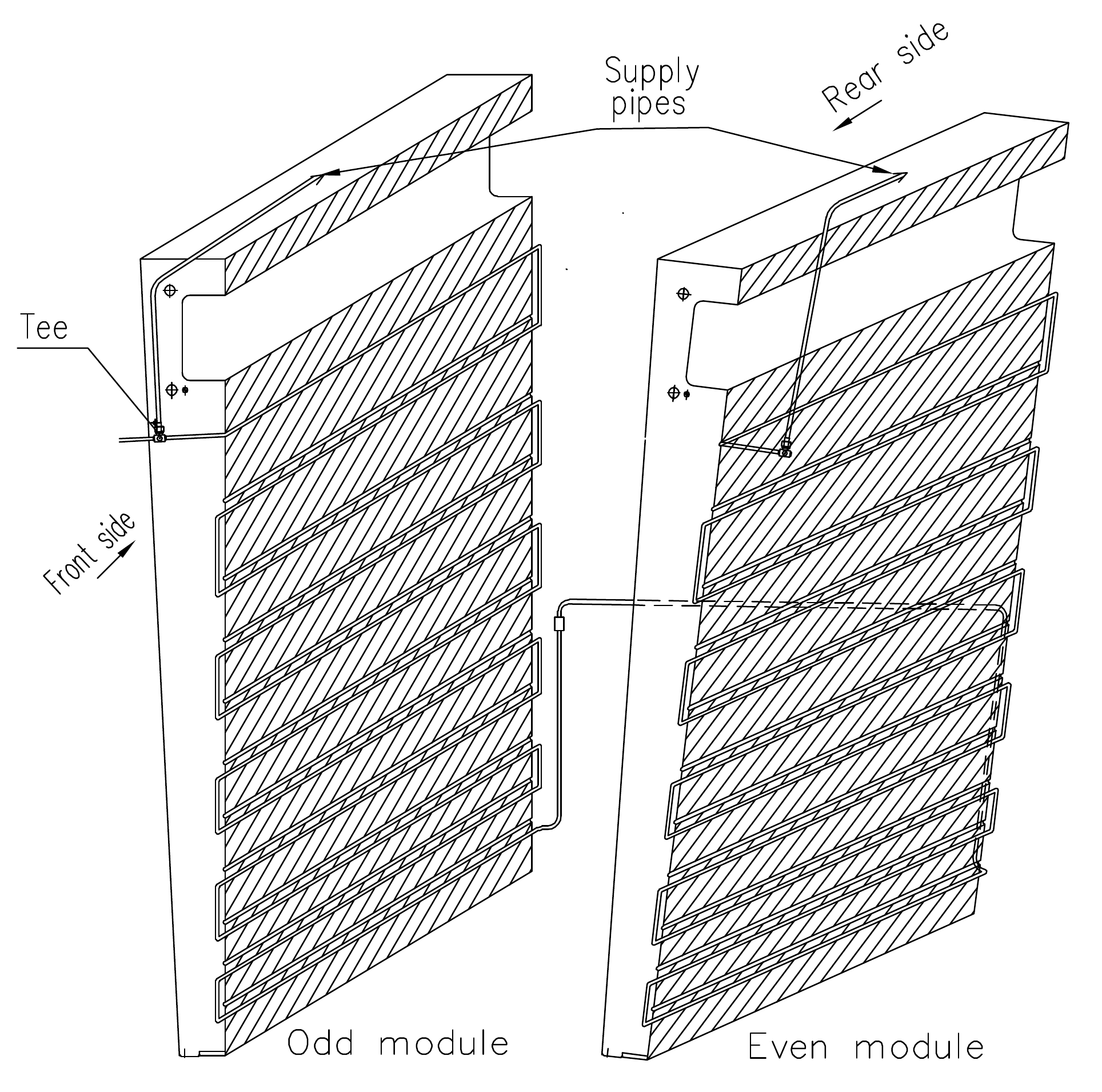}}
  \subfloat[][Extended barrel]{\includegraphics[width=0.57\textwidth]{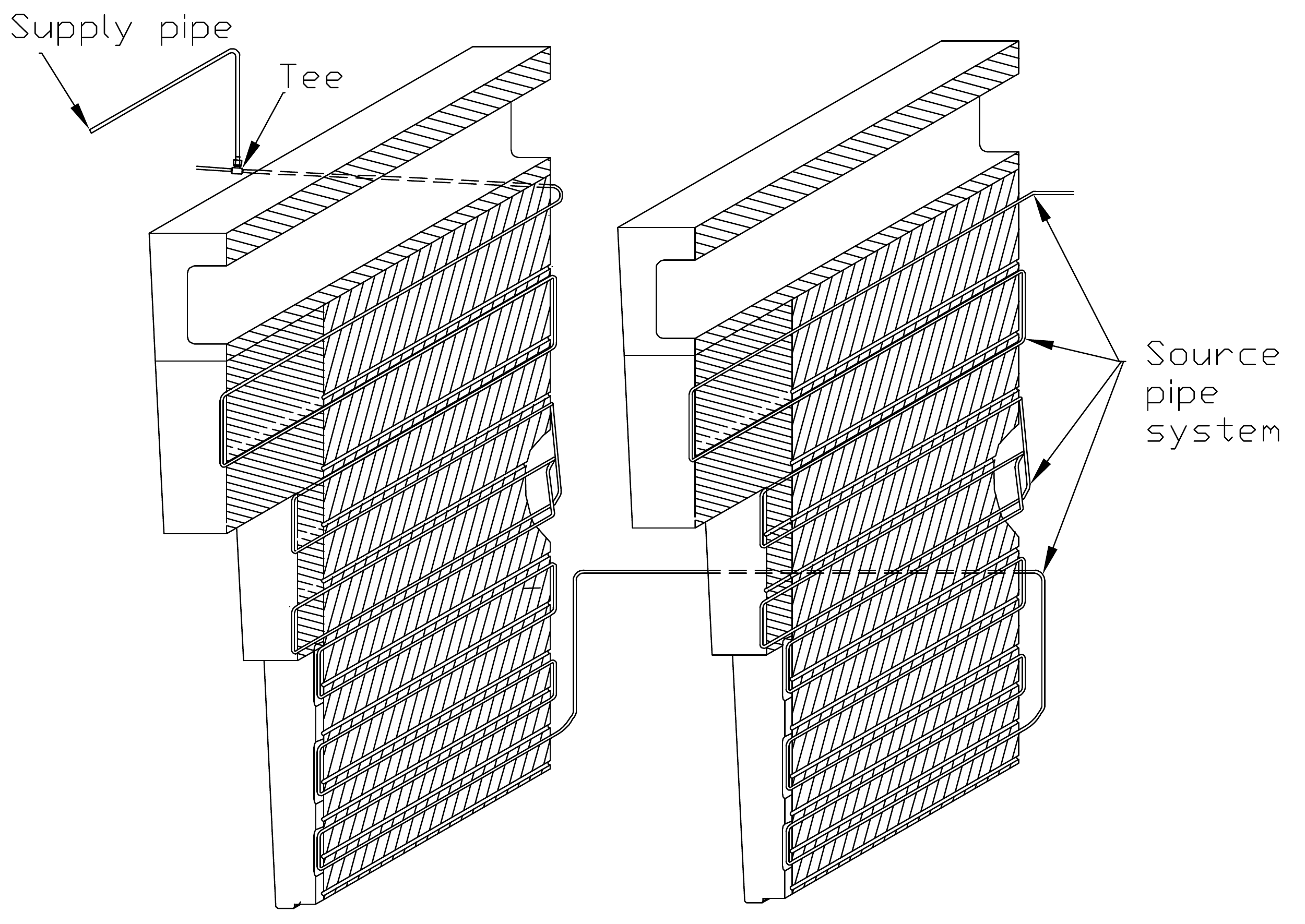}}
  \caption{Calibration tube layout in long barrel and extended barrel modules.}
  \label{fig:tubeslayout}
\end{figure}

The source circuit is realised with cold-rolled stainless steel tubes, \O6.0~mm~x~8.0~mm (ID x OD) consisting of straight, bent and interconnecting sections. The tube layout through a pair of neighbouring LB and EB modules is shown in figure~\ref{fig:tubeslayout}. The circuit through the LB is straightforward, while the one through the EB is slightly more complicated. 

This is because in an EB module both input and output tubes must be on the same side, and because of the peculiar geometry of the extended barrels at large radii also shown in figure~\ref{fig:tubeslayout}. Because of these constraints, the source passes twice through tile row \#7, using both holes. Therefore there are 11 straight tubes in an LB and 12 in an EB module. 

\subsection{Mechanical details}

\begin{figure}[!htbp]
 \begin{minipage}{0.5\textwidth}
  \includegraphics[width=\textwidth]{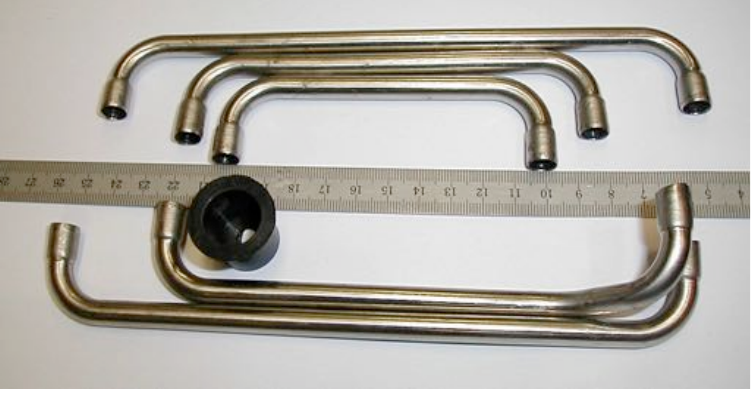}
  \caption{Examples of U-shaped tubes. The rubber cap is used to protect the calorimeter tiles from light, dust and glue during the interconnection of the tubes.}
  \label{fig:tubesbent}
  \end{minipage}
  \qquad
 \begin{minipage}{0.45\textwidth}
  \includegraphics[width=\textwidth]{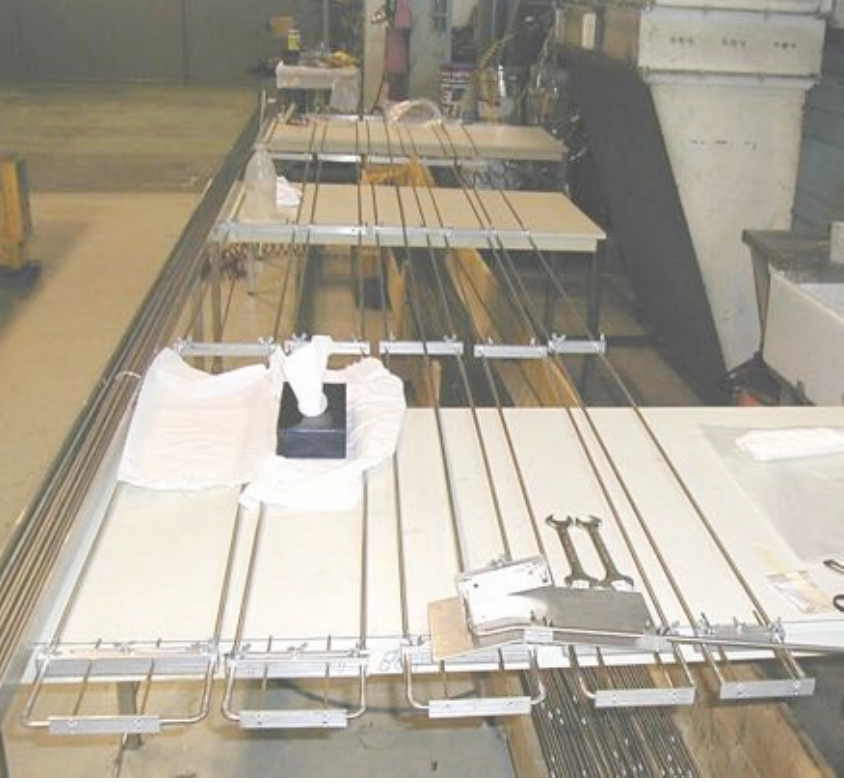}
  \caption{A few steps of the procedure to glue straight and  U-shaped tubes. }
  \label{fig:tubesglueing}
 \end{minipage}
\end{figure}

Specially made tubes, bent or U-shaped, interconnect the straight sections. These parts were bent with a technique that keeps the tube inner cross-section almost intact, thereby ensuring a smooth passage of the source (figure~\ref{fig:tubesbent}). The bent tube ends flare out for insertion into the straight tubes, and are joined with two-component Araldite 2011 epoxy-type glue. Glueing, rather than welding tubes together, is chosen in order to avoid damaging the plastic scintillating tiles during the assembly of the calibration circuit. The Araldite 2011 (former AW-106) glue is sufficiently radiation tolerant, at least up to 4 MGy~\cite{Epoxy}, while the maximum dose in TileCal after 10 years of LHC operation was not supposed to be more than 1 kGy.

All bent tubes have the same 15~mm radius of curvature, checked by several gauged dummy capsules just after fabrication, and at several stages of module instrumentation, in particular, after the glued joints are made. The dummy capsules have 6.00~mm outer diameter at fabrication and 5.85~mm at the final assembly stage. These dimensions guarantee passage of the 5.60~mm source capsule. Figure~\ref{fig:tubesglueing} shows some of the initial steps of the procedure of glueing together the different tube types.

More than 2200 U-shaped tubes of various lengths and configurations were used for 192 modules (and 2 spares), and more than 250 for module and source garage inter-connections. The total length of the source circuit is about 9~km for the three barrels: $\sim$4.3~km for LB and 2 times $\sim$2.4~km for EBA and EBC.

All 192 TileCal modules were delivered equipped with their calibration tubes at the optical instrumentation phase or just after it. At that time, a \Cs source was used to calibrate and certify them. Later on, the modules of each of the three sections (LB, EBA, EBC) were interconnected in situ, in the ATLAS cavern, creating three independent but functionally identical systems. Every step of tube insertion and interconnection was accompanied by cross-checks: calibrated dummy source passage, verification of geometry envelopes and air pressure tests up to 5~bar.

\begin{figure}[!htbp]
  \begin{minipage}{0.50\textwidth}
  \includegraphics[width=\textwidth]{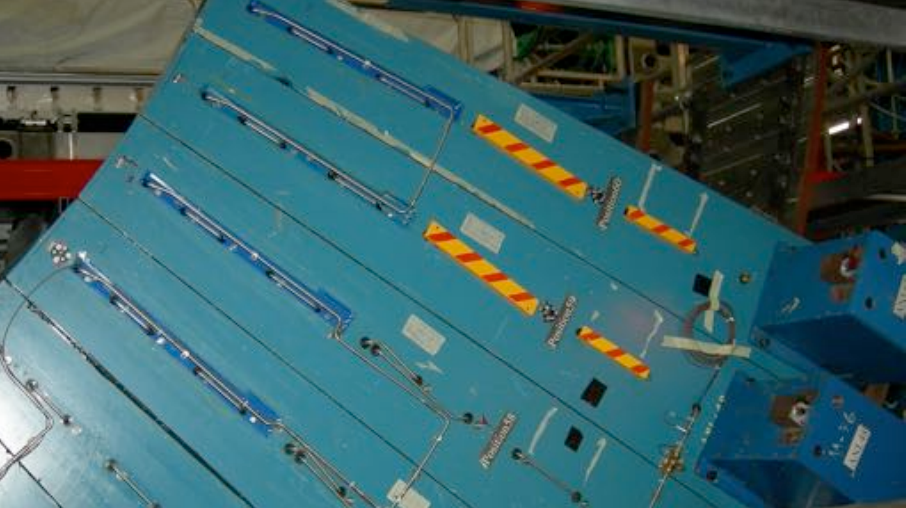}
  \caption{Layout of calibration tubes in a group of EBA modules during assembly in the cavern. One can see the U-shaped tubes connecting straight tubes, module inter-connecting tubes and tee-joints.}
  \label{fig:tubeseba}
  \end{minipage}
  \hfill
  \begin{minipage}{0.46\textwidth}
  \includegraphics[width=\textwidth]{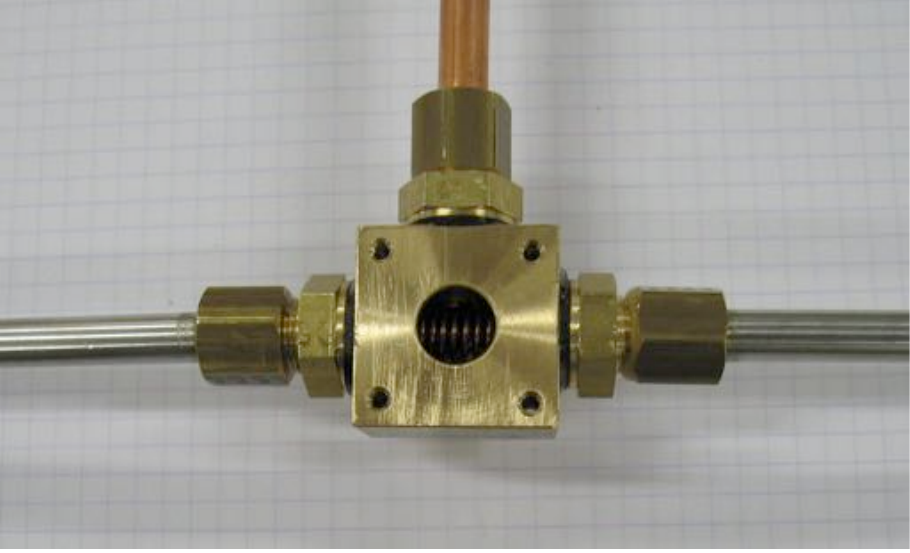}
  \caption{Tee-joint of calibration tubes (stainless steel) and supply pipe (copper). The brass spring guides the source capsule between calibration tubes.}
  \label{fig:teejoint}
  \end{minipage}
\end{figure}

The tube layout for a group of EBA modules during assembly in the cavern is shown in figure~\ref{fig:tubeseba}, in which one can see connections between straight tubes and between modules. 

\begin{figure}[!htbp]
  \centering
  \subfloat[][Long Barrel]{\includegraphics[width=0.50\textwidth]{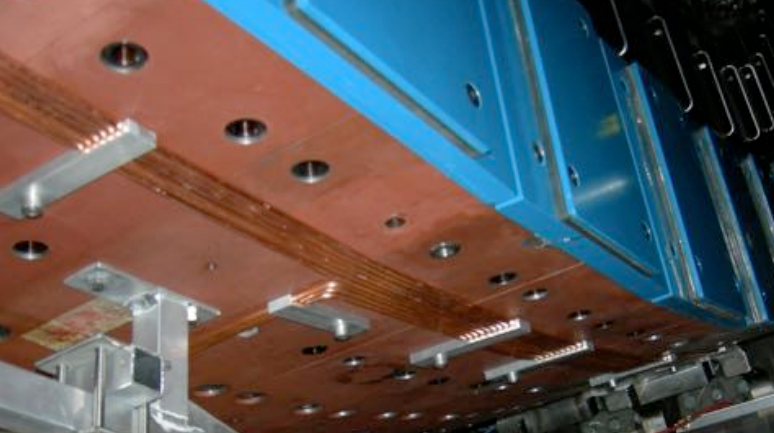}}
  \hfill
  \subfloat[][Extended Barrel]{\includegraphics[width=0.45\textwidth]{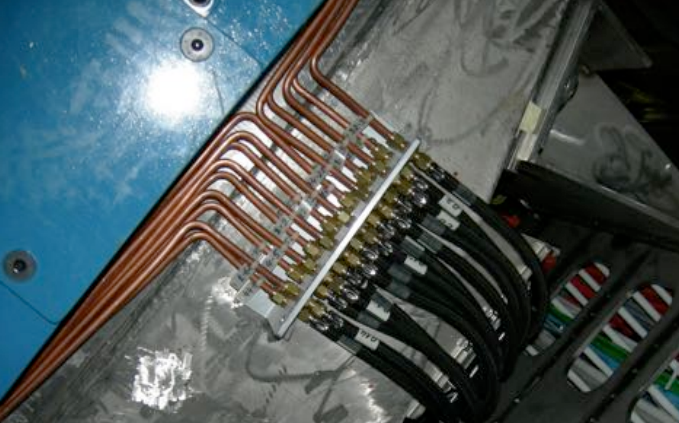}}
  \caption{Copper supply tubes at the bottom of the LB (a) and a patch panel of the EB, where copper tubes are connected to NBR pipes (b).}
  \label{fig:coppertubes}
\end{figure}

The rubber caps on the tubes serve as light seals. Part of the bent tubes lies in grooves machined in the module end-plates, in the area where the LAr cryostat supports, or its flange, are closest to the Tile calorimeter. 

To provide a free passage of the source capsule from one module to another and to connect the calibration circuit to a liquid supply pipe, a tee-joint is used. Figure~\ref{fig:teejoint} shows in detail the tee-joint with the top cover lid removed to show its internal structure. The lid also houses a pressure sensor, to be described later on together with other sensors.

There are 16 tee-joints in the LB barrel and 14 of the same design in each of the EBA and EBC, forming the corresponding number of circuit segments. All the tee-joints are located at the outer radius of the calorimeter, on the tubes that interconnect modules. 

Copper tubes of \O6.0~mm~x~8.0~mm (ID x OD) are used to carry the liquid from the drives to the calibration circuit at the detector's location. There are 16 copper supply tubes with lengths of 35 to 50 meters in the LB section and 14 in each of the EB sections. Flexible NBR (Nitrile Butadiene Rubber) of \O7.7~mm~x~13.5~mm (ID x OD) pipes, with a total length of 50 to 65 meters each (figure~\ref{fig:coppertubes}) connect to the EB tubes to allow moving the EB sections of TileCal when opening ATLAS. The total length of supply pipes is over 2~km for all three calibration subsystems.

The pneumatic part of the system consists of mechanical safety valves, electromagnetic valves built into the drives, and polyamide pipe (\O2.7~mm~x~4.0~mm ID~x~OD) distribution piping, interconnected with connectors made by LEGRIS.

The garage lock driving mechanism works with pressurised air, supplied via two pipes $\sim$70~m long. The same kind of pipes, a few hundred m long, supply air to the liquid storage vessels. Altogether the length of pneumatic pipes is close to 2~km. The garage locks are operated with pressures up to 5~bar, while liquid filling and draining are done at pressures under 2.6~bar. Both limits are controlled by calibrated safety valves.

\subsection{Sources and source capsules}
The Cs calibration and monitoring system uses several \Cs radioactive sources of the same design, from two different manufacturers (JINR, Dubna and Isotope Products, Prague). The sources are contained in stainless steel cylinders of outer dimensions of \O2.0~mm~x~9.0~mm, hermetically welded at both ends. The cylinder wall thickness is 0.2~mm, allowing an internal volume of nearly 10~mm$^{3}$ for the source itself. The \Cs isotope, obtained from the chemical purification of spent fuel rod arrays from nuclear reactors~\cite{Sources}, is sealed into a vitrified ceramic medium.

\begin{figure}[!htbp]
  \centering
  \subfloat[]{\includegraphics[width=0.48\textwidth]{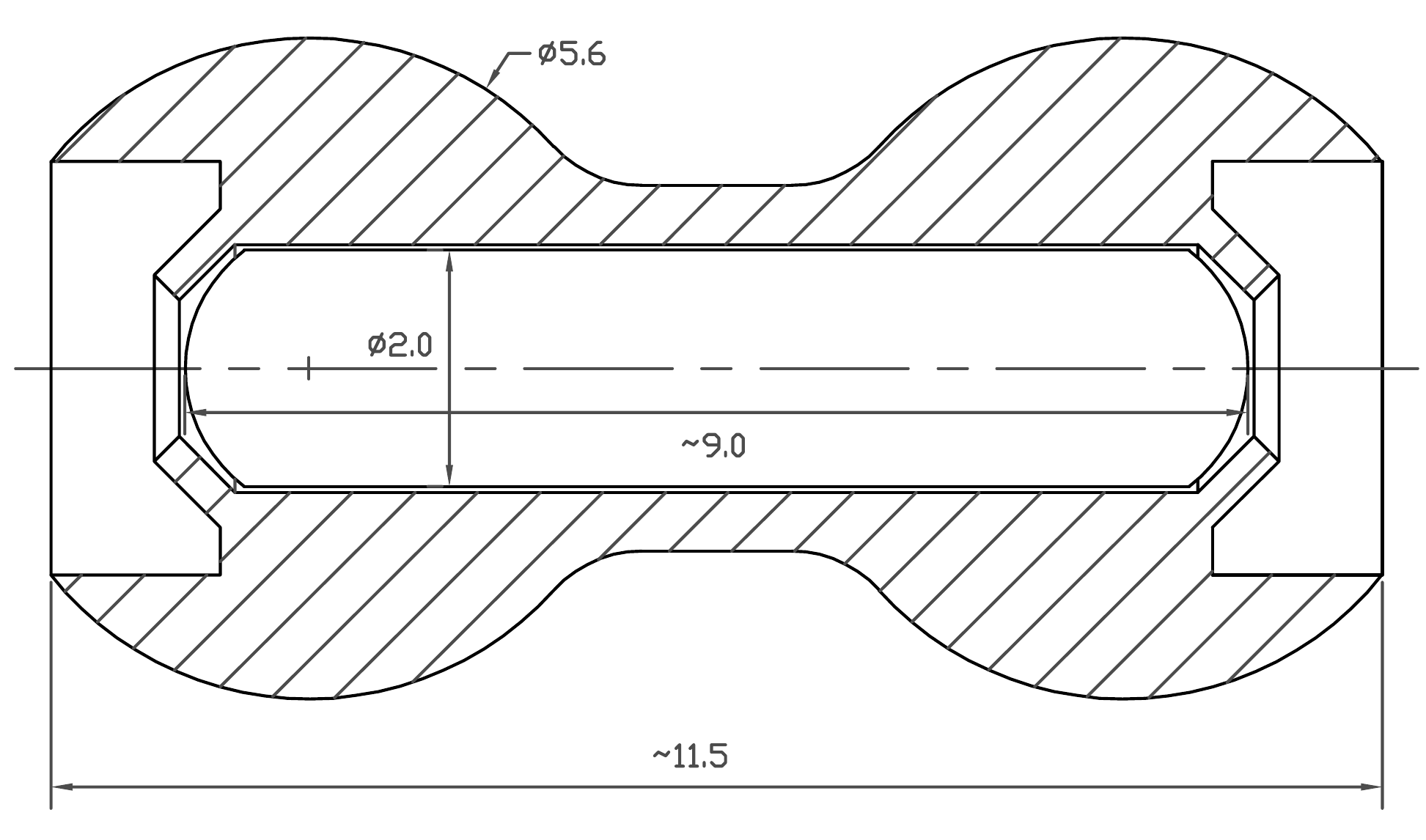}}
  \qquad
  \subfloat[]{\includegraphics[width=0.45\textwidth]{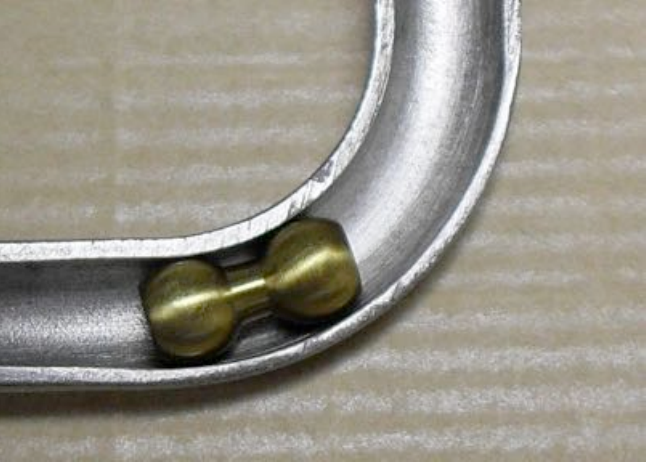}}
  \caption{(a) Source and protective capsule design. (b) Dumb-bell-shaped capsule in a bent segment of the calibration tubes. The capsule cross section is 5.60~mm, to be compared to the 6.0~mm inner diameter of the calibration tubes, leaving up to 0.4~mm clearance to the tube inner walls.}
  \label{fig:capsule}
\end{figure}

The source cylinders are mounted in dumb-bell-shaped capsules of a hardened aluminium alloy, plated with a 2 to 4 $\mu$m thick titanium nitride (TiN) film. Figure~\ref{fig:capsule} shows the design of the capsule. The diameter of the spherical end is 5.60~mm and the longitudinal dimension is slightly less than 12.0~mm. The source cylinder is fixed inside the capsule by crimping, using a specially designed tool. The shape and outside dimensions of the capsule allow it to be driven by the liquid flow through the bent tube sections with a radius of curvature down to 15~mm.

The capsules are not waterproof. They are designed for safe and reliable movement in the calibration circuit while protecting the sources from shock and friction during their travel, as well as from damage while they are handled. The titanium nitride coating decreases the friction with the stainless steel tube walls, thus improving the resistance of capsules to wear. The base material of the capsule is conductive, allowing it to be easily detected with inductive sensors. However, it is not chemically inert. Therefore precautions are taken to prevent corrosion of the capsule surface and of the source itself.

\begin{figure}[!htbp]
  \centering
  \subfloat[]{\includegraphics[width=0.3\textwidth]{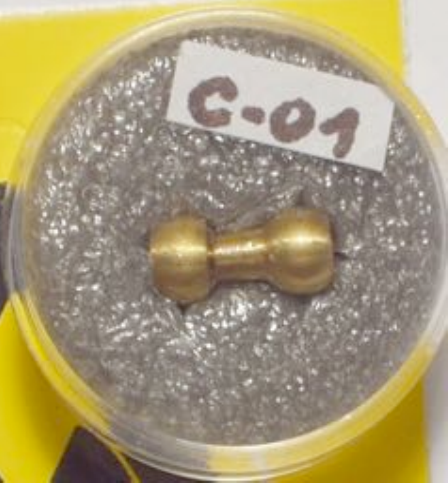}}
  \subfloat[]{\includegraphics[width=0.31\textwidth]{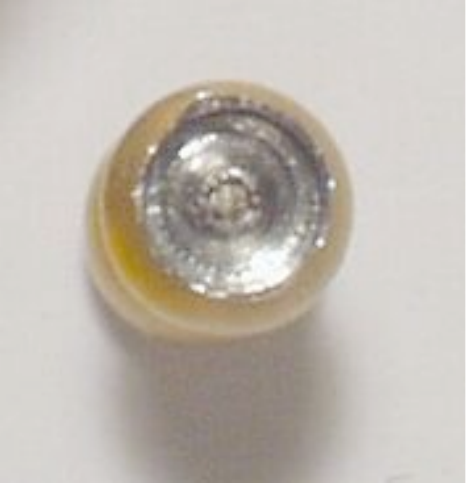}}
  \subfloat[]{\includegraphics[width=0.41\textwidth]{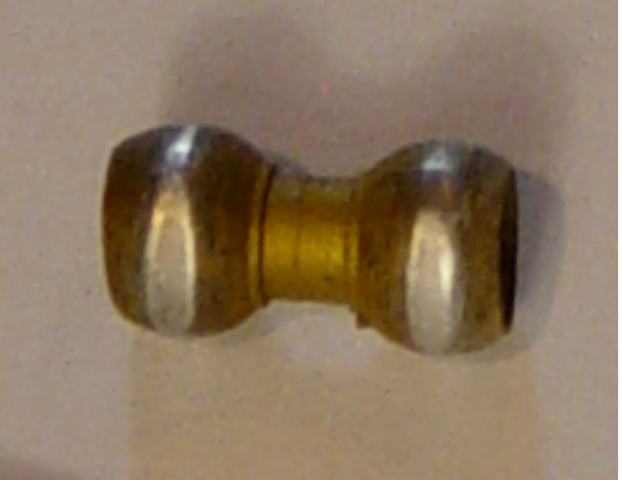}}
  \caption{From left to right: (a) a new encapsulated source; (b) the welded end of the source's stainless steel cylinder, after crimping it in the capsule; (c) a source capsule after passing through 250~km of calibration tubes over more than four years of use. The outer diameter was reduced from the initial 5.60~mm by less than 0.10~mm.}
  \label{fig:capsule2}
\end{figure}

In figure~\ref{fig:capsule2}, a new, unused source capsule is shown together with one that travelled more than 250~km through calibration tubes.

\begin{table}[!htbp]
 \centering
 \caption{\Cs sources used in the ATLAS Tile Calorimeter Cs calibration system.}
 \begin{tabular}{|c|c|c|c|c|}
 \hline
  Source & Estimated activity  & Measured response ratio & Produced at & Usage \\
  name   & (MBq $\pm$15\%)  & to 3713RP &  & \\
              & April 2009               & March 2009 &  & \\
 \hline
 3712RP & 319 & 1.2200 $\pm$ 0.0005 & JINR Dubna  & Instrumentation \\
 3713RP & 264 & 1.0000                        &                      & Test beam \\
 \hline
 RP4089 & 377 & 1.2180 $\pm$ 0.0007 & IP Prague & EBC monitoring \\
 RP4090 & 363 & 1.1590 $\pm$ 0.0005 &                  & EBA monitoring \\
 RP4091 & 372 & 1.1860 $\pm$ 0.0005 &                  & LB monitoring \\
 \hline
 \end{tabular}
 \label{tab:sources}
\end{table}

Five radioactive sources are used for TileCal response calibration. All had similar initial activities of 250 to 400~MBq, and were intercalibrated with several techniques, with a relative precision of about $5\times10^{-4}$. The nearly equal activity of these sources is convenient because it allows the responses of the three barrels to be the same dynamic range; in addition, switching the sources, if needed, requires fewer system changes. In addition to the usual activity measurements carried out by producers and the CERN Radiation Protection team, all sources were inter-calibrated in two campaigns (in 2009 and 2013) by multiple passes through the same modules (LB/EB). There were no differences between the obtained results of the two campaigns within the achieved accuracy, which suggests the same isotope composition of the sources. The source activities and the TileCal cell response ratios are listed in table~\ref{tab:sources}.

\begin{figure}[!htbp]
  \centering
  \includegraphics[width=0.7\textwidth]{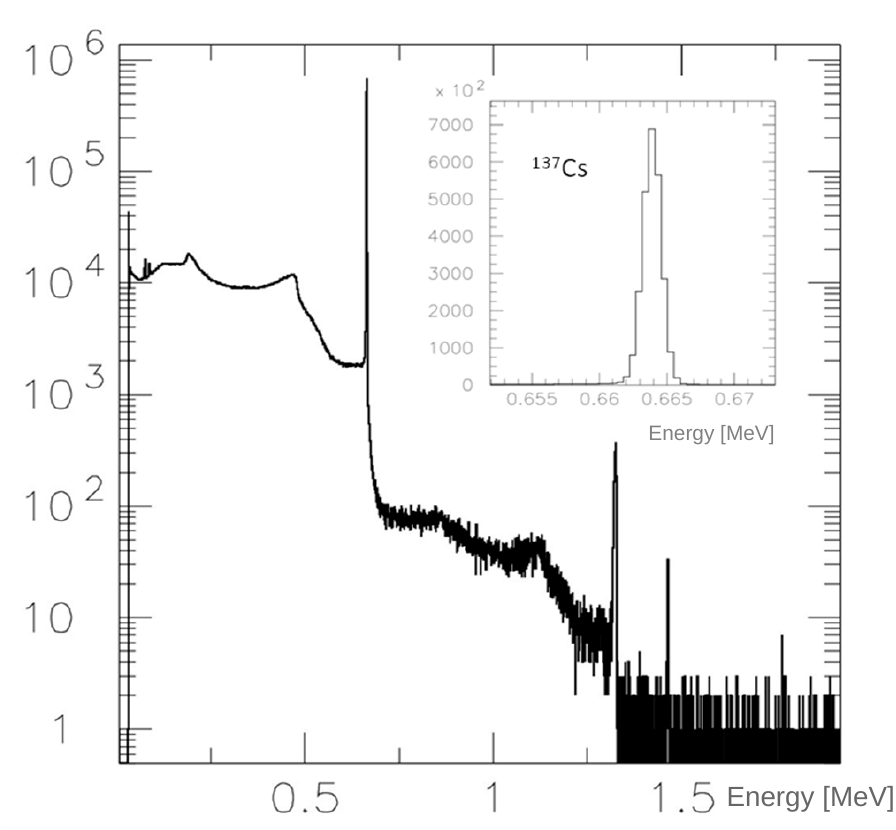}
  \caption{The energy spectrum of a \Cs source from data provided by the producer (Isotope Products), with zoom on the \Cs peak.}
  \label{fig:csspectrum}
\end{figure}

The typical energy spectrum of a source measured just after production is shown in figure~\ref{fig:csspectrum}. The \Cs photo-peak (0.662~MeV) dominates; any admixture of short-lived isotopes such as $^{134}\rm{Cs}$ (0.698~MeV peak energy, half-life~$\sim$2 years) would have been present at <$10^{-3}$ of the main isotope activity. During the recalibration campaigns, the sources' spectra were obtained and no differences were found between "old" (JINR) and "new" (IP) sources, and the purity of all the sources was manifested by the absence of other (short-lived) isotopes.

It is planned to re-encapsulate the sources currently used in ATLAS after the slow wear of capsules produces critical changes of their shape and size. After being used for about 5 years with no significant wear, the 3713RP and 3712RP sources were successfully re-encapsulated at Isotope Products in Prague, with no observed source material leakage.

Five years of use with regular monthly Cs scans in the ATLAS cavern produced no significant wear of these capsules. The sources are checked every few years during the planned long stops of the LHC machine, when ATLAS detector is open for maintenance.

\subsection{Source storage garages}

A ``garage'' is where a source resides between calibration runs. Nine garages, three for each subsystem (EBA, EBC, LB), were built and installed. The garages are uniformly distributed over the outer perimeter of the 64 modules in each TileCal section; they are integrated into the calibration tubes sequence of each subsystem. 

The garage lock driving mechanisms are operated with pressurised air, supplied via two $\sim$70~m long pipes, at the pressures mentioned in the previous subsection.

A source capsule can be loaded into a garage, extracted and moved in either direction, come to a stop in a garage coming from either side, kept there for any length of time and removed when needed (for instance, when ATLAS is opened). Functionally it is a detachable piece of the calibration tube circuit, radiation-shielded and equipped with two remotely operable locks and detecting sensors.

Each garage is supplied with a remotely operated electronic control card, that checks the status of the garage-locking mechanism, detects the presence of a metal capsule with a SIN sensor (described in section 4.2). On request, it can measure the capsule activity with a Geiger counter. The garage status is locally displayed with LEDs; this information is also sent to the control system either on request or every time the status changes.

Each source capsule normally resides in one of its three garages; the other two garages are usually unoccupied and are used as intermediate storage for the source during a normal scan or when a calorimeter scan has to be interrupted for some unforeseen reason, such as ATLAS operational priorities. At the normal speed of 35~cm/s, the source capsule travels between two garages (passing through 20--24 modules) in about 1 hour. This is known to be a reasonable time to decide whether or not to stop a full scan and let ATLAS proceed with its planned operations.

\begin{figure}[!htbp]
  \centering
  \subfloat[]{\includegraphics[width=0.31\textwidth]{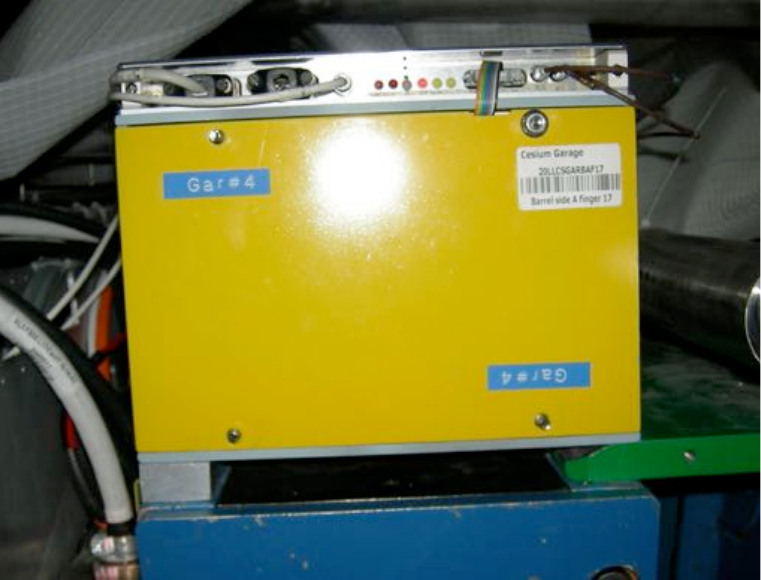}} \quad
  \subfloat[]{\includegraphics[width=0.31\textwidth]{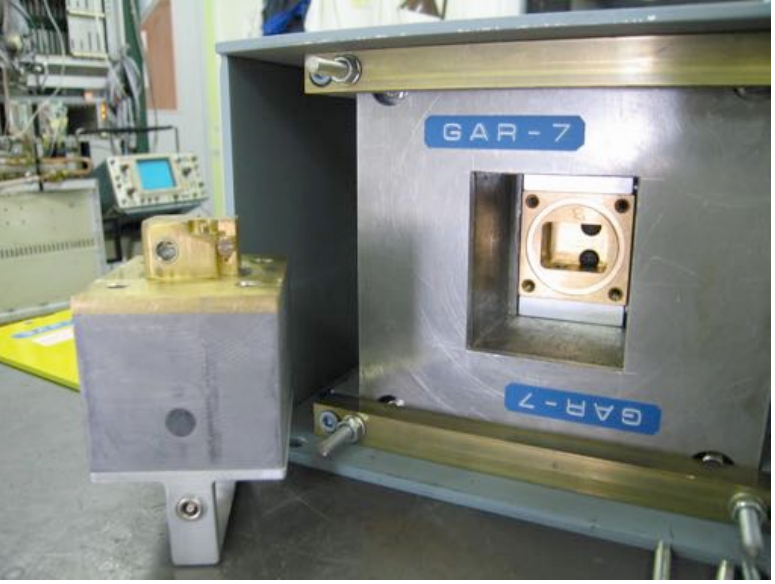}} \quad
  \subfloat[]{\includegraphics[width=0.31\textwidth]{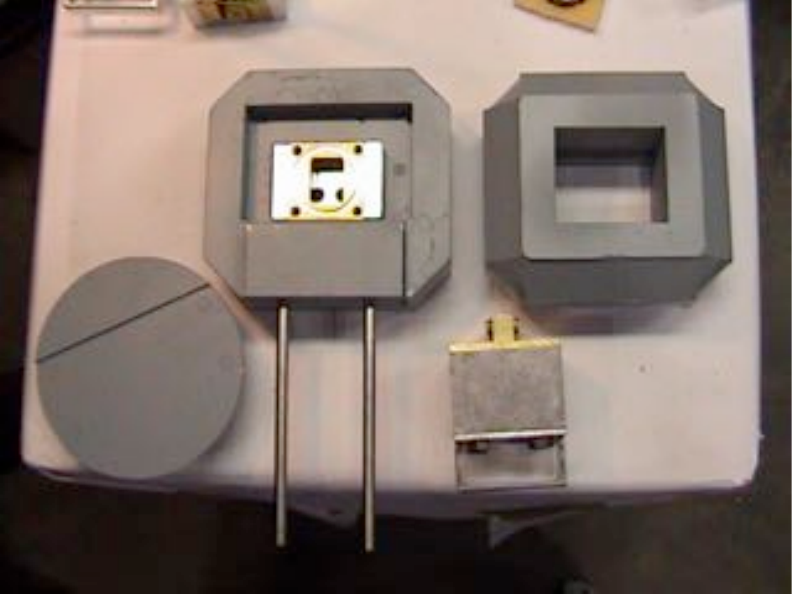}}
  \caption{Cs source garage: (a) Garage with the control unit, fixed at a module's outer periphery. (b) Garage case, body, detachable cassette. (c) Lead shielding components and lock mechanism.}
  \label{fig:garage}
\end{figure}

All the garages are identical and consist of an outer case, lead radiation shielding, the garage body itself with locks, driving pistons and a detachable cassette (figure~\ref{fig:garage}). The lead shielding thickness is 5~cm, and the dose rate at a 40~cm distance from the garage does not exceed 0.5~$\mu$Sv/h.

The source capsule is loaded into the cassette in a location providing an appropriate safety environment. The cassette lead shielding is the first level of operator protection; in addition, it is handled and transported to the garage location in a special container, specifically assigned to each cassette. The cassette installation into the garage body takes a few minutes, thereby limiting exposure of the operator to a safe dose. When the cassette is removed from the garage, it is replaced with a dummy one to close the calibration tubes circuit. The calibration pipe system must be drained and the garage volume must be reasonably dry before manipulating a cassette. 

\begin{figure}[!htbp]
  \centering
  \subfloat[]{\includegraphics[width=0.57\textwidth]{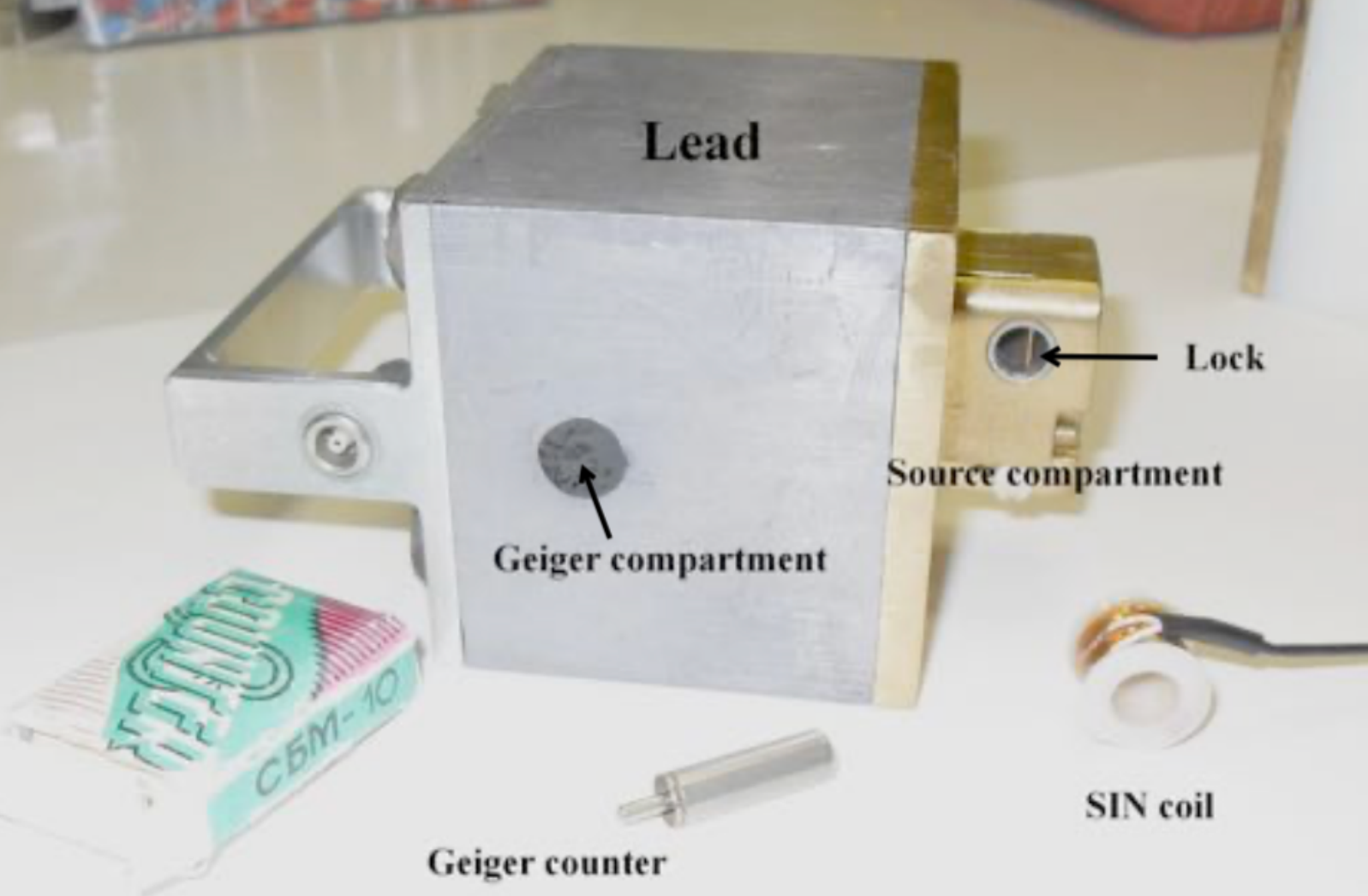}} \quad
  \subfloat[]{\includegraphics[width=0.37\textwidth]{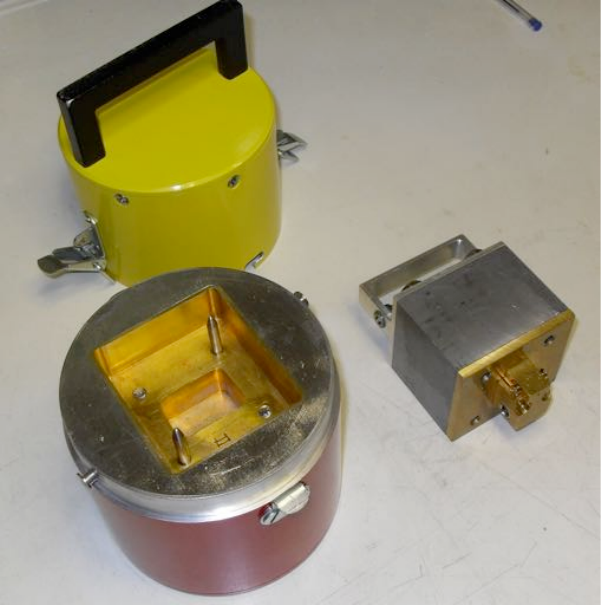}}
  \caption{ (a) Cs cassette and its components, showing the brass wire lock that keeps the source capsule in its compartment. (b) The transport container.}
  \label{fig:cassette}
\end{figure}

A source is locked in the cassette by brass wires, as shown in figure~\ref{fig:cassette}. These wires are moved by pistons operated with a 5~bar air pressure. The opening of locks allows the source capsule to move in the direction of the liquid flow. The pneumatically driven lock mechanism is insensitive to the ATLAS magnetic fields. The normal state of the locks is ``closed'', preventing unwanted capsule movements. If the capsule is inside the source compartment, it can exit only after the proper air pressure is applied and the liquid flow is provided.

There are two source location sensors in the cassette. An inductive sensor (SIN, see section 4.2) registers the presence of the metal capsule. The second is a Geiger counter that detects the presence and monitors the source's gamma activity. It is a Russian-made SBM-10 type Geiger counter, with a thin-walled metal case. The dimensions of the Geiger tube are \O6.0~mm~x~25.0~mm.  It is operated at an HV of 300--400~V and can take a rate of 700 pulses per second. Normally the counter HV is off; in order to increase the counter's expected lifetime, it is switched on only during source calibration runs, on request from the central control process.

To further decrease the rate load and thereby increase the Geiger counter's lifetime, the gamma-ray flux hitting it is reduced by embedding it in lead, within the cassette itself. With a 9~mCi \Cs source at a distance in lead to the detector of 25--35~mm, the measured counting rate is below 500~Hz.

The Geiger counter is operated under software control in either of two modes. Routinely, when the source reaches the garage from the calibration tube circuit, the presence of the source capsule, registered by SIN sensor, is confirmed by measuring the Geiger counter rate over a one-second interval. This measurement distinguishes a capsule containing the radioactive source from a dummy one. In the other operating mode, the activity of the source and general system operation are occasionally checked by taking sixty sequential one-second measurements. Over longer intervals, the Geiger counter data are an important check of the integrity of the source.

\section{Hydraulic system}

\subsection{Working liquid}

Metals of different properties are used in the system: regular steel, stainless steel, copper, copper-based alloys, aluminium alloys. As a consequence, care must be taken to avoid corrosion, especially of the sources and the capsules themselves. Oxidation would be very likely if the working liquid were plain or distilled water.

The dangers of corrosion are avoided by using as the working liquid a mixture of demineralised water (65\% by volume) and NALCO 00GE056 (later replaced by TRAC100) liquid (composition: water, disodium metasilicate 5--10\%, sodium molybdate 10--20\%, sodium tetraborate 1--5\%)~--- which is customarily used as a cooling agent in pipelines that include materials similar to those employed here. This liquid is particularly apt to suppress corrosion when different metals are present, is non-toxic and has low conductivity. It must be handled with protective equipment, because it may irritate skin and eyes.

\subsection{Liquid storage}

The total volume of the LB section's pipes is around 140~l, and slightly over 90~l for each EB. In practice, the total liquid volume circulated is 150--160~l in the LB, and about 100~l in EBA or EBC. Hence the total capacity of the storage system must be nearly 400~l; besides, this volume must be pumped in and out of this nontrivial pipe system. 

The height difference between the bottom of the ATLAS cavern and the top of the calorimeter is 18--20 meters, therefore the maximum top--bottom static pressure difference, taking into account the working liquid density ($\sim$1.15~g/cm$^3$) is about 2~bar. Pressurised air at 2.5--2.7~bar is used to inject the liquid into the piping system, or to return it into storage. 

\begin{figure}[!htbp]
  \centering
  \includegraphics[width=0.54\textwidth]{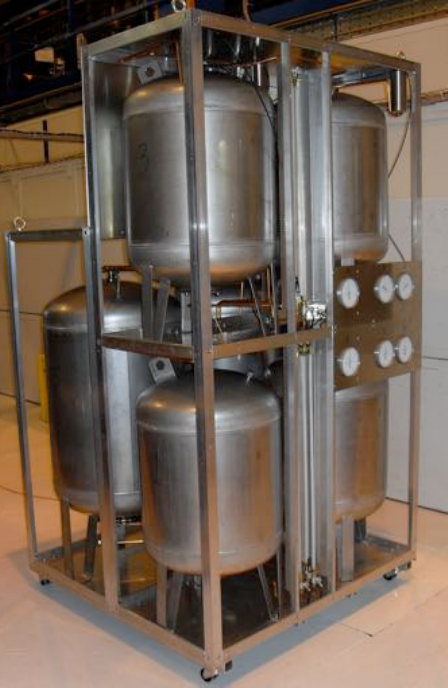}
  \caption{Liquid storage unit, made of two 230~l and four 130~l stainless steel tanks combined in pairs, for the liquid and  air volumes.}
  \label{fig:waterstation}
\end{figure}

The storage system (figure~\ref{fig:waterstation}) (water station, WS) consists of stainless steel tanks, one pair of 230~l for LB and two pairs of 130~l for EBA and EBC. For each couple of tanks, one is used to store the liquid and the other to store the air displaced during the filling operation. This air is eventually released into the atmosphere, carrying vapours which are absorbed by external filters. The volume of each system tank is slightly greater than the volume of the corresponding system piping.

\subsection{Hydraulic drive and control}

Three pumping units (``hydro-drives''), operated via 3U control crates located in the ATLAS cavern, attend to each calorimeter section. Two additional drives, one for further R\&D and a spare, are available. All drives are identical and exchangeable after appropriate configuration tuning.

The main purpose of the hydro-drive is to fill the piping system, to produce a stable and controllable flow of liquid in the appropriate contour -- be it in a pre-programmed or in an {\it ad-hoc} configuration -- and to drain the system. The maximum number of supported contours is 16, with a pressure difference of up to 4~bar at the drive in/outlets.

An additional task of the drive is to provide controllable pressurised air to operate and manipulate up to 6 garage locks and two liquid (or gas) storage units synchronised with the current operation procedure. All drive operations can be performed remotely as well as manually, using dual control features of the 3U crate.

\begin{figure}[!htbp]
  \centering
  \subfloat[]{\includegraphics[width=0.37\textwidth]{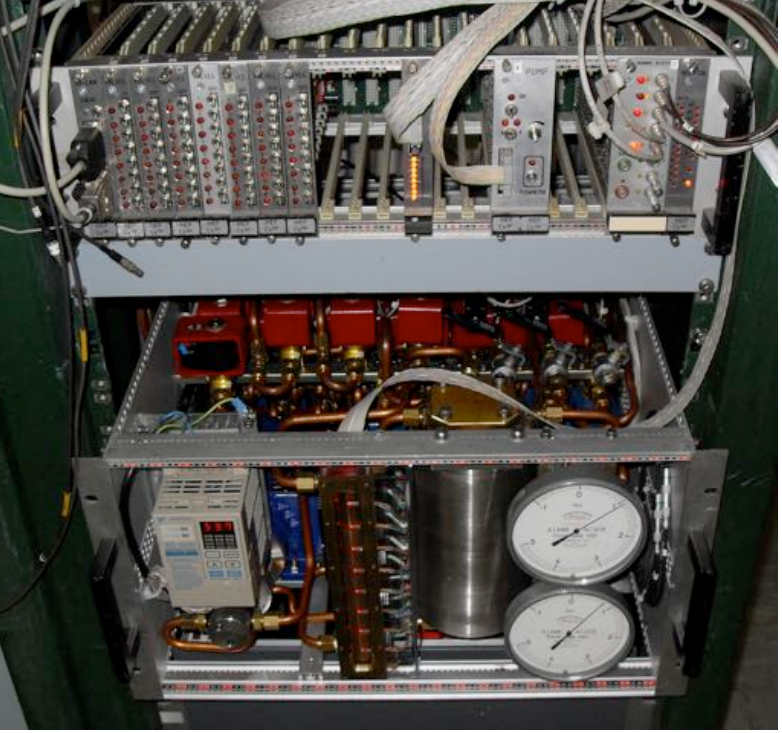}} \quad
  \subfloat[]{\includegraphics[width=0.6\textwidth]{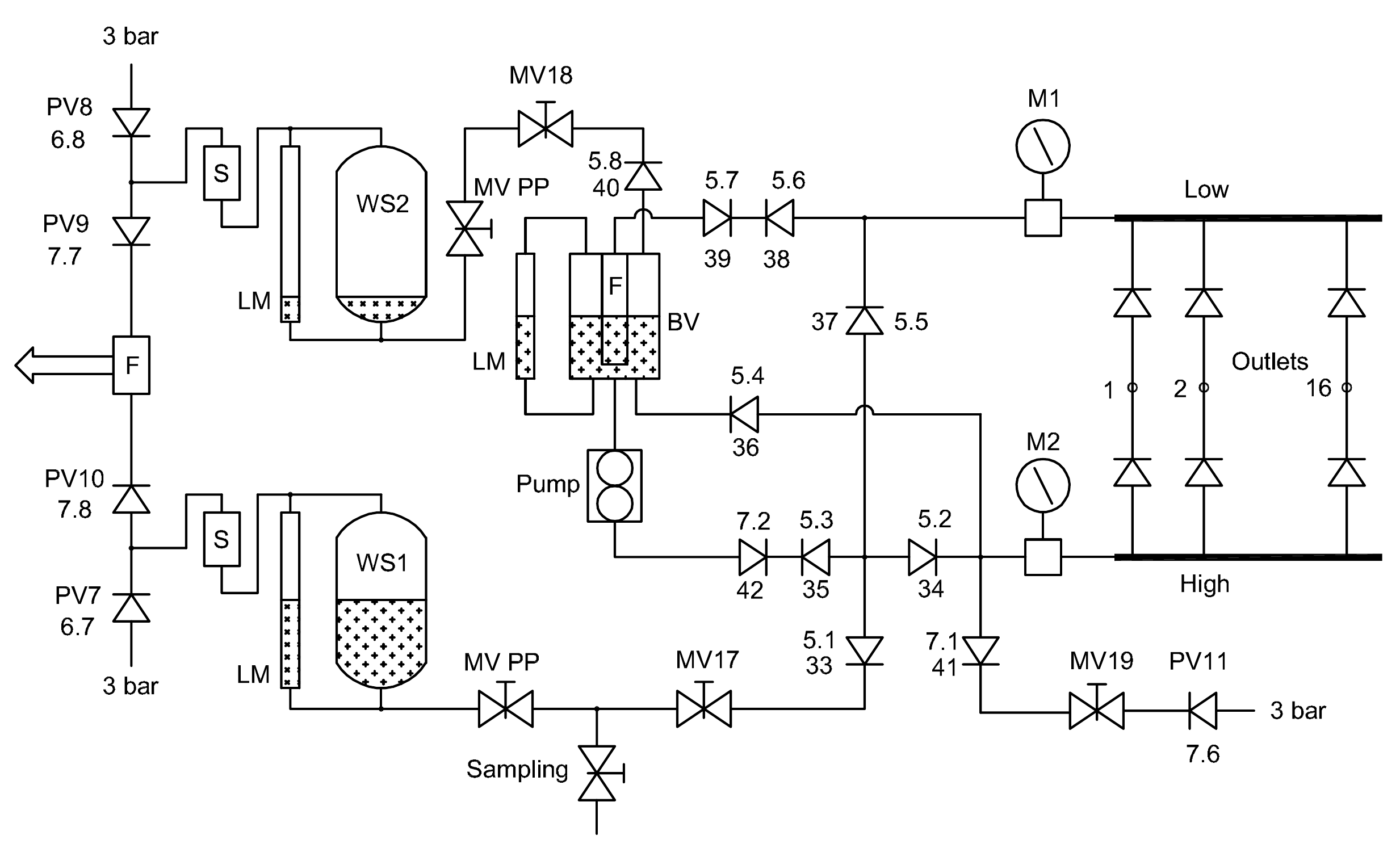}}
  \caption{(a) Hydro-drive with its control crate. (b) Hydro-drive schematics.}
  \label{fig:hydrodrive}
\end{figure}

Each drive unit (figure~\ref{fig:hydrodrive}) includes a magnetic gear pump (IWAKI  MDG-M2), a frequency-varying drive (YASKAWA VS mini C CIMR-XCACB) used as a controlled 200-watt power supply for the pump, 42 hydraulic (LUCIFER 121K01) and 11 pneumatic (LUCIFER 131M14) electromagnetic valves, a 1.8~litre buffer vessel with a level meter, pressure sensors and manometers;  additionally, a number of manual valves, tubes, pipes, cables, connectors, filters, etc. The drive occupies one full 6U euro-crate, weighs about 30~kg and operates at pressures of up to 5~bar.

The 3U control crate is equipped with a special-purpose local bus and contains a number of dedicated modules, which can be operated either remotely or manually. All the 3U crates and their modules are interchangeable. The main modules are listed here:
\begin{itemize}
\item A CAN 3U interface provides the communication between the 3U modules and the remote CPU, using a CAN bus interface and the 3U local special-purpose bus;
\item Up to 8 electromagnetic valve drives with 8 channels each. Both the hydraulic and pneumatic valves are opened with 24~V DC supply; the valve status is available at all times;
\item The pump drive itself, controlling the desired frequency and the power of the YASKAWA drive, hence the rotation velocity of the magnetic gear pump with the performance appropriate to providing the desired flow in the chosen contour;
\item The level meter control and display, which monitors changes of the amount of liquid quantity in the entire volume with an accuracy better than 200~ml;
\item The status display of the 3U-crate local bus, used for debugging purposes.
\end{itemize}

To drive the source capsule through the 6~mm inner-diameter tube with a steady velocity of about 35~cm/s, the liquid flow in the desired circuit section has to be about 10~cm$^{3}$/s. With these parameters the pressure drop in one LB module, containing $\sim$65 meters of calibration tubes, is 0.2~bar, and in one EB module ($\sim$35 meters of calibration tubes) is 0.15~bar. The number of modules in a contour varies from 4 to 6, therefore the applied pressure difference (positive or negative) to a contour has to be about or slightly over one bar above the local static pressure (-0.2+0.6~bar).

The hydro-drive is capable of pumping liquid with a finely controllable flow in the range of 5--30~cm$^{3}$/s, while providing a pressure difference of up to 4~bar. This pressure range corresponds to a capsule speed of 10--50~cm/s. The typical speed during early system tests was 25--30~cm/s, while currently, it is routinely 35~cm/s in all three calorimeter sections. The hydro-drive reaction time when run conditions change is adequate to the system piping, as designed.

\section{Sensors}

Several more types of sensors, together with the associated electronics, described in detail in Ref.~\cite{Electronics}, are used in the system to remotely control it either when idle or during operation. 

\subsection{Pressure sensors}

\begin{figure}[!htbp]
  \centering
  \subfloat[]{\includegraphics[width=0.50\textwidth]{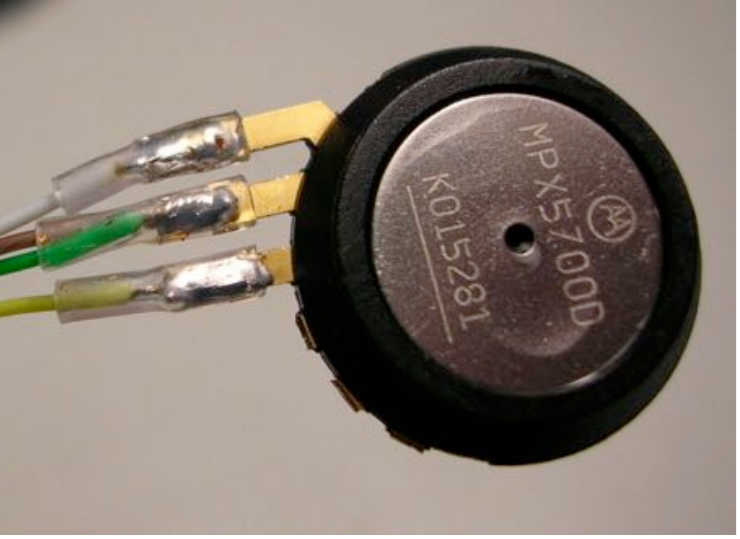}} \quad
  \subfloat[]{\includegraphics[width=0.43\textwidth]{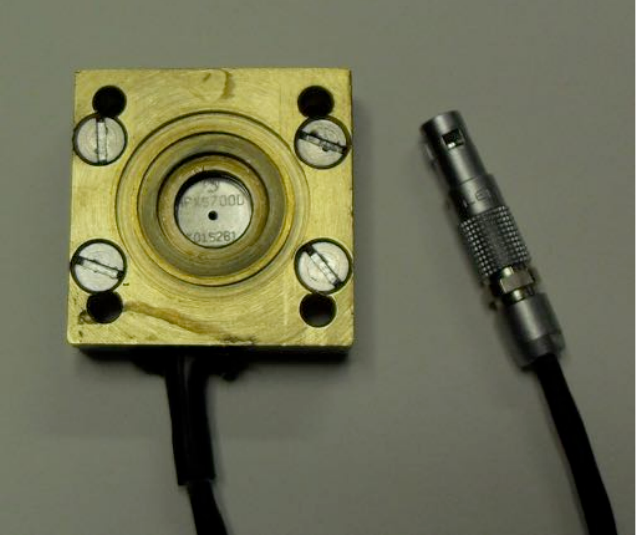}}
  \caption{Pressure sensor element MPX5700 (a) and its case (b). The main locations of the sensors are tee-joints, pumping units (hydro-drives) and liquid storage units (WS).}
  \label{fig:pressuresensor}
\end{figure}

The pressure sensors (PS, figure~\ref{fig:pressuresensor}) control the pressure in the calibration tubes circuit and liquid and gaseous supply vessels. More than 60 points are measured in the system, over the design range of~-1.0 to +5.0~bar). 

The PS elements are Motorola integrated monolithic silicon devices MPX5700D and MPX5700A. They are based on a piezo-resistive transducer on-chip signal, which is conditioned, temperature-compensated and pre-calibrated by the manufacturer with an accuracy of about 2.5\%. A silicon-rubber cover protects the integrated circuit from the working fluids.

Sixteen pressure sensors are installed on LB tees and 28 on EBA \& EBC tees in total. In addition, 15 sensors are part of the liquid pumping and storage equipment, giving a total of 59 units located in the cavern. 

\subsection{Inductive capsule sensor}

\begin{figure}[!htbp]
  \centering
  \subfloat[]{\includegraphics[width=0.40\textwidth]{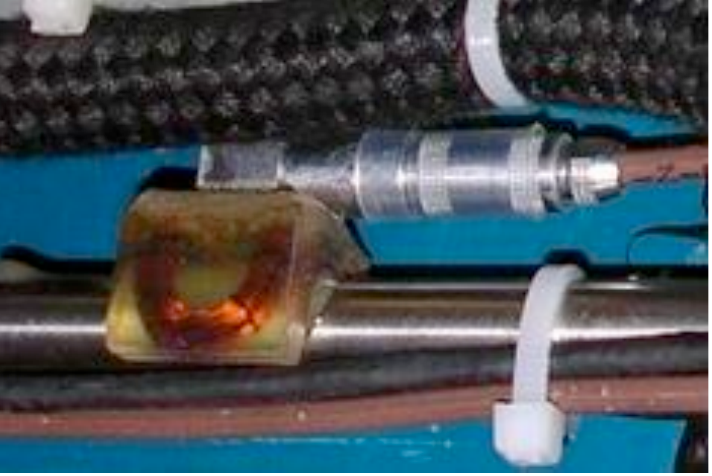}} \quad
  \subfloat[]{\includegraphics[width=0.53\textwidth]{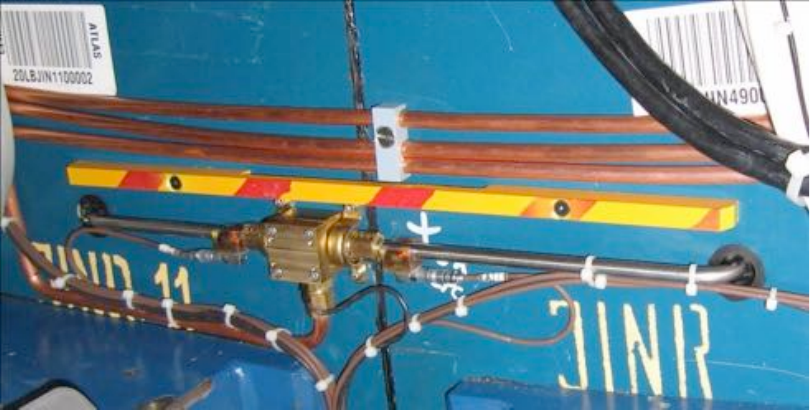}}
  \caption{SIN sensor installed on a section of the calibration tube (a) or around the tee connection (b).}
  \label{fig:sinsensor}
\end{figure}

The inductance sensor (SIN, figure~\ref{fig:sinsensor}) is designed to register the passage of the conductive body of a capsule. It is a continuously powered LC circuit in which the inductive element is a coil wound around a tube of the source travel circuit. The frequency of the generated electromagnetic flux is shifted by the change in inductance due to the capsule's conductive body. The frequency shift, hence the presence of the capsule, is detected by conventional electronics.

The SIN signal indicates to the control process the passage of the capsule from one module to the next, and when to switch the liquid flow to the next contour. Typically one SIN is located at the entrance of the liquid flow into a module's entrance and another one at the exit. In order to ensure the knowledge of the capsule location at all times, additional SINs are mounted at inaccessible zones of the calibration tube circuit, for instance under the Liquid Argon calorimeter cryostat's supports and flanges. More SINs signal the entrance and exit of the capsules from garages. 

SIN data allow to control online the capsule velocity, and to tune the pump speed, thereby steadying the movement of the capsule while the flow within any contour is varied. Altogether almost 500 SINs are used in the three calorimeter sections.

\begin{figure}[!htbp]
  \centering
  \subfloat[]{\includegraphics[width=0.48\textwidth]{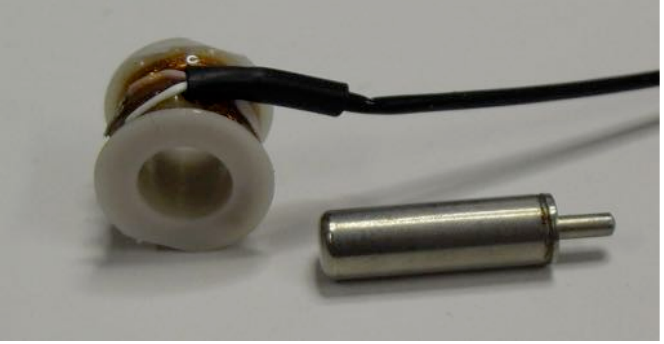}} \hfill
  \subfloat[]{\includegraphics[width=0.44\textwidth]{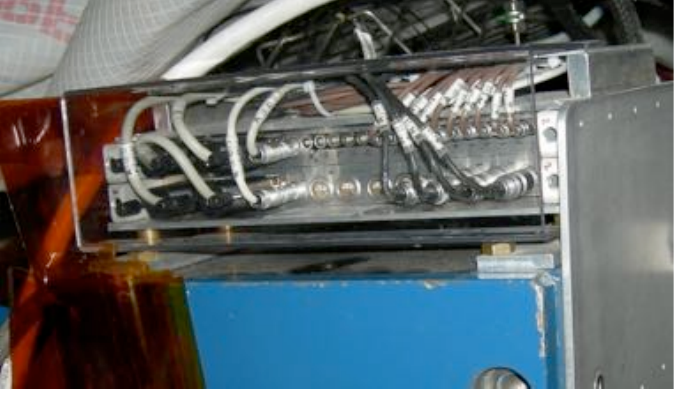}}
  \caption{(a) SIN sensor coil and Geiger counter used in a garage. (b) Mini-crate with SIN\_CAN (top) and ADC\_CAN (bottom) FEE units.}
  \label{fig:minicrate}
\end{figure}

SINs also register the presence of a source capsule in a garage (figure~\ref{fig:minicrate}a). In this case, the sensor coil has a slightly different shape but measure a frequency shift just like the others. The only difference is in the data treatment: while a conventional SIN only registers the passing of a capsule, the garage SINs are tuned to sense the change in conductivity due to the presence of a capsule. The corresponding parameters are saved in the memory of the appropriate garage module.

The front-end electronics (FEE) modules that collect and treat the pressure and SIN sensor data are located at the periphery of calorimeter modules and are contained in custom-made mini-crates (figure~\ref{fig:minicrate}b). One ADC\_CAN unit can handle up to 8 pressure sensors; one SIN\_CAN unit supports up to 16 sensor channels. Altogether, about 12 ADC\_CAN modules and 40 SIN\_CAN modules are installed. All data are transmitted to the central CPU under the CAN bus protocol. A more detailed description of the system's sensors and electronic modules is given in Ref.~\cite{Electronics}.

\subsection{Liquid level meter}

While being pumped, the propelling liquid passes through a buffer vessel in order to eliminate bubbles, which otherwise might create problems such as letting the pump dry up or losing control of the source movement. Continuous measurement of the liquid level in the buffer vessel (BV) (see figure~\ref{fig:hydrodrive}) allows detecting the presence of bubbles, especially while filling the system with liquid -- the time when bubbles are likely to be produced -- and thereby avoiding problems due to their presence.

To control the liquid flow during a Cs scan, the drive buffer vessel (BV) level has to be controlled in the presence of a magnetic field of up to 50~Gauss from ATLAS Toroid magnet.

\begin{figure}[!htbp]
  \centering
  \subfloat[]{\includegraphics[width=0.26\textwidth]{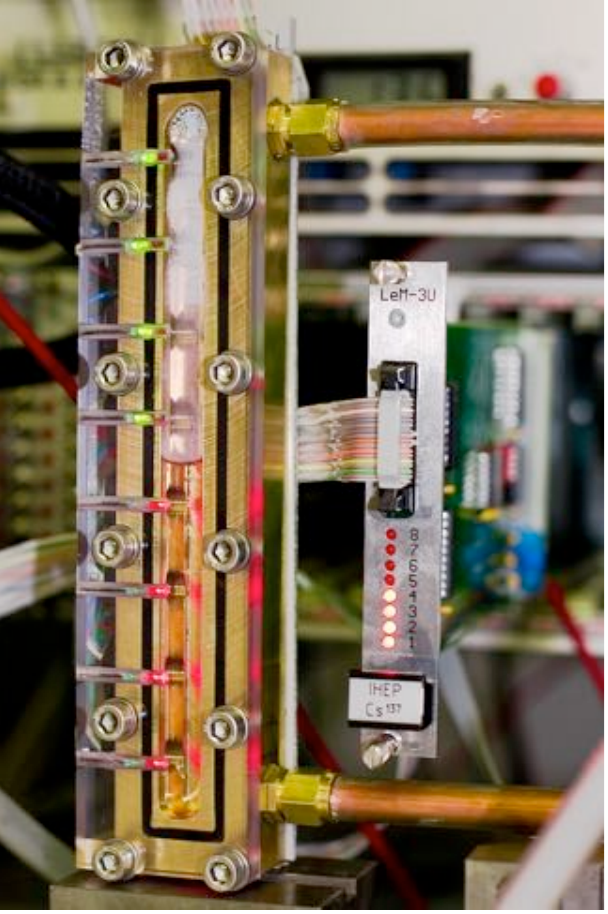}}\quad
  \subfloat[]{\includegraphics[width=0.70\textwidth]{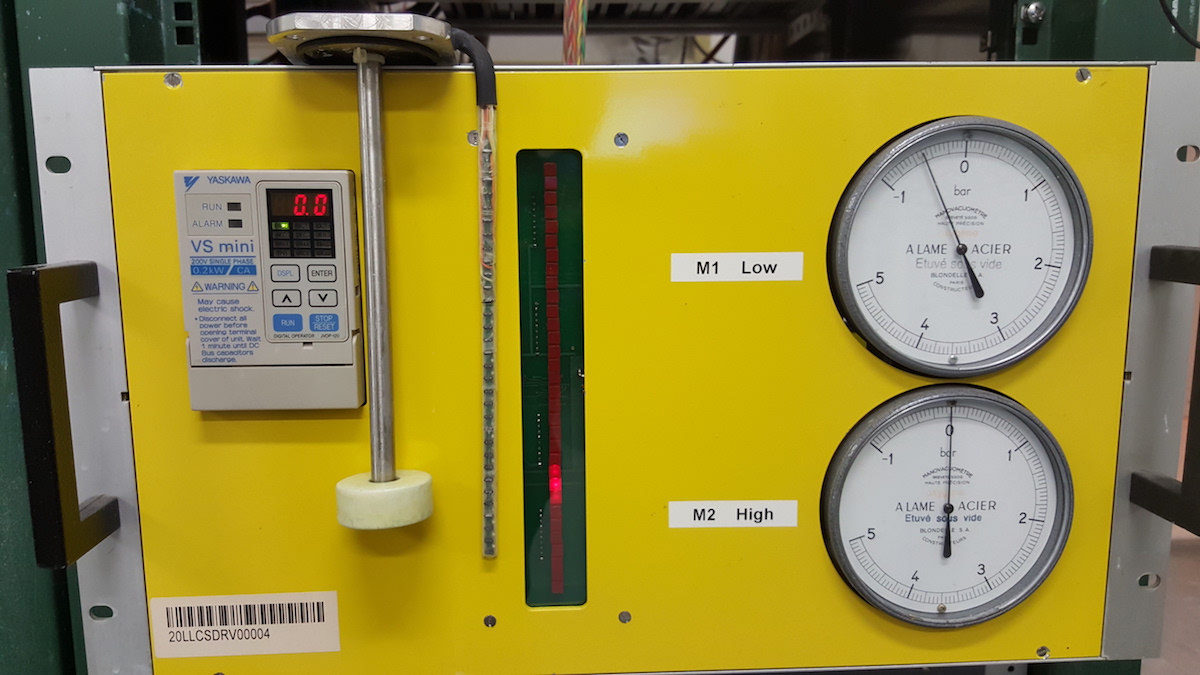}}
  \caption{(a) The buffer vessel's optical level meter with its control module. (b) The upgraded Hall-sensor level meter --- the float, the line of hall sensors and the display.}
  \label{fig:levelmeter}
\end{figure}

Originally, the level meter (LM) was based on optical measurements. A small test volume, directly communicating with the buffer vessel, was equipped with eight infrared optical sensors (Honeywell LLE101000). In these devices, an infrared light beam is transmitted through the outer lens surface when immersed in the liquid, or is reflected back and detected when not. The eight sensors, vertically arranged in the test volume, provided a coarse but sufficiently informative measurement of the level of the liquid. An assembled level meter with its 3U control unit is shown in figure~\ref{fig:levelmeter}a. Eight LEDs turned red or green depending on their location above or below the liquid level in the LM and changed colour whenever the liquid volume in the buffer vessel changed by about 200~ml. The control unit reflected the LED display state and sent digital information of the current level meter status to the central CPU, via CAN bus. The level meter worked fine, however, the precision was not enough to quickly detect possible liquid level changes and the optical sensors occasionally returned false readings.

To overcome these issues, the level meter was upgraded later on to a version sensitive to changes  of less than 25~ml, based on Hall effect sensors and a magnetic float. A narrow PCB of about 170~mm length with 32 Hall effect sensors (A1301EUA-T from Allegro MicroSystems) spaced by 5.5~mm is inserted into a stainless steel tube, equipped with a circular float with 4 embedded magnets. The sensors are read-out by the board with a micro-controller (STM32F205) that translates the readings of individual sensors into the liquid level, separating the signal by fitting the background. The results of the measurements, together with the raw readings are transferred into the 3U display card via SPI bus. For convenience, the float level is also displayed in an LED strip of the drive unit. The data from several sensors around the float position allows to subtract the background from the local environment's magnetic field and to improve the resolution of the level measurement beyond the step between individual hall sensors (20~ml vs 50~ml). The Hall effect sensor level meter float and the sensor board are shown in figure~\ref{fig:levelmeter}b.

\subsection{Liquid radioactivity monitors}

\begin{figure}[!htbp]
  \centering
  \subfloat[]{\includegraphics[width=0.25\textwidth]{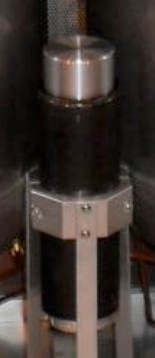}} \hfill
  \subfloat[]{\includegraphics[width=0.64\textwidth]{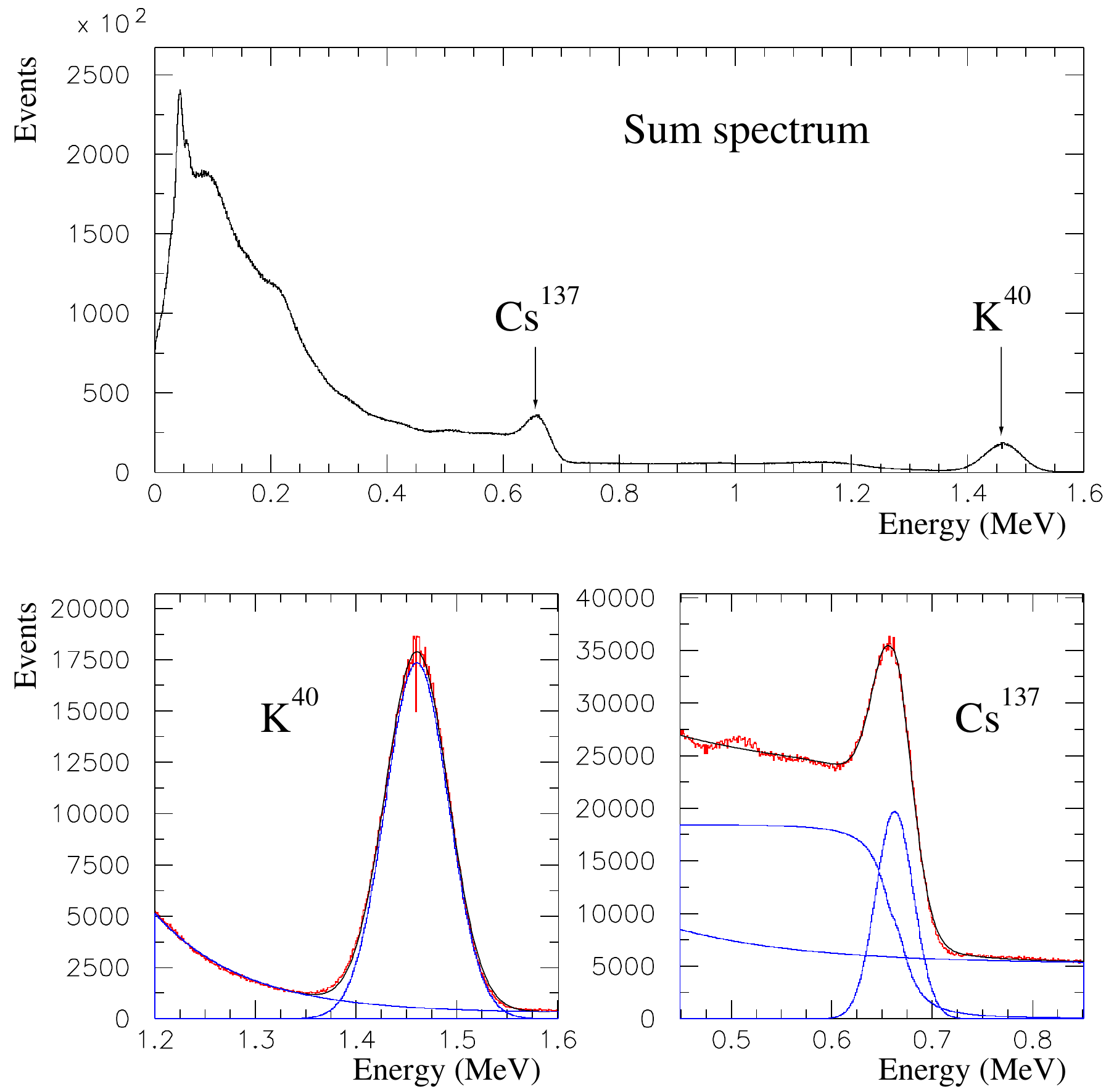}}
  \caption{(a) NaI(Tl) detector installed in the liquid storage system. (b) Energy spectrum measured in a test bench, with \Cs source in the vicinity, with zoom on the peaks showing the fit curves. The energy reference is obtained by the 1.46~MeV~$\gamma$-ray peak from $^{40}$K from the fertiliser in the NaI(Tl) detector assembly.}
  \label{fig:radmon}
\end{figure}

As a part of the radiation safety measures implemented in the Cs source system, an early warning of any leak from the Cs sources into the working liquid is provided by instrumentation that detects any unexpected radiation in the working liquid. 
Two radiation detectors installed in the liquid storage system (figure~\ref{fig:radmon}) record the gamma-ray energy spectra of the stored working liquid. One unit monitors the LB storage tank, the other the EBA and EBC. 

A \O2x2~inch NaI(Tl) crystal (\O50.8~mm with 50.8~mm thickness) produced by Bicron (Saint-Gobain), coupled with an ETI 9266B PMT is used as a gamma-ray total absorption detector. The signals are treated by a standard electronic chain consisting of a multiplexer, a CAEN N968 preamplifier and a CAEN N957 analyser. The PMT high voltage is provided by a standard CAEN N470 HV power supply. The analyser runs in self-triggering mode and the data are read out via USB.

The naturally occurring $\gamma$-ray signal at 1.46~MeV from $^{40}$K peak, obtained from a small sample of K$_2$O from commercial fertiliser, permanently deployed in front of the NaI(Tl) crystal, is used to calibrate the radiation monitor's energy scale. In figure~\ref{fig:radmon}, a $\gamma$-spectrum with the two peaks is shown, when the detector assembly was put in the vicinity of the \Cs source on the test bench.

\clearpage

In addition to online monitoring by the early warning system, periodic chemical analyses of the system liquids are carried out every three months by the CERN Radiation Protection group. No significant radioactive contamination of the drive liquid was detected by either method over the decade-long operation of the Cs monitoring system. Because of the lack of a signal, a precise determination of the sensitivity of the early warning system is not yet available.

\section{DAQ and online software}

\subsection{DAQ architecture}

\begin{figure}[!htbp]
  \centering
  \includegraphics[width=\textwidth]{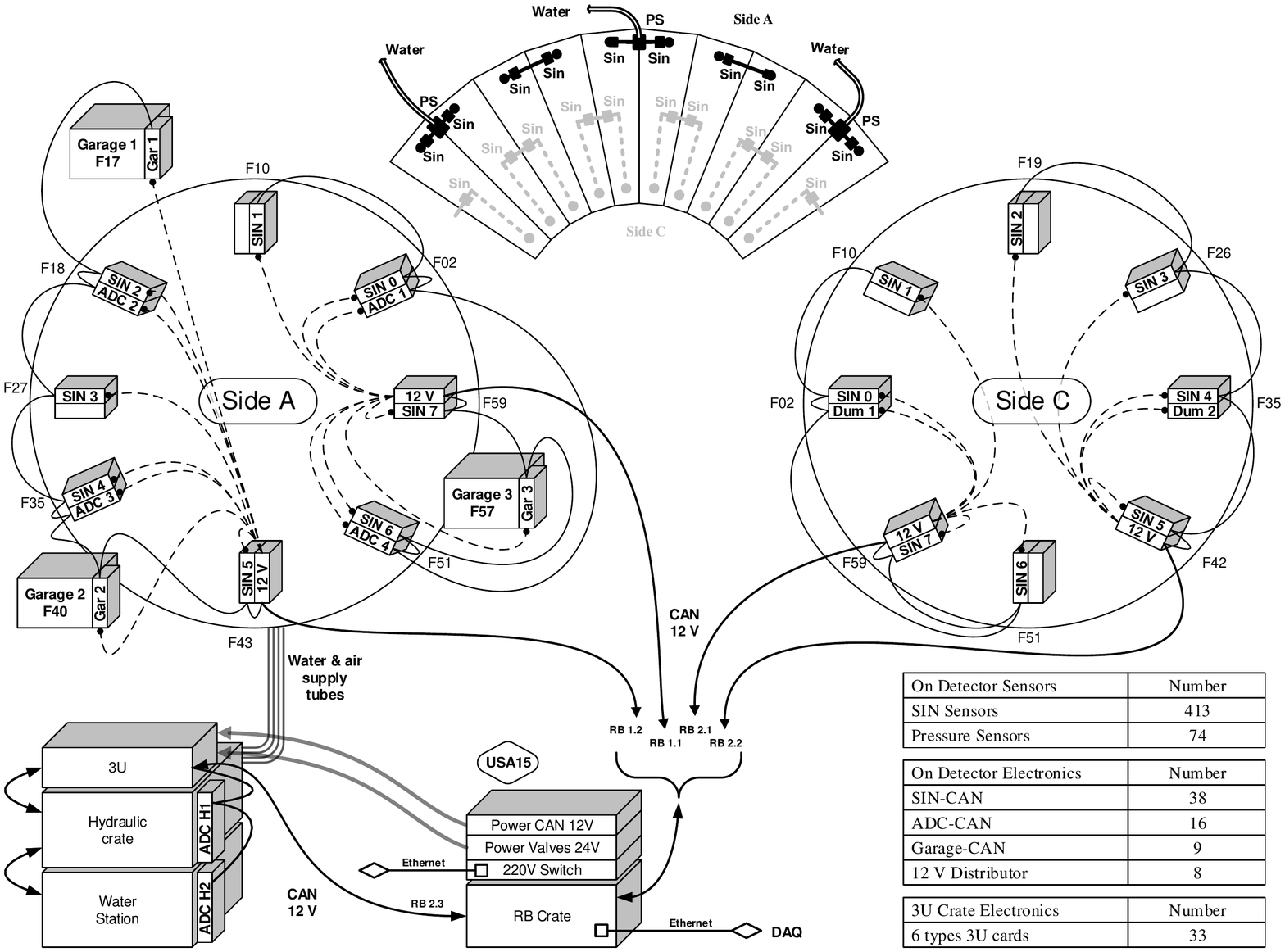}
  \caption{The schematics of the control and operation equipment structure of the LB TileCal subsystem. The CAN bus daisy-chain is represented by solid lines, while the dashed lines show power supply connections. The table shows the total numbers of sensors, FEE modules and 3U-control for the entire calorimeter.}
  \label{fig:csarch1}
\end{figure}

As already pointed out, the three source subsystems are functionally independent and have identical sets of sensors on the source path, garages, hydraulics drives, distributed front-end electronics (FEE) modules and power supplies. As a result, they have very similar architecture. Figure~\ref{fig:csarch1} schematically shows the layout of the control structure and read-out of one subsystem.

\begin{figure}[!htbp]
  \centering
  \subfloat[]{\includegraphics[width=0.45\textwidth]{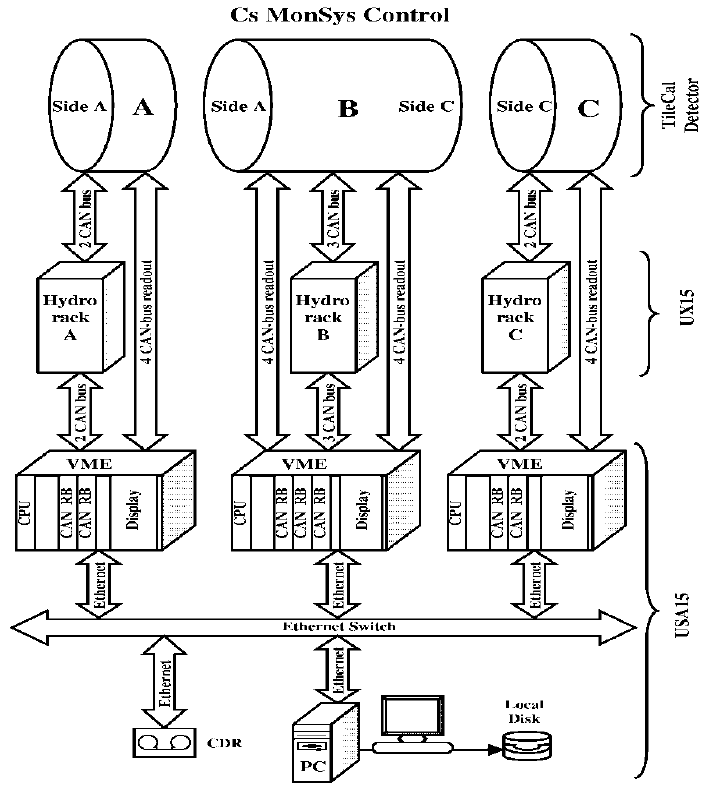}} \quad
  \subfloat[]{\includegraphics[width=0.52\textwidth]{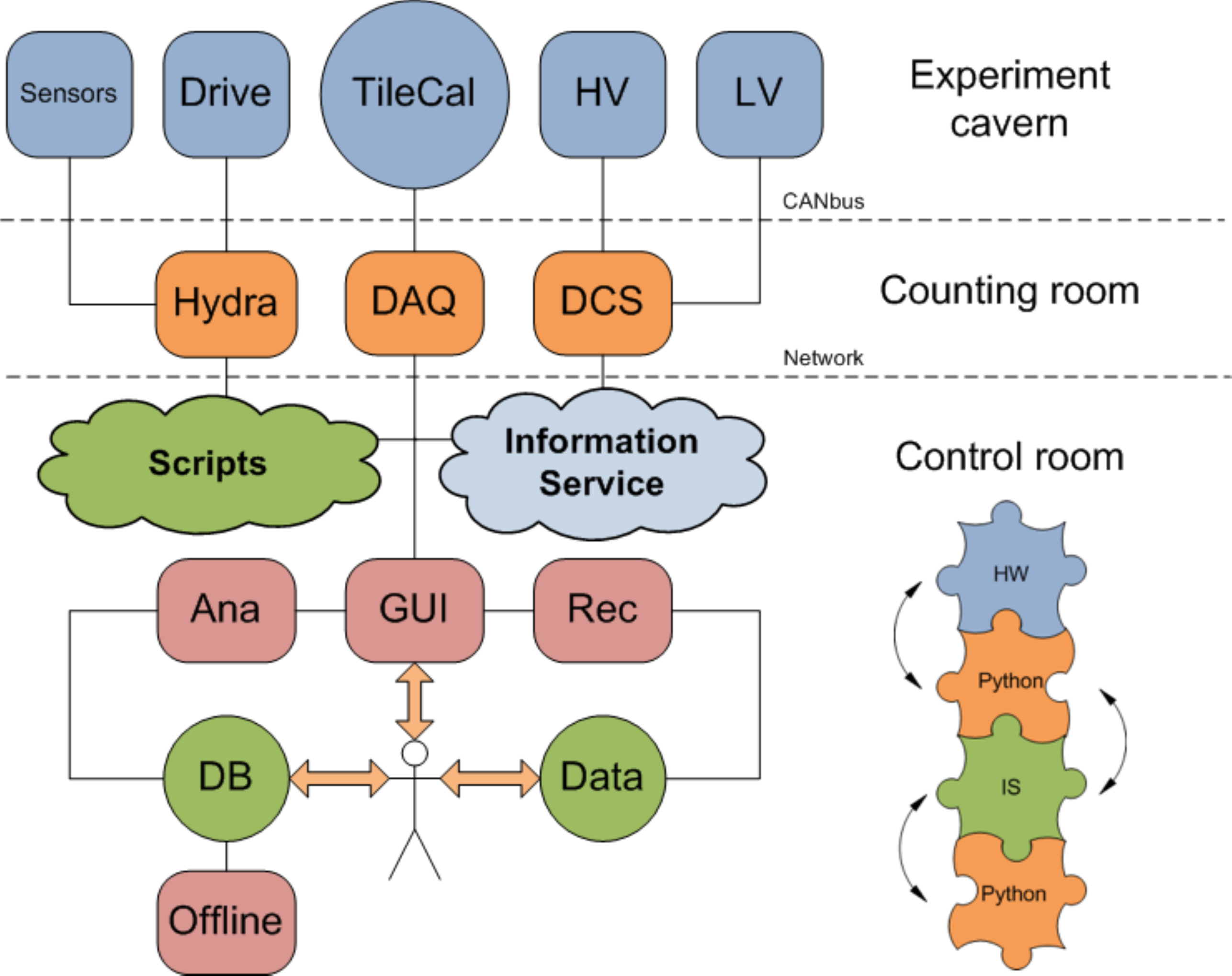}}
  \caption{(a) Schematics of the data flow. (b) Online software structure. Hydra, DAQ and GUI are the main processes active during data taking~\cite{Software}.}
  \label{fig:csarch2}
\end{figure}

The sensors are read out by the FEE modules, distributed over the TileCal body, and the service crates, located on the cavern floor.  They are interconnected via 50 kBaud CAN bus daisy-chains to the CAN bus Read Out Buffers (RBUF) located in the ATLAS USA15 control room. During system operation, the changing hardware configuration, sensor hits and registered PMT responses record the time-dependent status of the system, including the status of the hardware, the latest changes of its parameters, the conditions of the fluids, the direction, speed and location of the sources, etc. Data flow and communications between equipment components are shown schematically in figure~\ref{fig:csarch2}a.

\subsection{Online software}

The collected status information is recorded and analysed by a number of custom-developed control processes~\cite{Software}, written in C++ and Python, outlined in figure~\ref{fig:csarch2}b. They use libraries provided by the ATLAS TDAQ~\cite{TDAQ}, namely, VME libraries~\cite{VMElib} for hardware access, IS~\cite{IS} for data exchange and ERS~\cite{ERS} for error reporting. The HYDRA process takes care of all mechanical operations: it runs the module scan according to the pre-programmed scenario via a specific interface to the 3U-control crates and synchronises the relevant readout subprocesses. The DAQ process cyclically reads the currents of the PMTs, module by module, at the design frequency of 90~Hz according to the position of the source, provided by HYDRA, and sequentially stores the PMT data. 

In an auxiliary mode, a Detector Control System (DCS) process, in common with the entire TileCal DCS~\cite{DCS}, retrieves the essential information on the HV and LV for each module, using the DIM communication package~\cite{DIM}, and publishes the data and the state of the sensors. 

In addition to the already mentioned HYDRA, DAQ and DCS processes, the software includes other general components:
\begin{itemize}
\item The ATLAS Information Service, used for communications;
\item Embedded scripting facilities for program logic and planned actions, containing descriptions of the standard and/or specific actions;
\item GUI~--- a graphical user interface that allows the operator to communicate with the running sub-processes and to visualise the state of the system, the operator's actions and its results;
\item Analysis (Ana) and data recording (Rec);
\item The common TileCal Data Base;
\item The CAN bus branches that control the system's drives and sensors.
\end{itemize}

The total number of sensors runs up to 500, while the number of FEE cards is about 80 altogether in all three sections of the system. Each of three 3U crates contains 11 specially designed cards of 6 types to control the communications, pump, valves, level meters, etc. 

On-line operations and procedures of the system are designed maintaining full independence of mechanical and readout functions. Specifically, any capsule can be run through the entire calorimeter independently of the calorimeter readout status, with all the attributes of a data-taking run: full capsule movement control, visualisation and final data-file recording. The source signal readout sequence can be included into the mainstream data flow according to the source movement data reference or following a pre-arranged program. In turn, readout procedures can be run without source movement, for checking and testing purposes, including integrator calibration and pedestal measurements, as described later.

Synchronisation between run control and readout procedures is based on module entry and exit SIN signals: the module entered by the source is included into the readout chain, and when the output SIN sensor gives an exit signal its readout stops. In the normal scan preparation procedure, all modules to be read out are initialised for data acquisition. Normally just one module is read out while the source is inside it -- there are only a few cases in which two neighbouring modules are read out together because SINs could not be installed between them due to mechanical restrictions in the zones where the electromagnetic calorimeter cryostat supports did not allow to do so.

\subsection{Data recording}

The initial, current and final hardware status, the data flow and the integrator readout information are recorded as a raw data file with ROOT structure. This file contains time stamps that document all changes in all the relevant parts of the system: drive, garages, SINs, PSs, all modules channel responses and general information such as run conditions and constants like source ID, integrator gain, readout frequency, HV, LV and drawer internal temperatures, a list of bad channels, etc.

Integrators are read out via 250~kBaud CAN bus at 90~Hz. This readout rate is fast enough to observe the tile-to-tile structure clearly and is slow enough to switch readout to the next channel. The data flow is below 25~kBytes/s and does not pose any significant requirements on the infrastructure.

The raw data structure is organised to allow retrieving the past run procedures. This ``history'' option, together with the corresponding data flow, has been very helpful while developing and debugging the hardware and software.

The data are stored in separate files, corresponding to a run between two garages. The file size ranges from 5~MBytes for Extended Barrel up to 30~MBytes for Long Barrel runs.

\subsection{Scan operations}

Driving the source at $\sim$35~cm/s allows to scan an EB in 3 hours, and the LB in 5 hours.  To prevent substantial data losses if a scan cannot be completed, the full scan of a TileCal barrel consists of three separate sub-scans of 20--24 modules each. The physical limits of the sub-scans are determined by the garage locations so that if a scan must be interrupted, the source can be safely stored in a garage in no more than one hour, reversing the source direction if convenient. The source velocity is easy to change: for instance, when passing through tees, the speed is programmed to increase up to~40~cm/s.

In the LB the source moves in one direction in one run, and in the opposite direction in the next one. EBA and EBC are scanned every time in both directions. The reason for such a scanning mode is that two modules in EBA and EBC can be read only in one direction. Being smaller, EB scans require just over one half of the time of LB scans, therefore scanning EB modules in both directions requires about as long as one LB scan in one direction. 

Usually, all three calorimeter sections are scanned in parallel, requiring 6--8 hours altogether including scan initialisation and finalisation procedures. In principle, all scan procedures are fully automatic, but they are performed under operator supervision in case human intervention appears necessary. 

\subsection{User interface}

\begin{figure}[!htbp]
  \centering
  \includegraphics[width=0.95\textwidth]{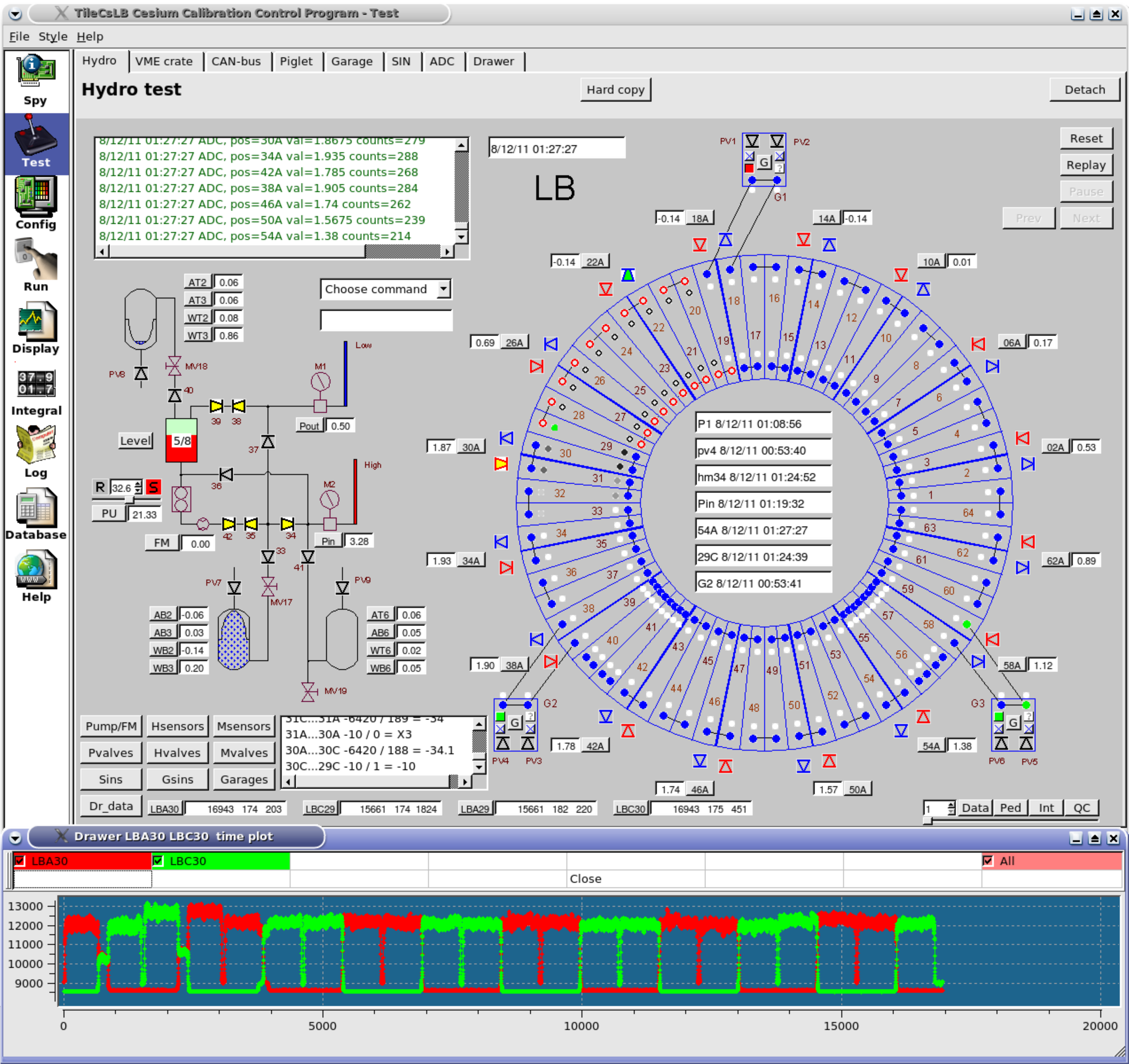}\\
  \includegraphics[width=0.95\textwidth]{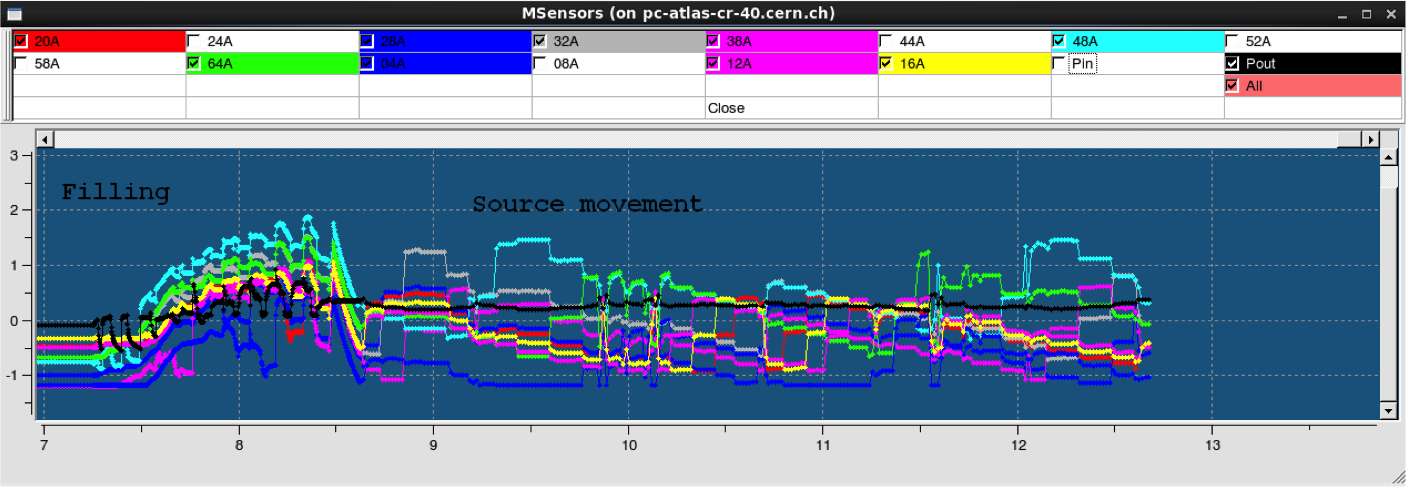}
  \caption{An example of GUI display screen during a Cs-137 scan. The source is moving inside module 30 between garage 2 (G2) and garage 1 (G1). 
  The middle panel shows the sum of raw signals from all PMTs of the module being read out. The bottom panel is a display of pressure readings in different locations of the calorimeter calibration tubes during several operation phases.} 
  \label{fig:gui}
\end{figure}

The operator controls the execution of the scan using the graphical user interface (GUI) written in C++ and using the Qt toolkit. Figure~\ref{fig:gui} shows a screenshot of the interface during the on-going source scan. One can see here:
\begin{itemize}
\item The status of the drive pump and valves;
\item The pressure in the drive and at the tee-joints around the calorimeter;
\item The status of the garages, including the state of garage locks and possibly the presence of the source capsule;
\item The status of SIN sensors, represented by circles at the module entrances and exits: empty circles indicate no hit, coloured circles indicate hits with colour coded number of hits and an indication of the time passed from the hit, coded by the level of grey colour;
\item The window giving information on the SIN hits, the direction and the speed of the source;
\item The command window, that provides to the operator the list of scripts to be used to perform specific manual interventions;
\item A display of the responses of the PMTs in the module being scanned as a function of time.
\end{itemize}

\clearpage
\section{Offline processing}

\begin{figure}[!htbp]
  \centering
  \subfloat[]{\includegraphics[width=0.48\textwidth]{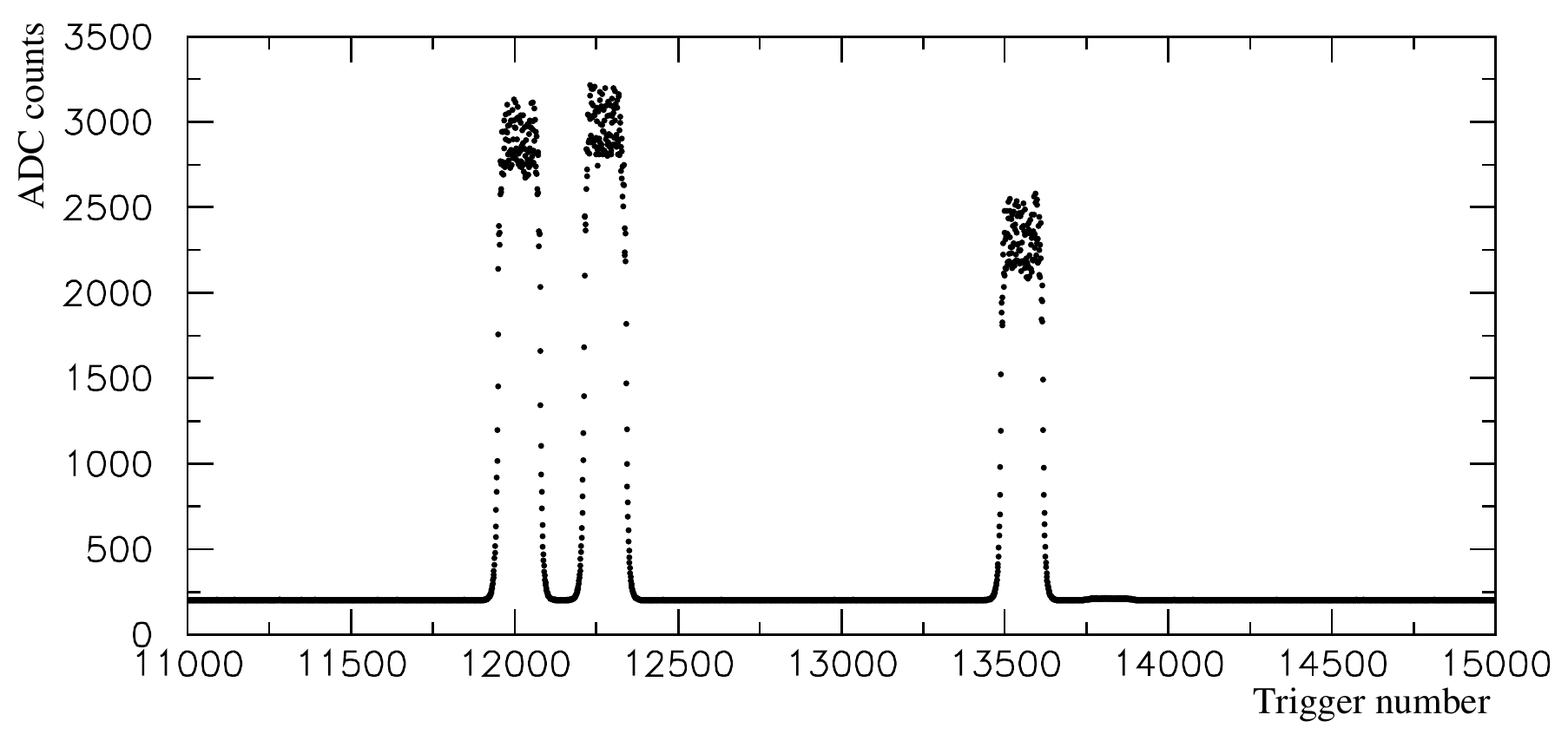}} \quad
  \subfloat[]{\includegraphics[width=0.48\textwidth]{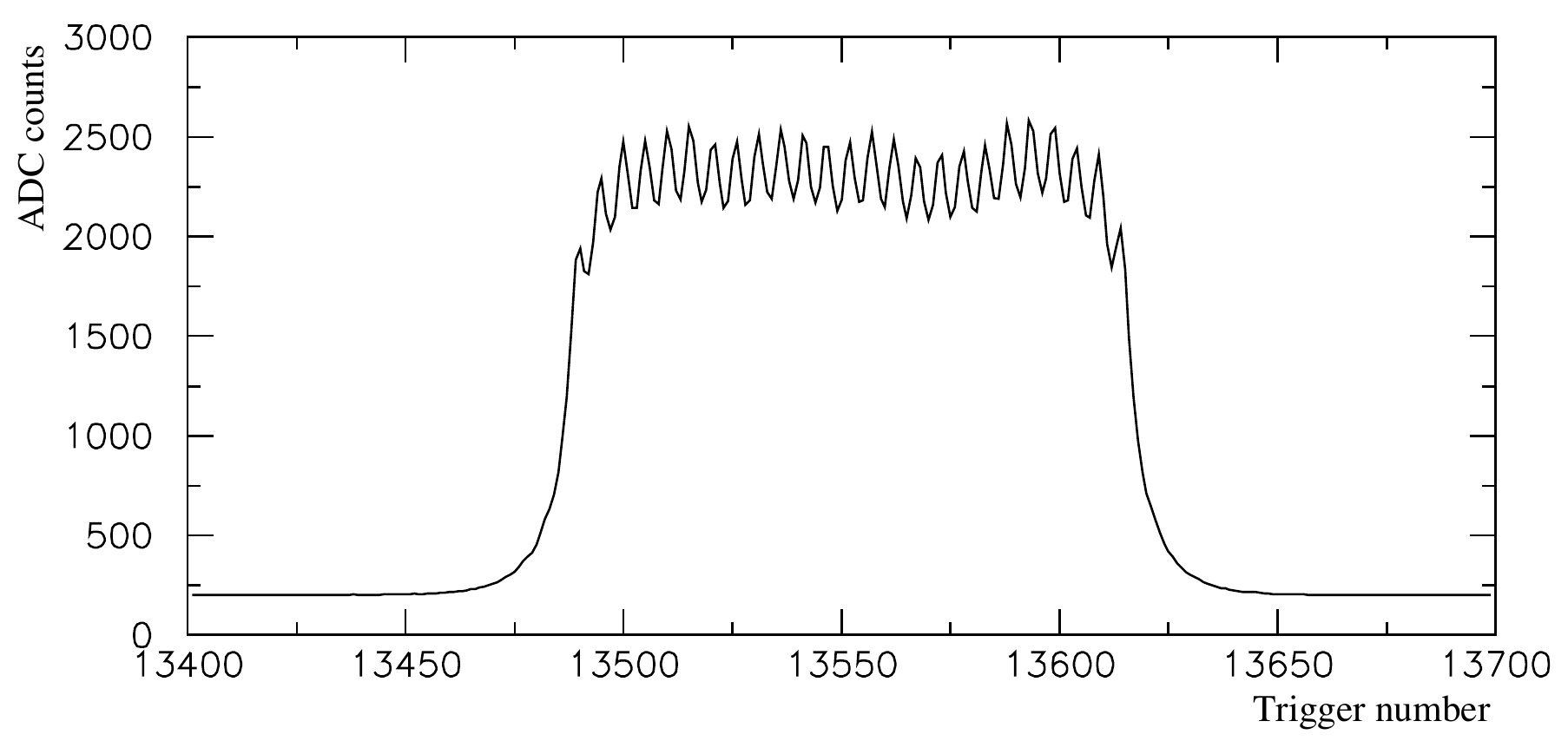}}
  \caption{(a) A typical response sequence when the source traverses three tile rows of a cell and (b) the signal from one of the PMTs, covering part of a tile row. The pedestal and signal regions are clearly visible, as well as the regular tile structure in (b). On the abscissa, the number of readouts (at the 90~Hz rate) is shown.}
  \label{fig:pmtresponse}
\end{figure}

The source signals from a readout cell sequentially represent the response of the tile rows in a cell traversed by the source. As the source passes through successive tile rows, the PMT responses in turn display pedestal and signal regions, as shown in figure~\ref{fig:pmtresponse}. Appropriate processing of the raw source data provides an accurate estimate of the cell response to the source.

\begin{figure}[!htbp]
  \centering
  \begin{minipage}{0.56\textwidth}
    \subfloat[]{\includegraphics[width=\textwidth]{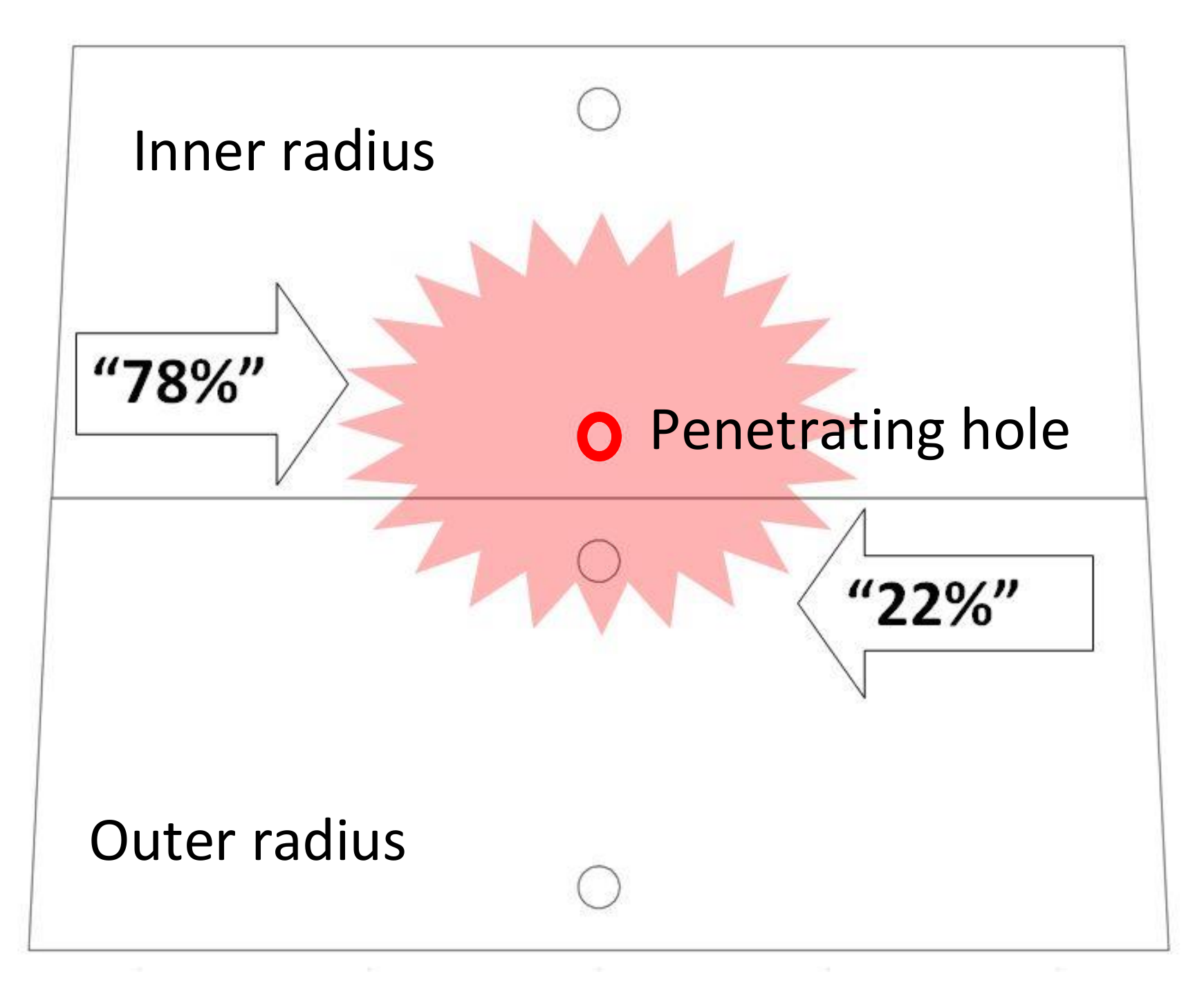}}
  \end{minipage}
  \begin{minipage}{0.40\textwidth}
    \subfloat[]{\includegraphics[width=\textwidth]{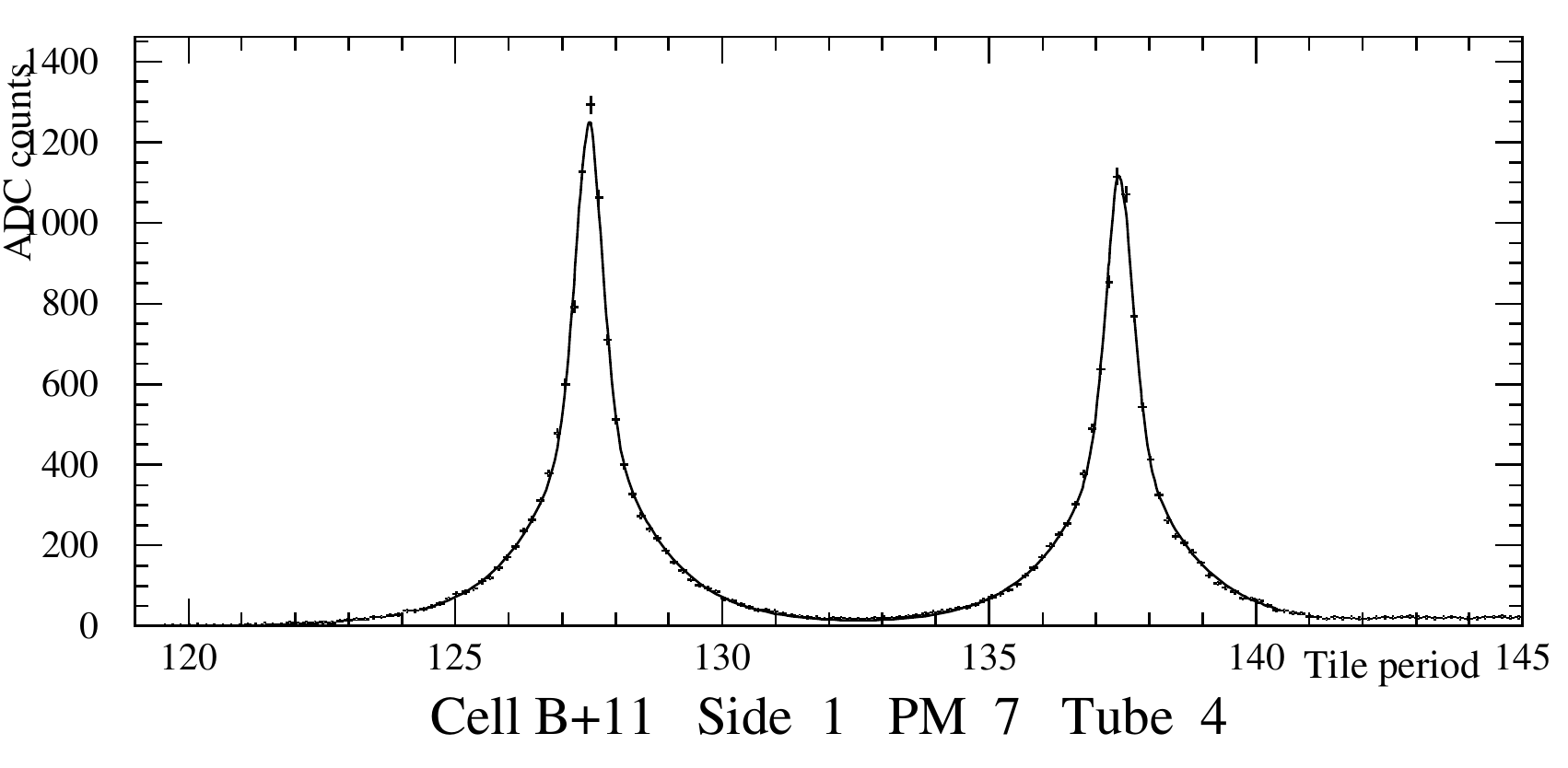}} \\
    \subfloat[]{\includegraphics[width=\textwidth]{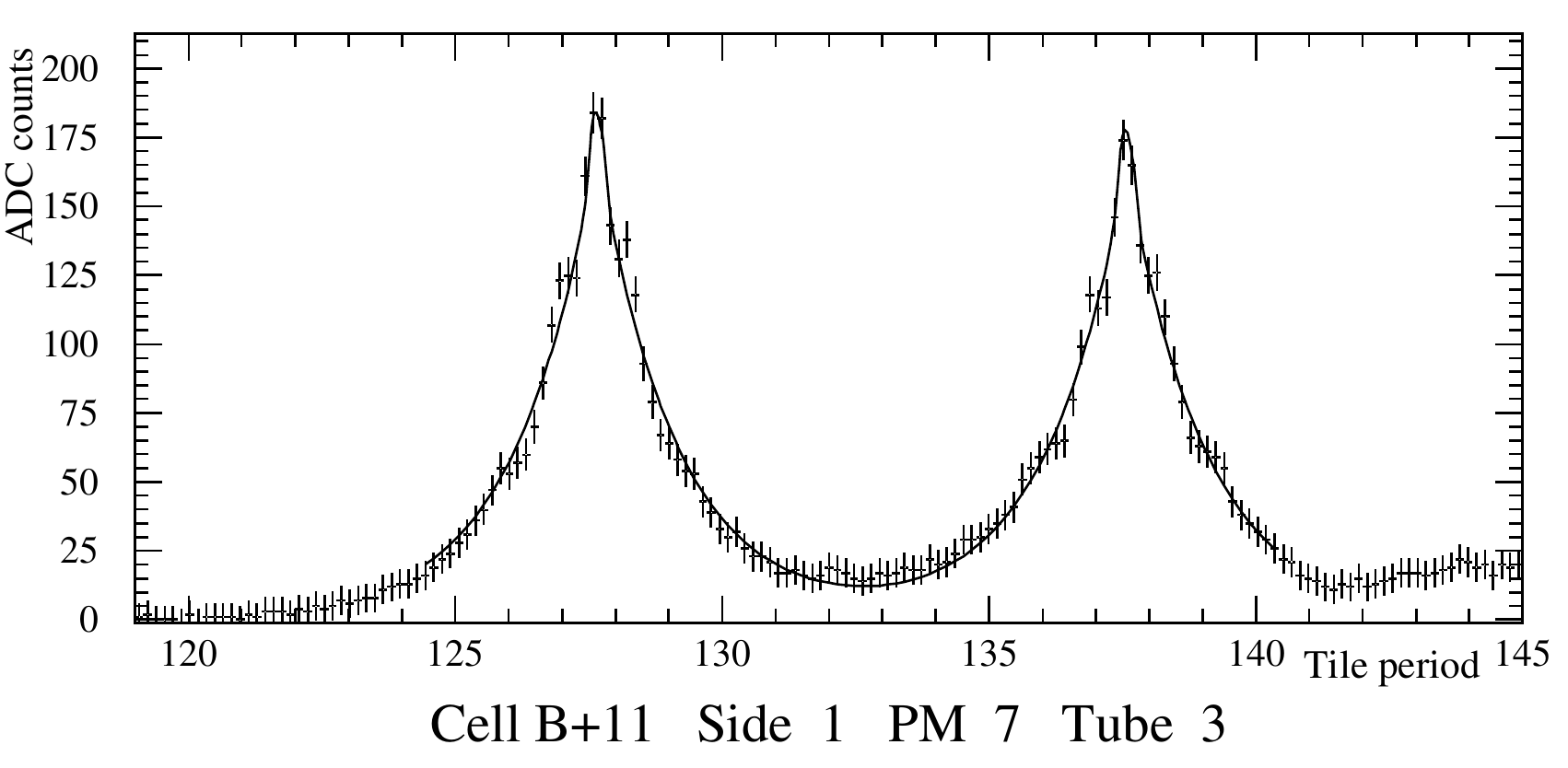}} 
  \end{minipage}
  \caption{Gamma-ray energy sharing between two adjacent tile rows (a). An example of the response measured from two tiles separated by ten 18.2~mm periods: the ``78\%+22\%=100\%'' case, showing the response from the same two tiles when the source passes through a tube within the two tiles (b); the ``22\%'' case, when the source passes in a tube located in the adjacent tile row (c). The responses are fitted with a sum of a Gaussian and an exponential.}
 \label{fig:radsharing}
\end{figure}

A conceptual picture of the sharing of the gamma-ray energy between a pair of adjacent scintillator tile rows in the same cell of any module is shown in figure~\ref{fig:radsharing}a. Due to the geometry of the tiles and the location of the source tube holes within tiles the energy of the \Cs gamma rays is deposited into adjacent tile rows in a ratio of about 22/78. In the figure, 78\% of the gamma-ray energy is deposited in the top tile row (at a smaller radius from the colliding beams) and 22\% in the bottom tile row. The cell signal consists of the sum of the two tile rows. A particular case occurs when the tile row at the larger radius is the last one in a cell; then the cell signal contains on average only 78\% of the energy because 22\% is deposited outside the cell limits. A detailed discussion of this effect and of the procedure adopted to correct it is given later in this section.

The response of an individual tile, traversed by the source, is accurately parametrised by a sum of a Gaussian and an exponential:
\begin{equation} 
g\times\exp^{-0.5((x_0-x_i)/\sigma)^2}+(1-g)\times\exp^{-\lvert x_0-x_i \rvert/\lambda}
\label{eq:csfit}
\end{equation}
where $0<\mathrm{g}<1$ is the fraction of the integral of the parametrising function represented by a Gaussian, $x_i$ is the instantaneous coordinate of the source capsule and $x_0$ is the coordinate of the centre of the tile.

The fit parameters were evaluated with specially-designed tests wherein a specific set of WLS fibres were coupled to the tiles whose individual response was under study. As an example, figures~\ref{fig:radsharing}b and \ref{fig:radsharing}c present the case where the source passes in the tile row at the next smaller radius with respect to the tile row being read-out (``22\%'' case) or through it (``78\%''), together with the fitted response functions, given in units of the TileCal periodic structure of 18.2~mm. Note that several data points are recorded for each 18.2~mm period.
They are the PMT response measurements corresponding to the position of the source along the calibration tubes, at a distance given by the source velocity divided by the readout frequency of 90~Hz, i.e. 35/90 = 0.4~cm. 

The first correction applied to the raw data calculates to a good approximation the instantaneous response of the PMT current, mostly eliminating the distortion caused by the front-end electronics, a charge amplifier with a low-pass RC circuit with a time constant typically of about 10~ms. 

\begin{figure}[!htbp]
  \centering
  \includegraphics[width=0.5\textwidth]{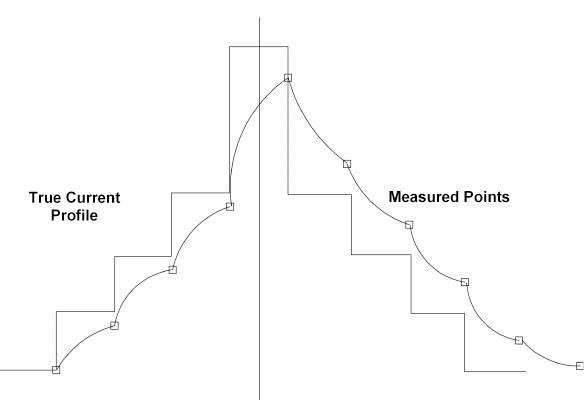}
  \caption{Time dependence of the response of a single tile: ideal and recorded by the integrator's low-pass RC input circuit.}
  \label{fig:taucorr}
\end{figure}

Figure~\ref{fig:taucorr} schematically shows the distortion of the signal from a single tile. The effect of the change of the peak's shape, due to the integrator's RC circuit, shifts the coordinates of the signal's centre of gravity, thereby producing a bias that depends on the direction of the source movement and the characteristics of the RC circuit. 
The correction of the tile line shapes is performed using a simplified formula that reverses the effect of the RC filter by correcting the signal by an amount proportional to its derivative at every measured point. The correction is:
\begin{equation}
A_i \rightarrow A_i + \delta\times(A_i-A_{i-1}) 
\label{eq:taucorr}
\end{equation}
where the $A_i$ and $A_{i-1}$ are measurements adjacent in time and $\delta$ is set empirically, depending on the actual RC circuit value and the readout frequency.
The asymmetry of the responses also shows up as asymmetric tails at the beginning and the end of the sequence of signals from each PMT, depending on the direction of the source movement. 

Numerous high-statistics test scans using the actual front-end amplifiers led to a fitted value for  $\delta$ of 0.7$\pm$0.1 at the default 90~Hz readout rate. This correction makes the response curve symmetric to a good approximation, independently of the direction of the source movement. 
Most important, the correction sharpens the peak/valley ratios of the PMT response curves with respect to the raw data shown in figure~\ref{fig:pmtresponse}b, hence it turns out to be very useful to precisely evaluate the response of individual tiles. 

After this correction, pedestal values, recorded when the source is out of the cell being measured, are calculated and subtracted. After pedestal subtraction, any slightly negative values are set to zero, and the data are subjected to further treatment.

\subsection{Pattern recognition and tile row response calculations}

Calculating tile row responses is the first step following the ones just described. Then a cell's mean response is calculated averaging the tile row responses.   

\begin{figure}
  \centering
  \includegraphics[width=0.9\textwidth]{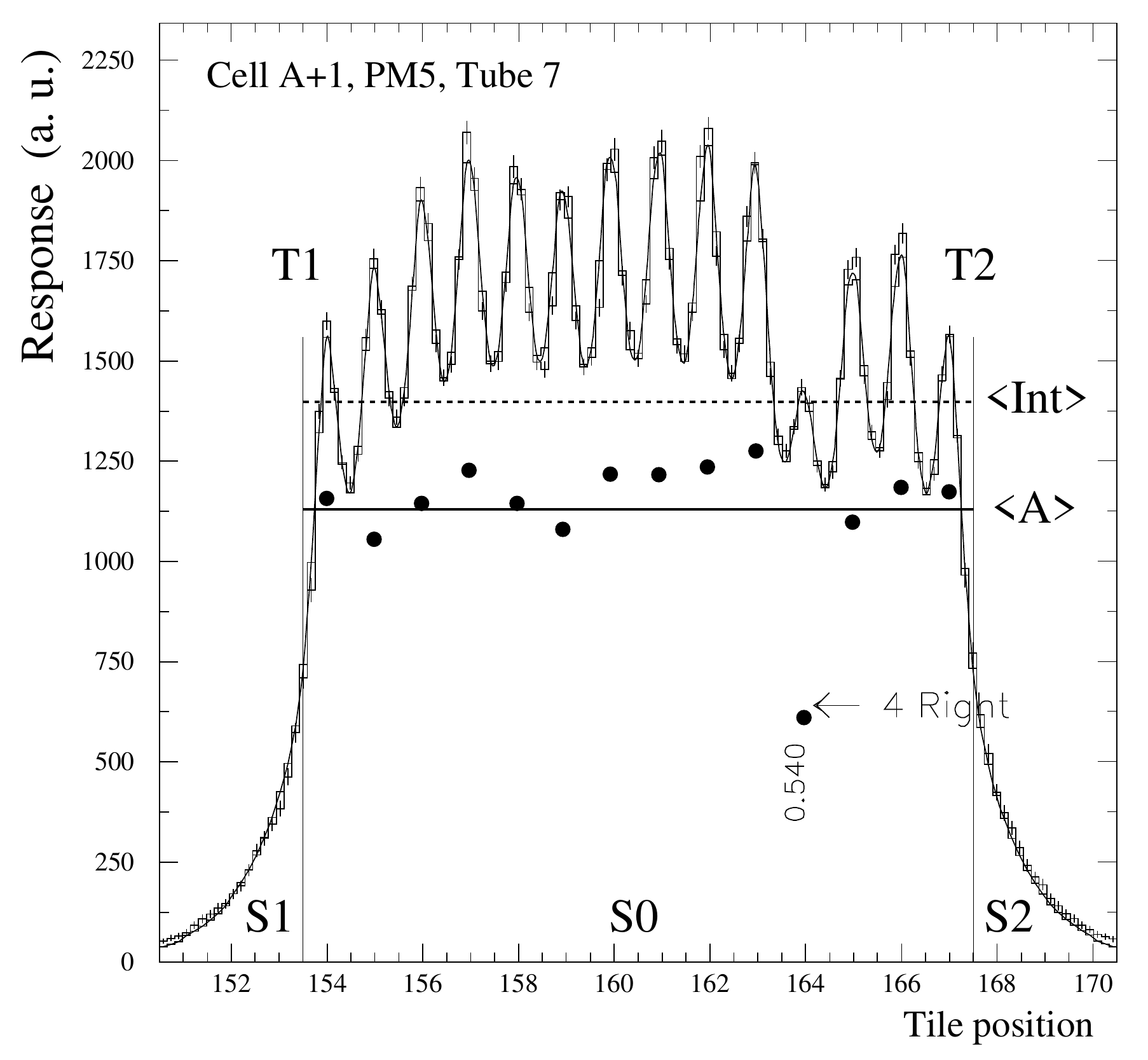}
  \caption{An example of a tile row response sequence taken at the 90~Hz readout frequency and presented here as a function of tile position. The line labeled <Int> gives the mean response calculated with the ``integral'' method and the line labeled <A> is the response obtained with the ``amplitude'' method.}
  \label{fig:cstilerow}
\end{figure}

A typical readout sequence from the third row of an A-sample cell (78\% case) is shown in figure~\ref{fig:cstilerow} after applying all corrections described in the previous subsection. The units on the abscissa are the calorimeter 18.2~mm periods with one scintillating tile per period.

The cell edges are shown by the vertical lines; each of the 14 tiles of this tile row displays a clearly identifiable peak. One can see that the total response is the sum of the overlapping individual tile responses. In physics runs, the tile row response corresponds to the energy lost by particles in this tile row.

Two complementary approaches are used to evaluate the tile row responses. They are referred to as ``integral'' and ``amplitude'' methods. The two methods and their results are illustrated in figure~\ref{fig:cstilerow}.

The integral approach works by adding all the response points in the three regions labelled S0, S1 and S2; the sum is divided by the width of the distribution, labelled T2-T1. The proper setting of the cell edges is based on an appropriate determination of the tails and the correct count of tile peaks. The result of the integral method calculation is a mean response (<Int>) of the tile row of this cell. The method is fast and in most cases, the results are stable at the level of 0.2--0.3\%, estimated from repeatability tests.

The amplitude approach uses the coordinates of each tile in a tile row and calculates the sum of overlapping individual tile responses, determined as follows. A MINUIT fitting procedure, using the equation~\ref{eq:csfit} is applied to the sliding regions that cover in successive steps intervals of 5 to 10 periods with step-to-step overlaps of 3 to 5 periods. The individual tile responses (the ``amplitudes'' evaluated at the positions of their peaks) are shown by the black points in figure~\ref{fig:cstilerow}. Their mean is the mean row response labelled <A>. These calculations are rather complicated and CPU-intensive, besides needing good response parametrisation, but produce detailed pictures of the calorimeter's internal structure and of the positions of tile row edges, which are useful in case of module edge effects and tiles of special shape (cut tiles).

Indeed, properly reconstructing the Cs source signals in a tile row is not always as simple as shown in the previous figure. As already mentioned, about 22\% of the source gamma-ray energy leaks into the tile at a larger radius. The light is collected by the WLS fibres coupled to the tile sides and transported to the PMTs, where the light from all the tile rows belonging to the same cell side is added up. If the cell has 3 tile rows, in two cases the cell picks up this 22\% leak, but when the sources goes through the row at the outermost radius, the leak goes to the other cell. Also, if all the tile rows had exactly the same leak of the Cs radiation to the next-outer tile row, the overall cell response could be calculated quite easily, but in most cases, it is not so.

Furthermore, adjacent tile rows may have different numbers of tiles, leading to an additional distortion of individual peaks, and of the overall picture. Another feature of the tile layout that must be taken into account when reconstructing individual tile responses is that tiles in adjacent rows are shifted by half of a period (9.1~mm).

\begin{figure}[!htbp]
  \centering
  \subfloat[]{\includegraphics[width=0.48\textwidth]{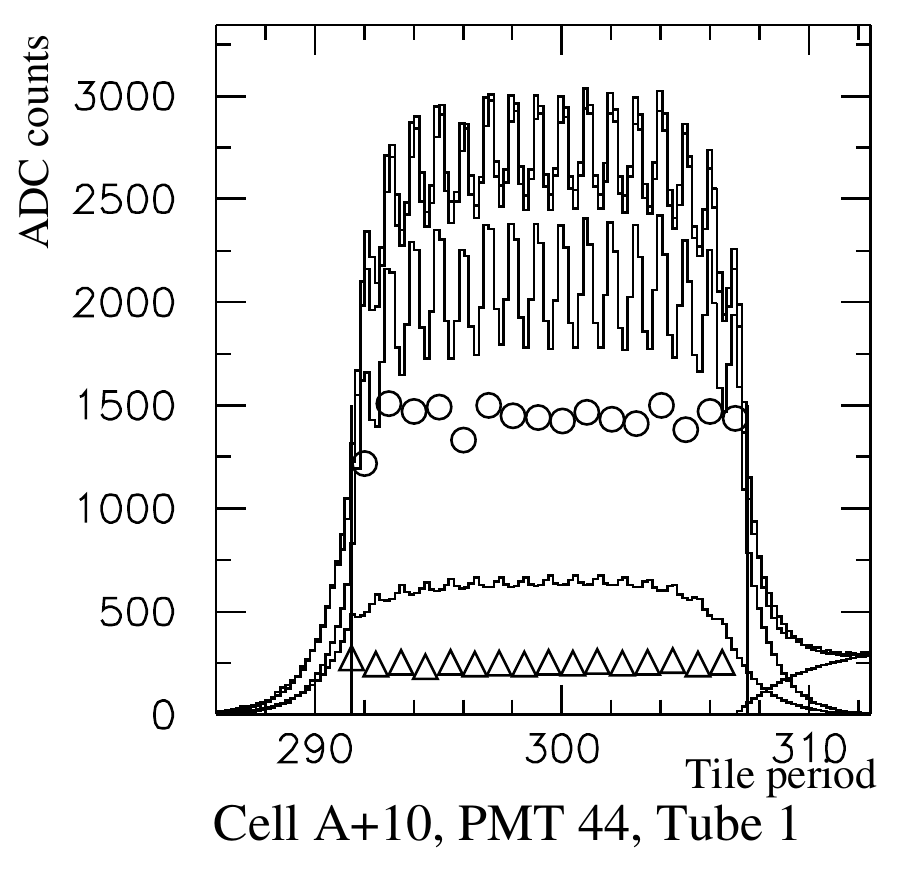}} \quad
  \subfloat[]{\includegraphics[width=0.48\textwidth]{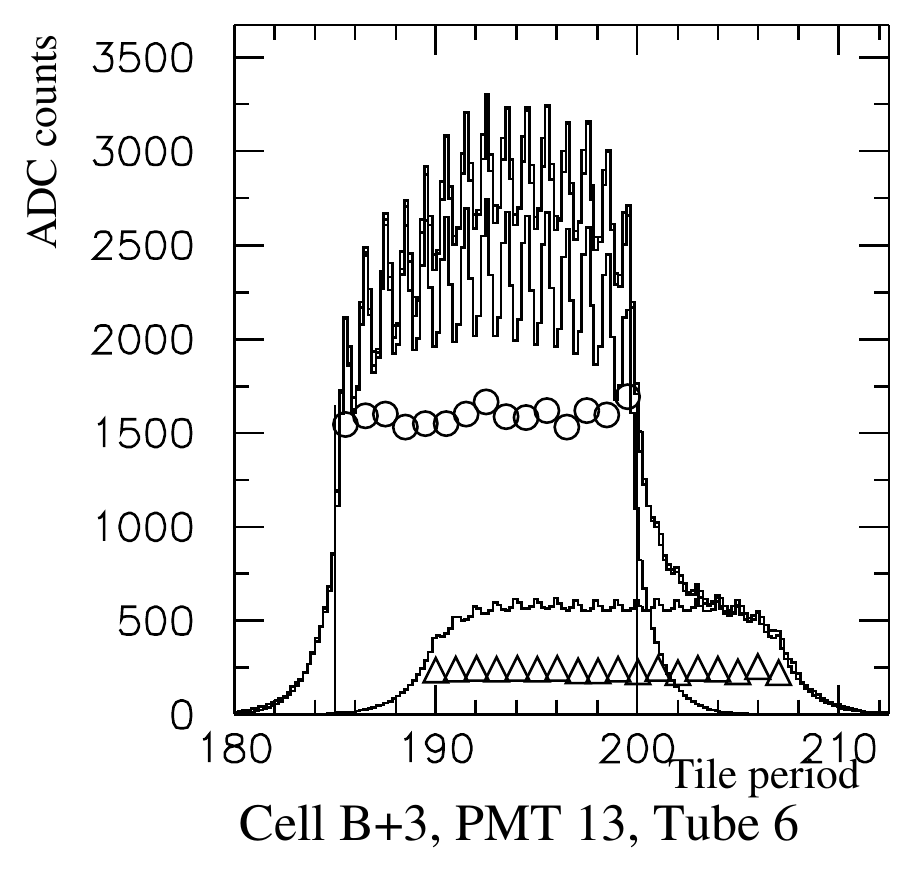}}
  \caption{Two examples of sum distributions from adjacent tile rows where the ``22\%'' leak takes place. The X-axis shows the 18.2~mm tile/steel period number. The top two curves show the measured and the fitted cell response, while the middle curve shows the fitted curve with subtracted ``22\%'' response from the adjacent tile row, denoted by the bottom curve. Circles denote the individual tile reconstructed amplitudes, triangles show the reconstructed ``22\%'' response from the adjacent tile row. (a) The effect of the 2~cm thick end-plate (EP) is shown, as the rising curve at the right bottom. (b) Row displacement coming from the cells structure (B and C rows) causes a large distortion of the resulting response curve, visible as the "hump" on the right, making the determination of the cell edge location very difficult and the responses balance problematic.}
  \label{fig:tilerow7822}
\end{figure}

Figure~\ref{fig:tilerow7822} presents two typical cases in which these features must be considered in order to properly extract the response of each tile row from the Cs scan data, eliminating the complications due to the geometry of the particular cell. The first case is when the cell has a thick steel end-plate, causing the distortion of the result at the edge of the cell. The second case shows the influence of the adjacent tile row that is shifted in Z versus the tile row where the capsule is flowing, making the disentanglement and correct amplitude determination particularly complicated.

For this purpose, measurements were made in order to measure precisely the ``78/22'' ratio for different tile rows, and how to correct for it when calculating the tile row responses with the integral and the amplitude methods.
The measurements were made setting up special Cs source signal readout configurations for all eleven tile rows. In addition, the data from all modules underwent specific analysis procedures in order to obtain complementary information from each tile row.

\begin{figure}[!htbp]
  \centering
  \subfloat[]{\includegraphics[width=0.45\textwidth]{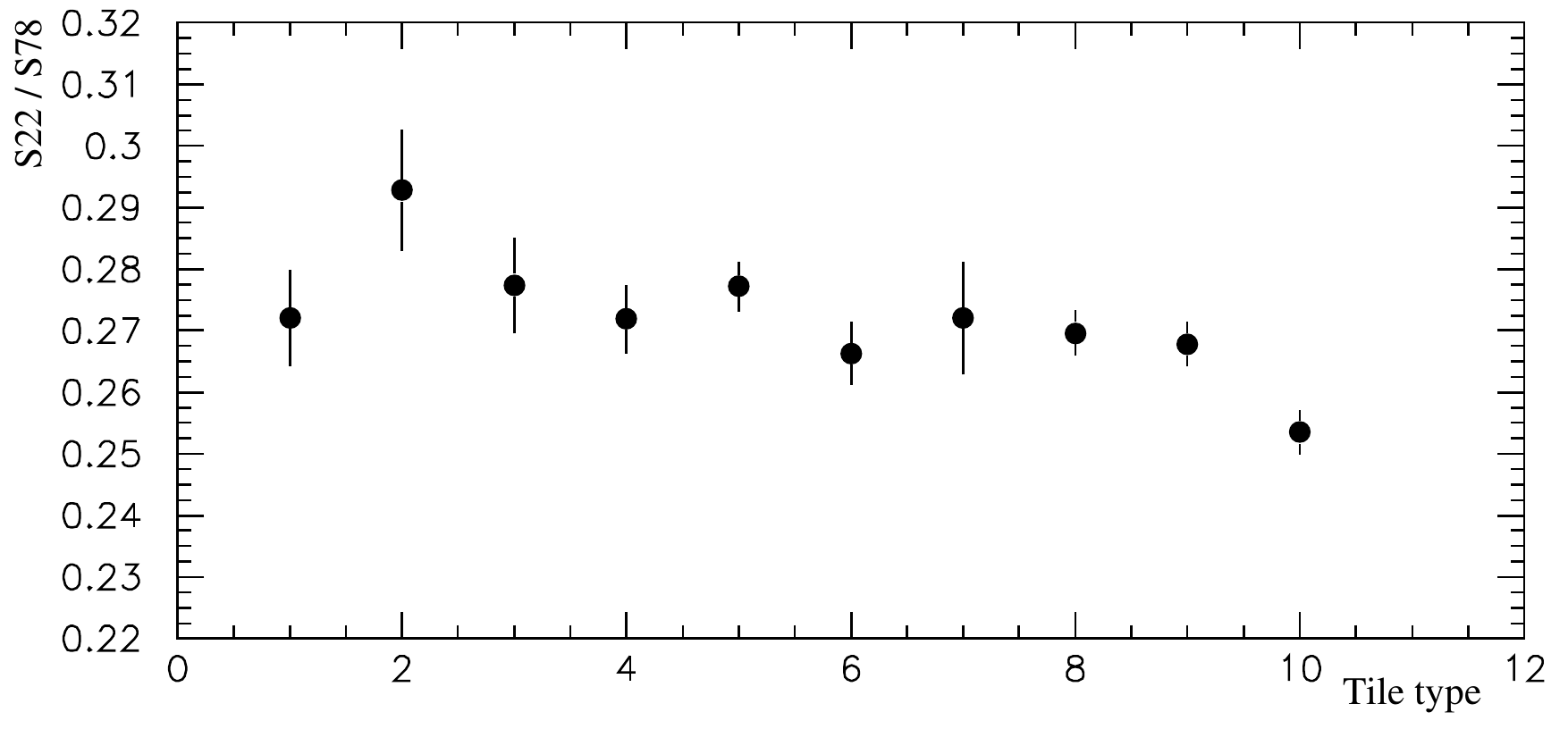}} \quad
  \subfloat[]{\includegraphics[width=0.46\textwidth]{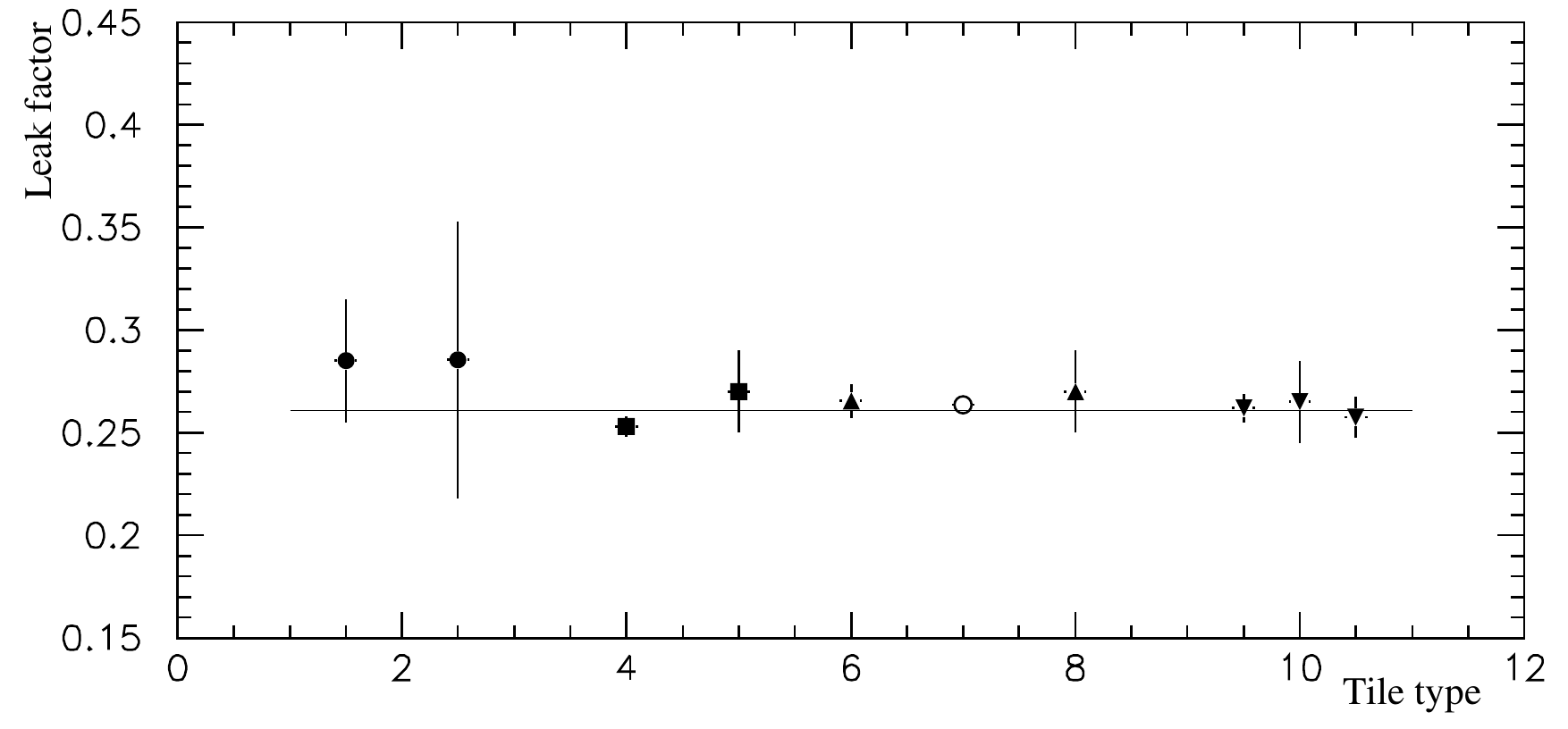}}
  \caption{Experimentally measured ratios of the leaked and absorbed gamma ray energies. (a) The ``22/78~ratio'' measured using a specially equipped module with single tile readout. (b) Calculation of the energy leak factor using data analysis of the production modules in the detector.}
  \label{fig:tilerow7822-2}
\end{figure}

In figure~\ref{fig:tilerow7822-2} ``integral'' ratios are shown for each tile row in which the two signal components are present and were separated. In panel (a), the ratios were extracted from Cs source runs through a specially instrumented TileCal module. In (b), the ratios were extracted from normal source data from all 64 LB, EBA and EBC modules.

The first approach provides a direct measurement of the ratios but is affected by systematic uncertainties arising from the optical properties of the particular module used. The second has much larger statistics, but here the ratios are evaluated from an overall fit, subject to the errors due to the algorithms used.

The ratios for smaller tiles (\#1-3) look slightly higher than for the other ones, however, the differences are within the statistical and systematic errors. The effect may be associated with the tile volume, because of a smaller ``78\%'' absorbed fraction in the smaller tiles. It is encouraging that the two methods give very similar results, compatible within an errors corridor. The mean value of the leak over ``B9'', ``BC'' and ``D'' cells, where rather direct measurements are available, equals 0.261$\pm$0.012, corresponding to  split ratio of 20.7\%/79.3\%.

When using the integral approach, the ``22/78 ratios'' depend on the tile row, while with the amplitude approach the variation with tile rows is smaller than the errors, hence it is reasonable to use the fixed 0.261 ratio. The two correction methods are complementary: with the integral method, the cell edge is sometimes hard to precisely locate, whereas, with the amplitude method, based on a constrained fit function, there is no such problem. Furthermore, the latter method's precision in determining single tile responses is about 2\%, which is amply sufficient to estimate the uniformity of the full modules.  The amplitude method is CPU expensive but doesn't offer significant improvement with respect to integral method it is significantly better only for C10 cells (with only 5 periods and without the end-plate), therefore the integral method is normally used for calibration and equalisation. On the other hand, the amplitude method was very useful during instrumentation, because we could see immediately which fibre was broken and had to be fixed, while CPU required to reconstruct signal just in one module is not significant.

\subsection{Cell response evaluation}

After taking care of the corrections just described, the response of a readout cell is simple to calculate. As shown in figure~\ref{fig:radsharing}, the source typically traverses a tile through the hole at the larger calorimeter radius, depositing the ``78\%'' fraction in that tile row. Therefore within each cell, the signal in the tiles at the larger radius is not mixed with the signal from the adjacent tile row, which belongs to the next cell. It is convenient to adopt this signal for every tile, subtracting the calculated ``22\%'' leakage from the tile at a larger radius. After obtaining the ``78\%'' signals for every tile row in a cell, the overall cell response is calculated as the mean of these signals weighted by the volume of each tile row. There is no loss of information in not correcting for the ``78\%'' fraction in calculating the cell responses because they do not propagate any further than uniformity estimates.

\begin{figure}[!htbp]
  \includegraphics[width=\textwidth]{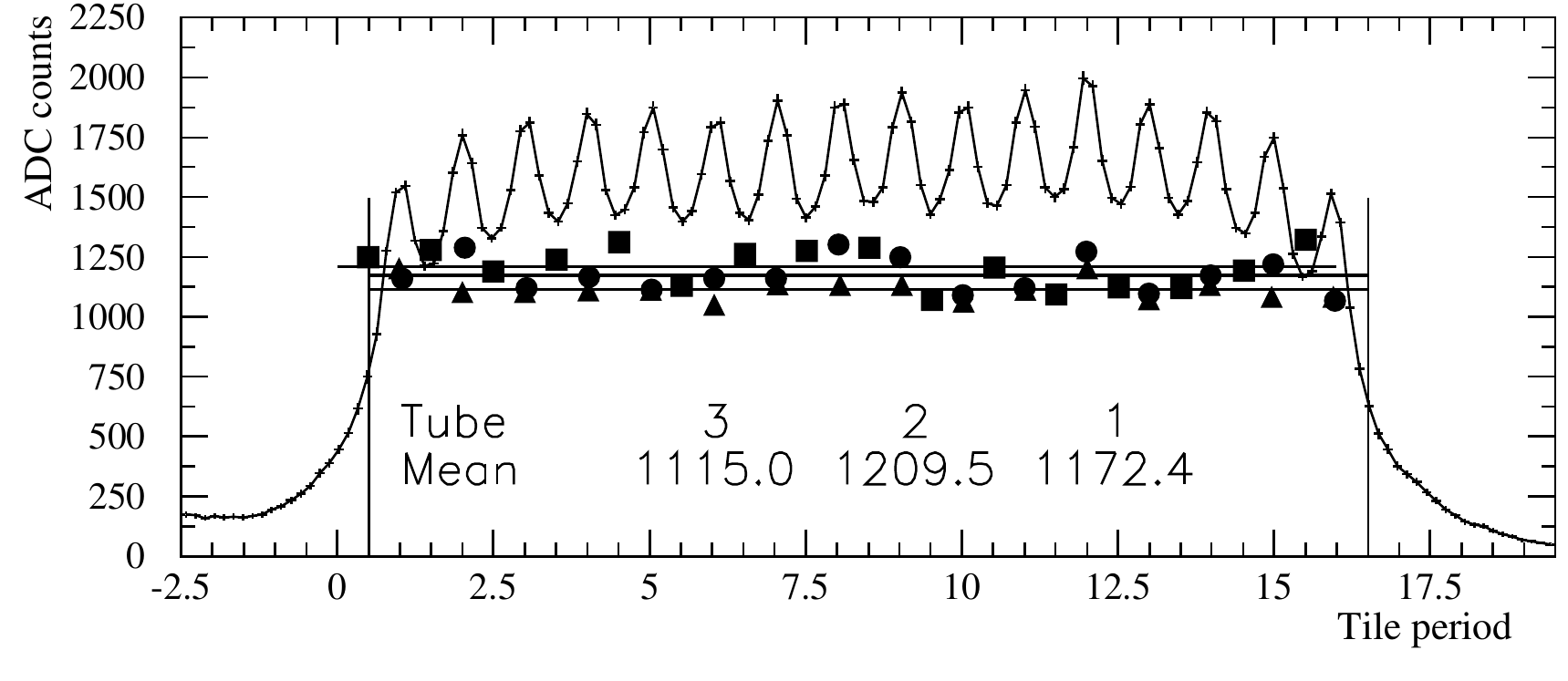}
  \caption{An example of the uniformities observed in an A-sample cell: tile responses belonging to one of the 3 tile rows are shown separately by squares, triangles and circles. The three horizontal lines are the averages in each tile row. From the variation of the response of all individual tiles in a cell, the instrumentation uniformity is calculated. From the values of the three horizontal lines, the physical uniformity of cells in a module is calculated. Here and in similar figures, the tube number is the tile row number.}
  \label{fig:tilecellresponse}
\end{figure}

It is useful to parametrise the uniformity of the response of tiles in a row, of rows with a cell, and of cells within a module. For the purpose of quality checks, two different parameters are found to be useful. They are referred to as ``instrumentation'' and ``physical'' uniformities. The instrumentation uniformity is the RMS/mean value of the responses, obtained with the amplitude method, of all the individual tiles in a cell. The physical uniformity is the RMS/mean value of the mean tile row responses within a cell; it is relevant to the quality of the response of the calorimeter cells in a module to hadronic showers. Figure~\ref{fig:tilecellresponse} shows an example of the data used to perform these calculations.

\section{System use and results}

The Cs calibration and monitoring system was used to check the quality of the prototype and production modules of TileCal and to set the electromagnetic energy scale (EM-scale)\footnote{In a calorimeter only some part of hadron shower energy, deposited in the sensitive parts of the detector, is responsible for the formation of
calorimeter response. This part is called visible energy, or energy in the electromagnetic scale (EM-scale).} of the calorimeter and the dynamic range of its readout system (1996--2004)~\cite{EM}. Since the installation of TileCal in the ATLAS cavern, together with the overall TileCal monitoring system, the Cs system is used to monitor and analyse the slow variation in time of the response of its various optical and readout components. In addition, it was used to study certain expected or unexpected phenomena, such as the effect of the ATLAS magnetic field on the TileCal signals and a slow drift of the PMT response, observed while testing the early module prototypes.
These uses of the Cs system are illustrated in this section.

\subsection{Checks of the module optical instrumentation}

The optical instrumentation stage in the construction of modules is described in detail elsewhere~\cite{Instrumentation}. Its steps are briefly recalled here to clarify the extent of the quality checks performed with the Cs system. This stage consisted of the insertion of tiles and fibres into the module steel structure, defining the readout cells by combining selected fibres into bundles, and coupling the fibre bundles to PMTs. The LB modules' mechanical structures, assembled at JINR, were delivered to CERN and fully instrumented in the latter laboratory, while the EBA modules were assembled and instrumented at Argonne National Laboratory (ANL) and Michigan State University (MSU) in the United States, and the EBC modules were assembled and instrumented at IFAE in Barcelona, Spain.

The main goal of the optical instrumentation checks was to verify the proper quality of the entire optical path, tiles to fibres to PMTs, and the correctness of the pattern of cells. It included the following steps:
\begin{itemize}
\item \Cs $\gamma$-source runs through the module with the system's hydraulic equipment;
\item Calculating all the individual tile responses, thereby obtaining a maximally granular  picture of the module and of its quality;
\item Checks of the fibre positioning in the cells, of the polishing of fibre bundles, of fibre glueing, of scintillator quality, of fibre quality -- looking for cracks, etc.; 
\item All deviations of >25\% of the individual tile responses from the tile row mean were repaired, when possible;
\item After all repairs, Cs runs were repeated; 
\item  To document the final certification step, the final module response map and overall uniformity figures were stored in the database together with information on any observed faults or irregularities. 
\end{itemize}

\begin{figure}[!htbp]
  \centering
  \subfloat[]{\includegraphics[width=0.30\textwidth]{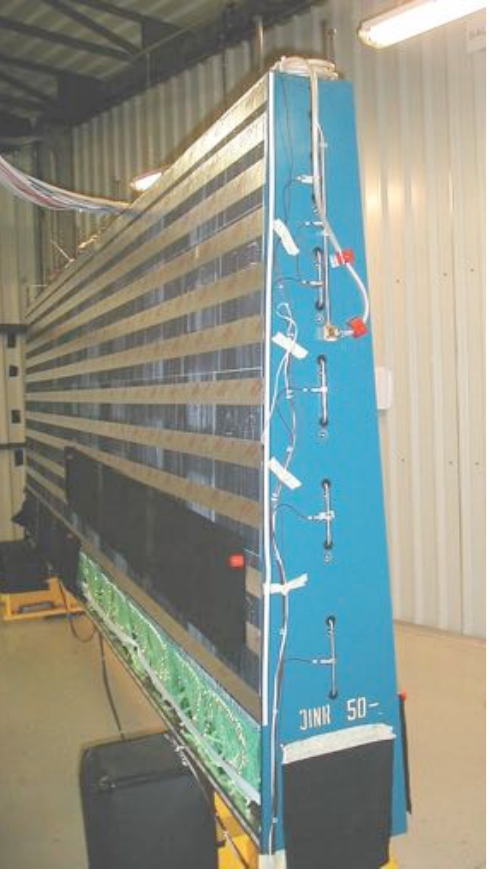}} \quad
  \subfloat[]{\includegraphics[width=0.60\textwidth]{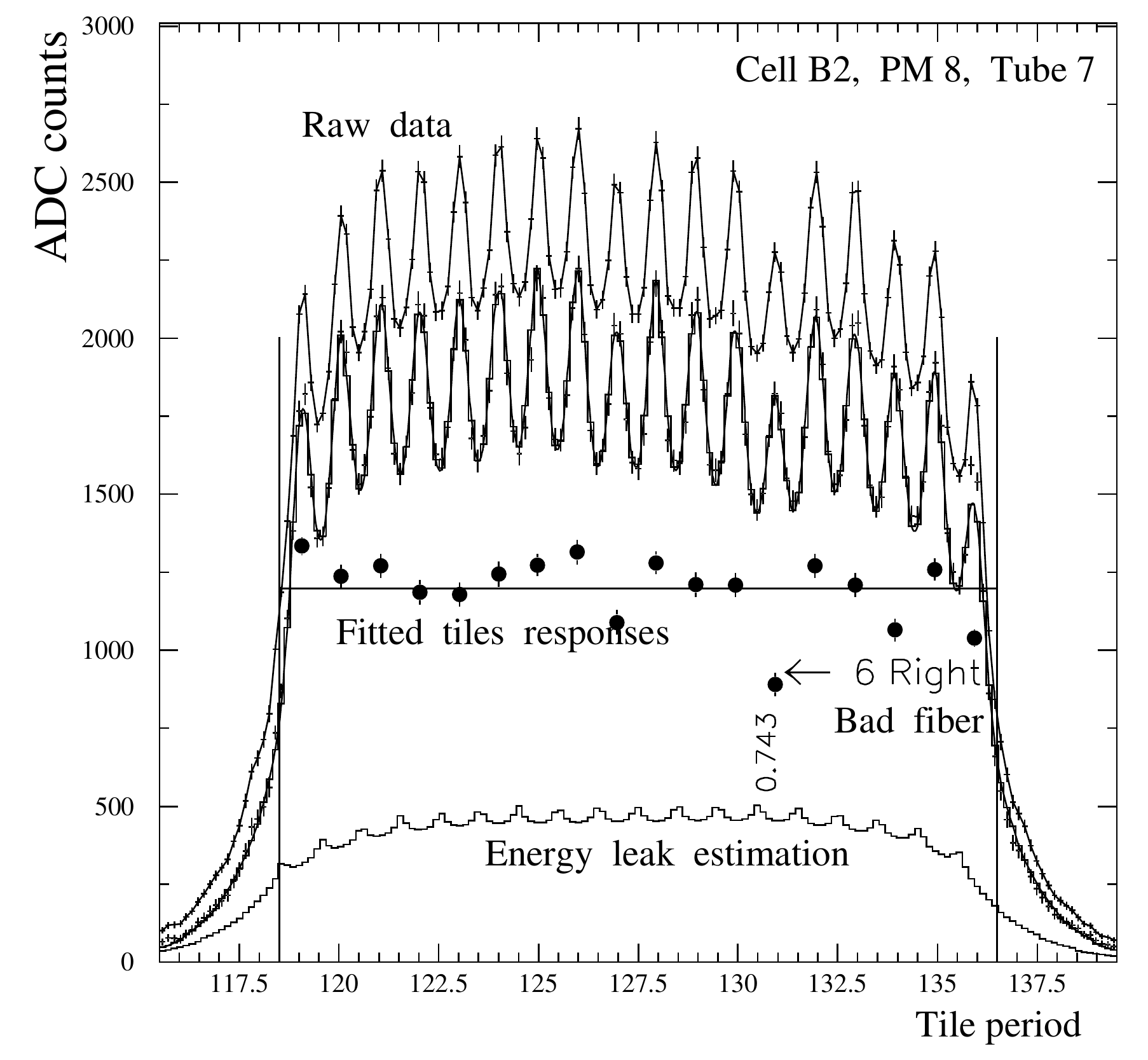}}
  \caption{LB module at CERN instrumentation site under tests with Cs source (a). The calibration tubes and SIN sensors are visible. Example of a defective tile-to-fibre coupling (b)~\cite{Instrumentation}.}
  \label{fig:instrumentation1}
\end{figure}

Figure~\ref{fig:instrumentation1} shows an LB calorimeter module ready for a Cs scan and an example of quality check plot that reveals a bad fibre coupling, worthy of repair.

\begin{figure}[!htbp]
  \includegraphics[width=\textwidth]{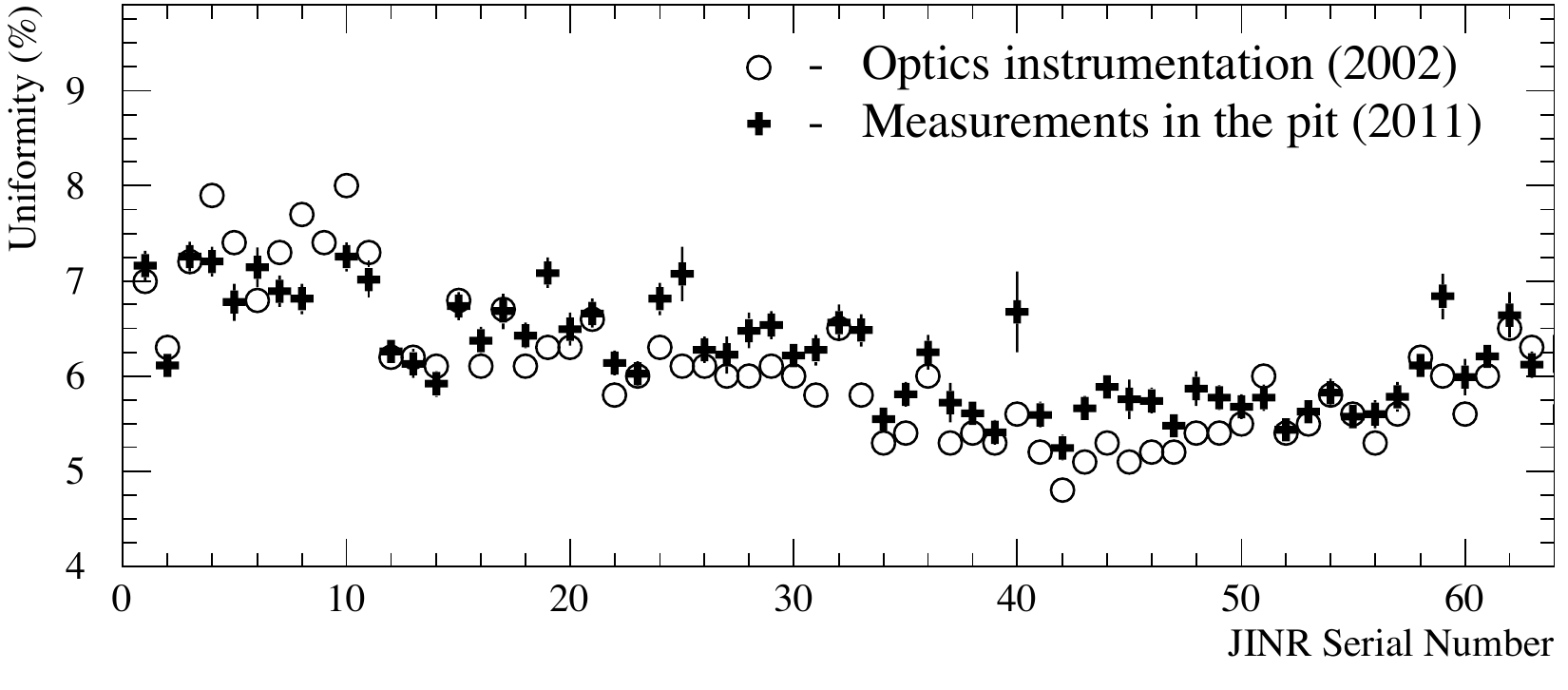}
  \caption{LB modules uniformity measured with the Cs system, just after the final certification (circles) and ten years later (crosses). The physical uniformity of the cells of all modules is better than 10\%~\cite{Instrumentation}.}
  \label{fig:instrumentation2}
\end{figure}

These checking and correction procedures noticeably improved the quality of instrumentation and helped to achieve the overall goal~--- an RMS/mean spread of physical uniformity numbers in any module better than 10\%. In figure~\ref{fig:instrumentation2} the physical uniformity of LB modules {\it vs.} their sequential production number is shown. Its value was kept well below the 10\% goal, despite appreciable drifts of the quality during the instrumentation time span (1999--2002). TileCal was assembled in the cavern in 2004--2006. These LB modules uniformity measurements were repeated in 2011, showing good general agreement with results at the time of module production and excellent optical integrity of the LB.

\subsection{Test beam calibrations: setting the EM energy scale}

The first hydraulic Cs source system was tested in 1996, on the prototype LB module. In 1997, two prototype EB modules (EBA and EBC) were under beam tests and the Cs system was used to calibrate and monitor these modules. After the regular TileCal module production started, in 1999, one of every eight modules was exposed to electron and hadron test beams, in order to measure their response to high-energy particles. In parallel, extensive response stability checks were also performed over the entire prototype and production module construction period (1996--2004).

\begin{figure}[!htbp]
  \centering
  \subfloat[]{\includegraphics[width=0.48\textwidth]{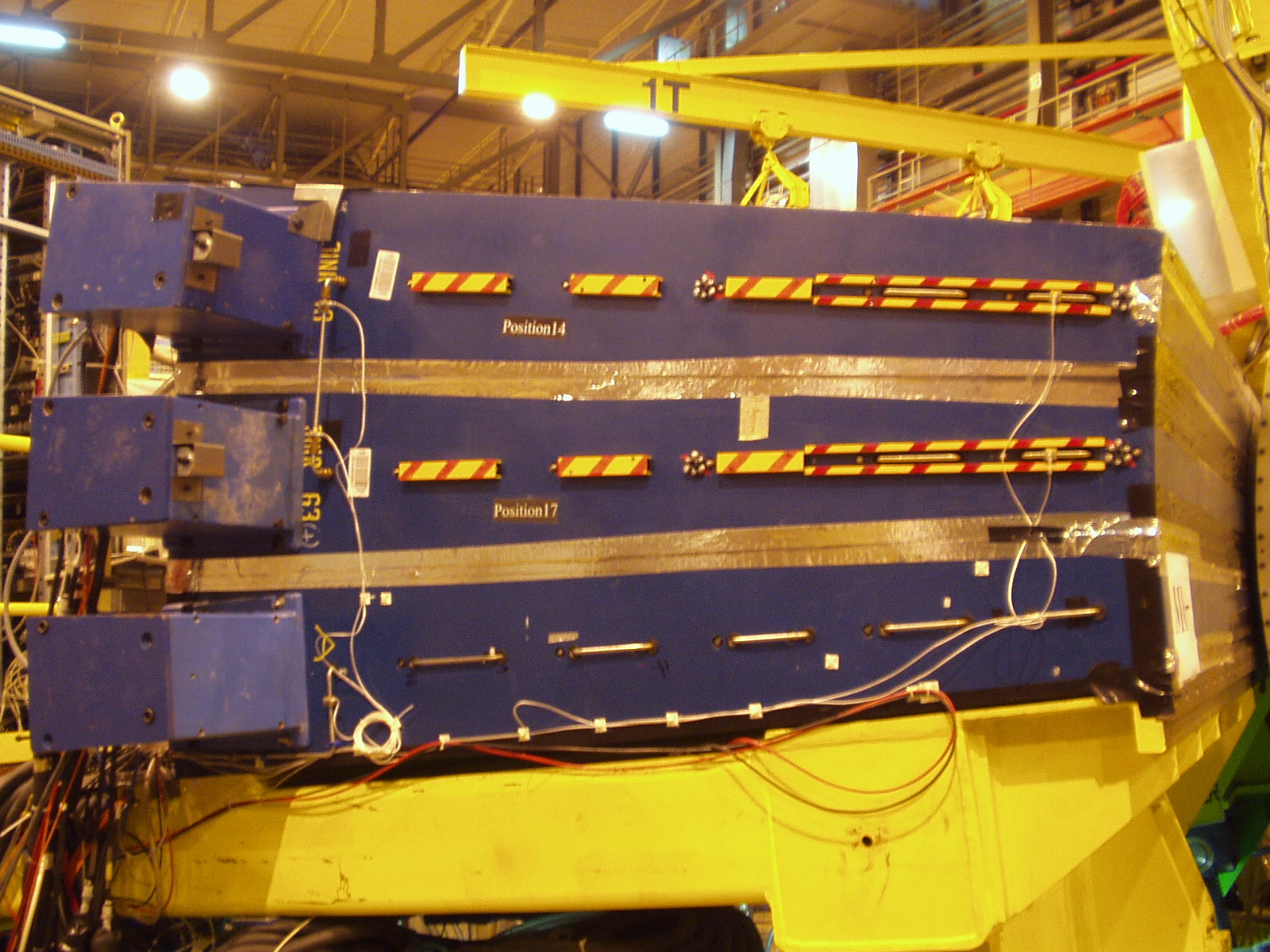}} \quad
  \subfloat[]{\includegraphics[width=0.48\textwidth]{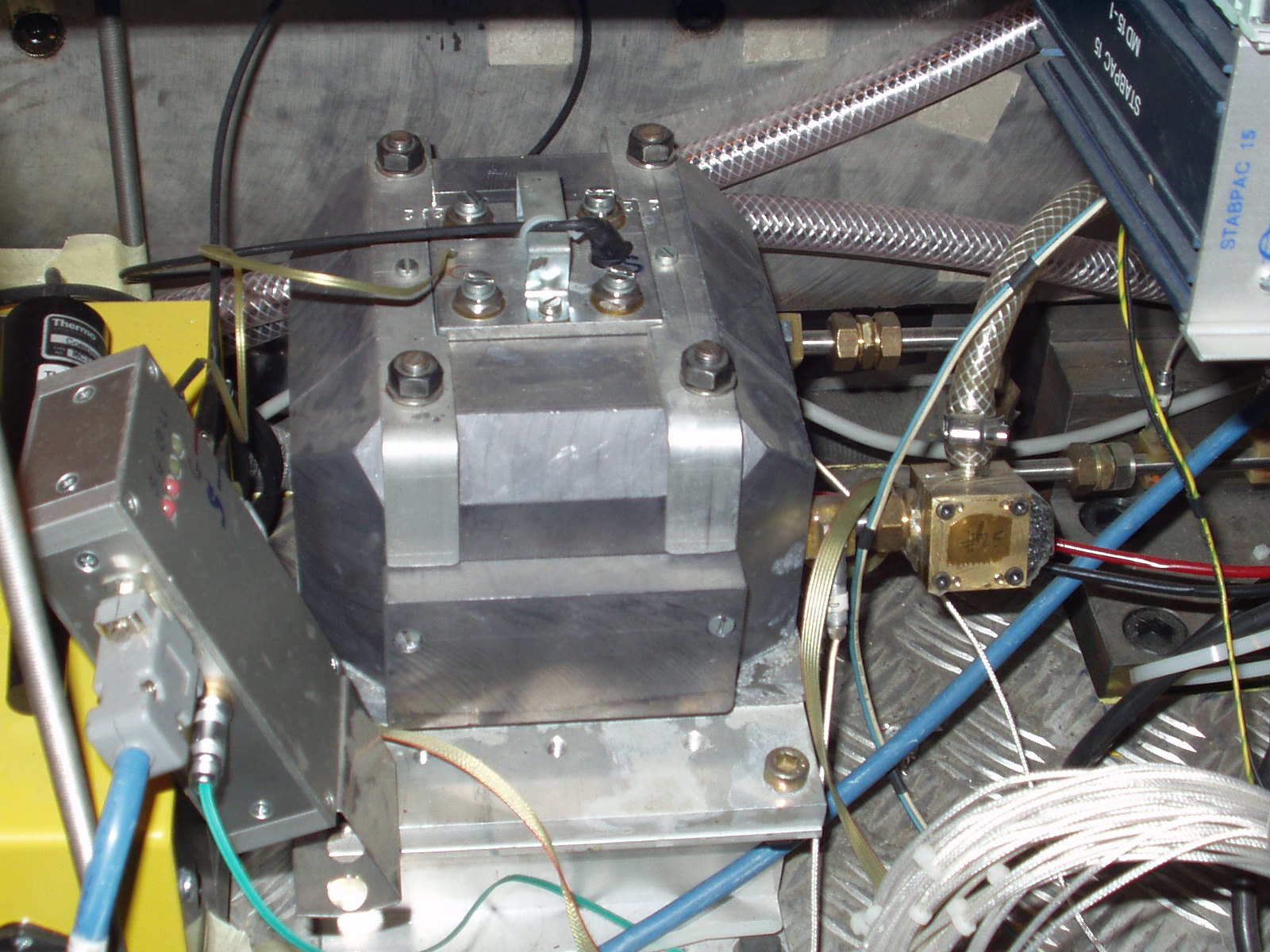}}
  \caption{Tile Calorimeter modules on the scanning table at the ATLAS test beam, equipped with the Cs source calibration system (a). The Cs garage prototype at the test beam (b).}
  \label{fig:testbeam1}
\end{figure}

Figure~\ref{fig:testbeam1} shows the Tile Calorimeter modules on a scanning table at the ATLAS combined test beam during the 2004 runs, with the hydraulic Cs scanning system. Calibration tubes, sensors and source garage are clearly visible.

The electron test beam runs on one out of every eight LB, EBA and EBC modules~\cite{Testbeam} allowed to set the scale of the response of TileCal modules to electromagnetic energy deposits, by connecting the charge measured by the readout system with the energy deposited in the calorimeter. The mean charge-to-energy conversion constant is 1.050$\pm$0.003 pC/GeV.
Scaling by the response of each module to the Cs sources, the energy calibration from the modules measured at the test beam was extended to all modules.

\subsection{Studies of module response stability}

Long-term monitoring of the TileCal prototype modules with Cs system allowed detecting a gradual signal loss, at a rate of about 1\%/month, much in excess of the \Cs source decay rate. Subsequent intensive tests and source scans helped to pin down the effect to a loss of photocathode response with accumulated light~\cite{PMT}. Reporting this effect to the manufacturer, Hamamatsu Photonics, allowed to develop more stable version of the R7877 PMTs, ahead of the full detector construction.

\subsection{Response equalisation and monitoring}

\begin{figure}[!htbp]
  \centering
  \includegraphics[width=0.95\textwidth]{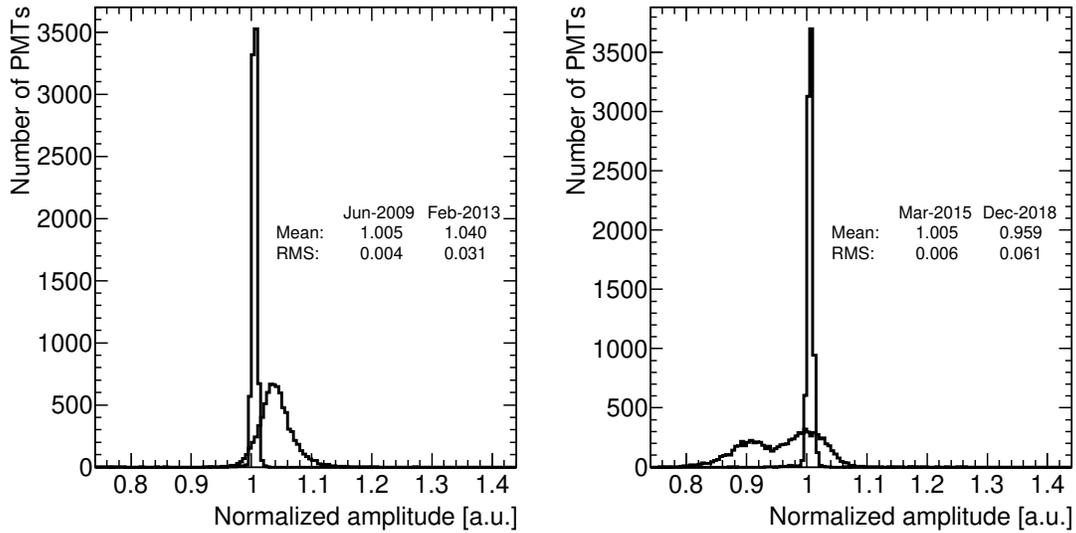}
  \caption{Entire TileCal normalised response distributions just after the initial equalisation (2009)  and four year later (left), the same distribution after another equalisation (2015) and three years later (right). One can see the shift of the mean and widening of the distributions due to the changes of the PMTs performance and detector response.}
  \label{fig:equal}
\end{figure}

The TileCal PMT voltages are set so that their linear dynamic range extends to 2~TeV/cell. In the process, running the Cs source through the entire volume of the calorimeter allows equalising all cell responses to Cs signals at a level of about 1\%. The process requires 2--3 iterations, in which measured cells responses are used to recalculate the next HV corrections in order to precisely get the desired response for all cells. The different activities of the sources used in the calorimeter sections are taken into account. Figure~\ref{fig:equal} shows the distributions of the normalised responses of all the readout channels, except for a few dead ones, at different times: when the equalisation was made, and the distribution of responses measured several years later. In between these two measurements, the PMT HV settings were not changed, but the response of the entire TileCal was monitored with Cs source runs and the slowly changing response measurements were used to update the calorimeter's energy scale. 

Just after the initial equalisation step, channel responses are all within about 2\%, while an overwhelming majority of them are within 0.5\%. After a few years, the dispersion of responses has grown to almost 3\%. The mean of the response, normalised to the decay of the source, is also observed to change, partly corresponding to the so-called "up-drift" of the PMT response. The average value immediately after the equalisation is slightly above 1.0 because the equalisation was done without the magnetic field, that affects the scintillator performance (see section~\ref{sec:mfe}), and the measurement shown on the plot was done when magnetic field was switched on.

As both the scintillator and PMT change their response in time due to the ageing, the irradiation damage in scintillator and WLS fibres, and charge collection in the PMT anode, it is required to keep monitoring the performance of the calorimeter with the Cs system. By taking several Cs scans per year, one can study the evolution of the response of the TileCal cells in time.

\begin{figure}[!tbp]
  \centering
  \includegraphics[width=0.85\textwidth]{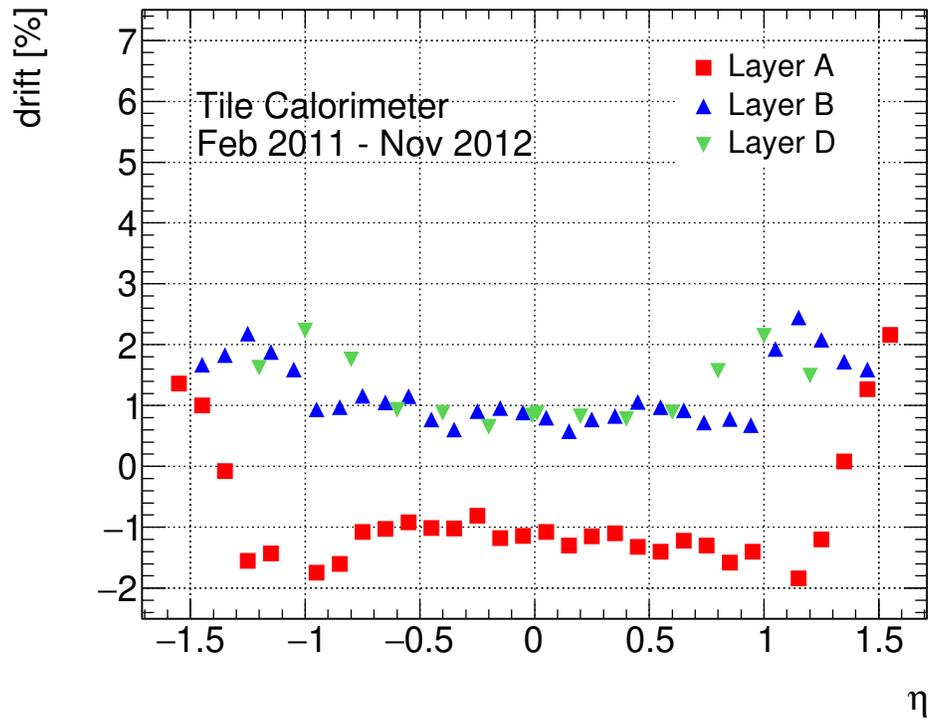} \\
  \includegraphics[width=0.85\textwidth]{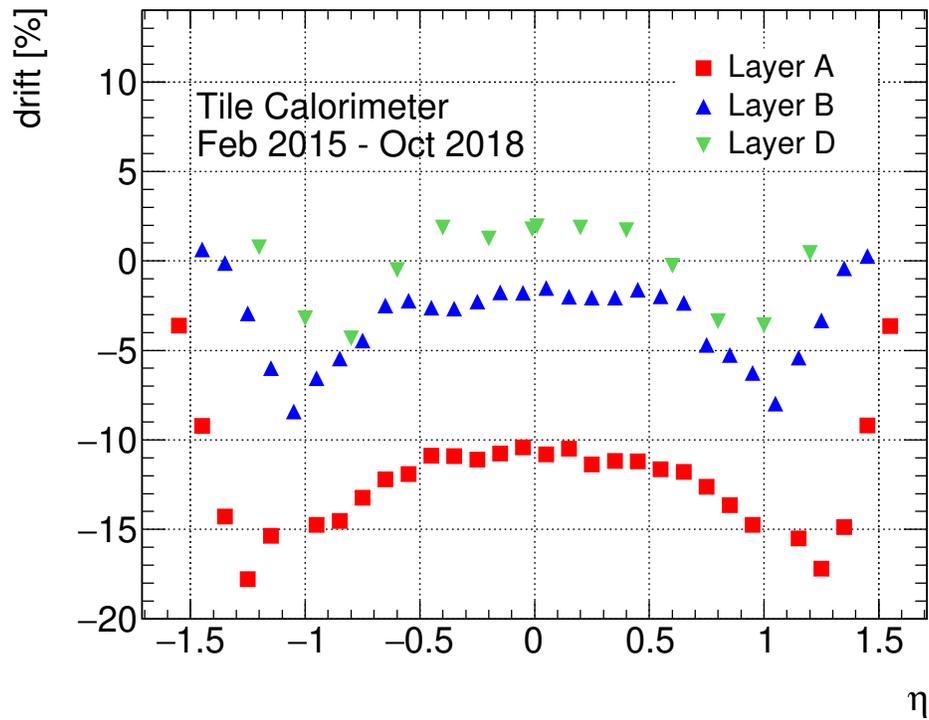} \\ 
  \caption{The drift from the expected response of the cell to \Cs source vs. eta of the cell for two run periods. 2011--2013 (Run 1) at the top, and 2015--2018 (Run 2) at the bottom. Most of the drift is coming from the change of the PMT response.}
  \label{fig:stability}
\end{figure}

Comparing the behaviour of the responses between relatively low luminosity in 2011--2012 (Run~1~\cite{TileRun1}) and much higher luminosity conditions in 2015--2018 (Run~2) in figure~\ref{fig:stability}, one can see a much higher spread and higher degradation of the cells' responses in the latter case. There is an evident dependence of the amount of degradation on the layer and $\eta$ of the cells that corresponds to the distribution of the particle flux, hence the amount of charge seen by PMT. The most affected part is layer A, closest to the beam pipe, and the region between the Long and Extended Barrels.

\begin{figure}[!tbp]
  \centering
  \includegraphics[width=0.9\textwidth]{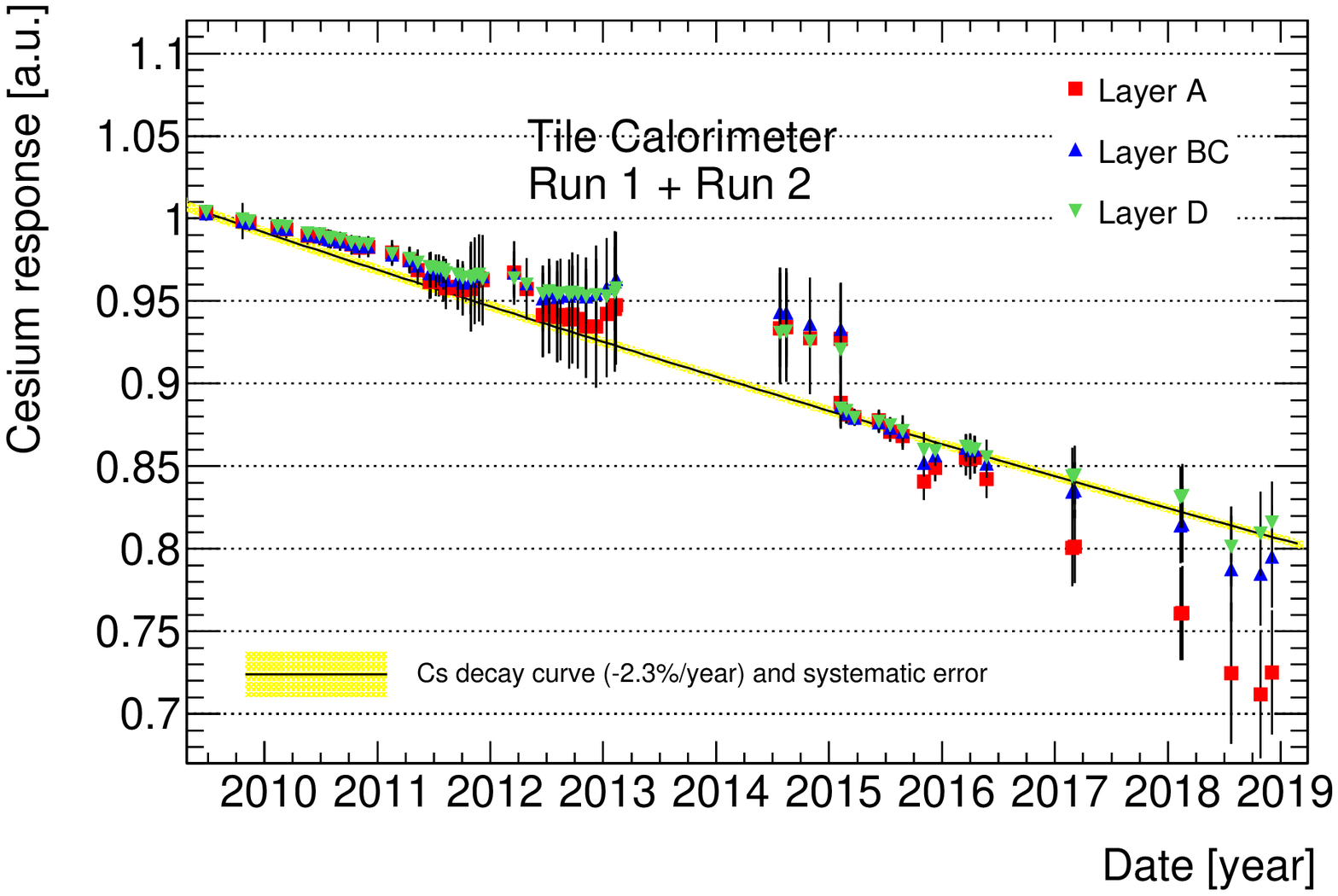} \\
  \includegraphics[width=0.9\textwidth]{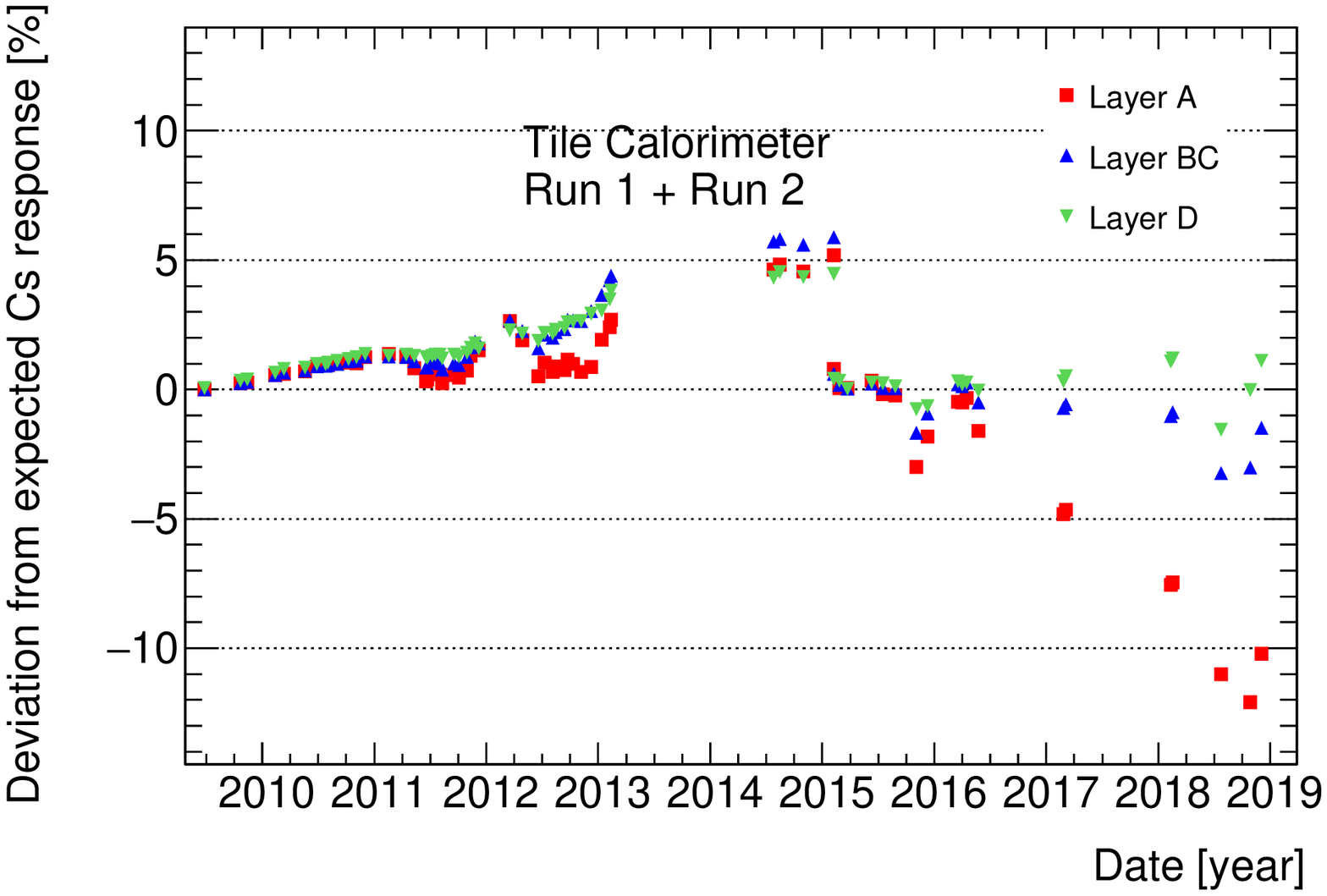} \\ 
  \caption{The behaviour in time of the mean responses of TileCal cells to \Cs and their dispersions in three longitudinal layers (top) and the deviation of the average cell response from expected values (bottom).}
  \label{fig:stability2}
\end{figure}

A closer look at the variation in time of the mean and deviation of the calorimeter responses of layer A, closest to the beam pipe, in figure~\ref{fig:stability2} shows that both the mean and the deviation values change more significantly during the higher luminosity periods (bottom plot). A more detailed analysis of this performance is presented in~Ref.~\cite{TileRun1}.

This kind of reaction argues for the need of regular monthly Cs calibrations of the entire calorimeter, to adequately correct the drift of its response to particles from LHC collisions.
A detailed study of calorimeter response was performed in combination with other TileCal calibration systems~\cite{Calibrations}.

\clearpage
\subsection{Magnetic field effect}
\label{sec:mfe}

Effects of the magnetic field on the calorimeter's response may arise from two separate phenomena: changes in the response of the PMTs or changes in the light yield of scintillators~\cite{MagnetB}, \cite{MagnetM}. In TileCal, the former is expected to be small or not observable, because PMTs were carefully shielded from the residual magnetic field at their location. The observed significant magnetic field effect is likely to arise from the second phenomenon.

The Cs monitoring system allows detecting response variations at the row or cell level with an accuracy of 0.5\% or better. With sufficient statistics from runs with and without magnetic field, it was possible to detect the effect of the ATLAS magnetic field.

\begin{figure}[!htbp]
  \centering
  \subfloat[]{\includegraphics[width=0.95\textwidth]{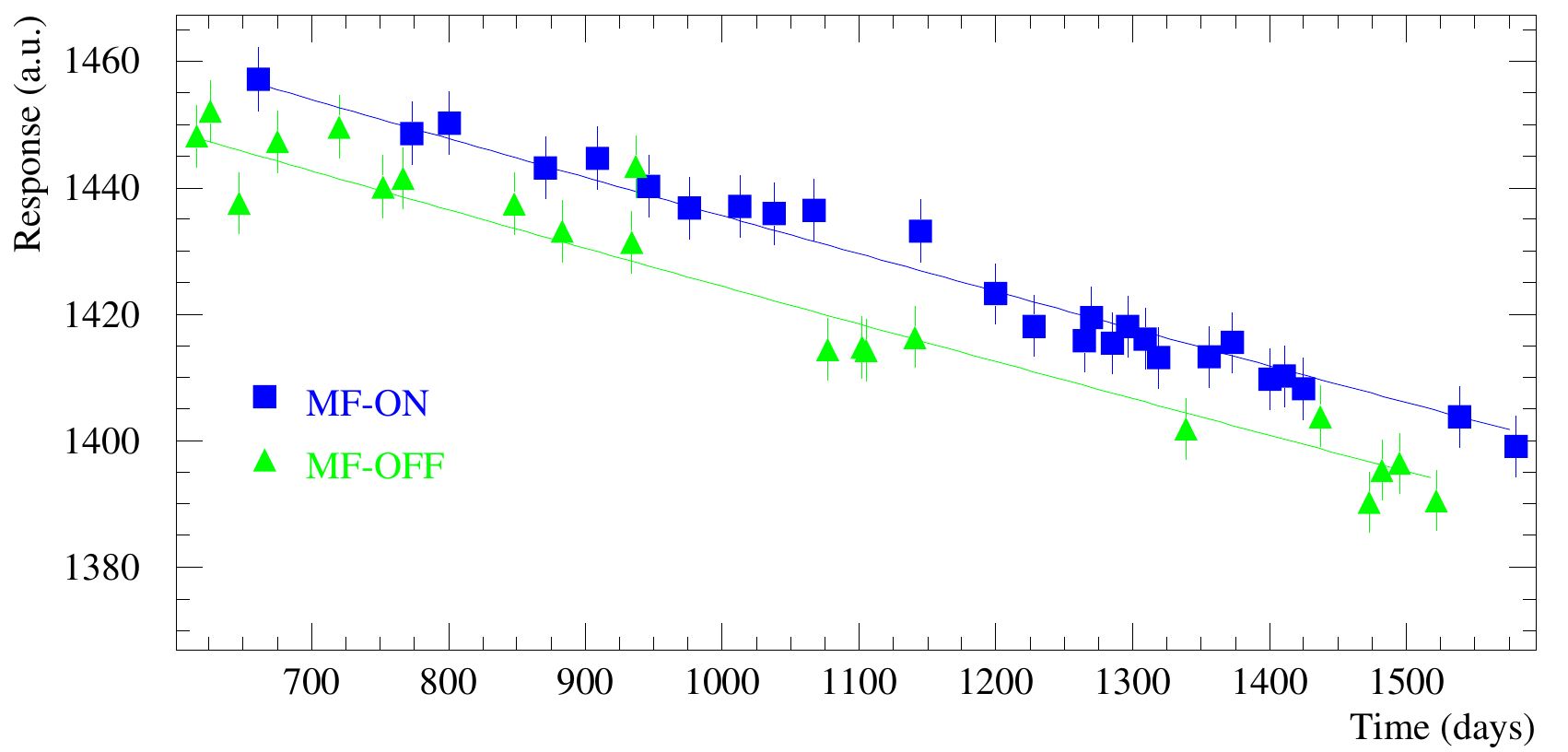}}\\
  \subfloat[]{\includegraphics[width=0.95\textwidth]{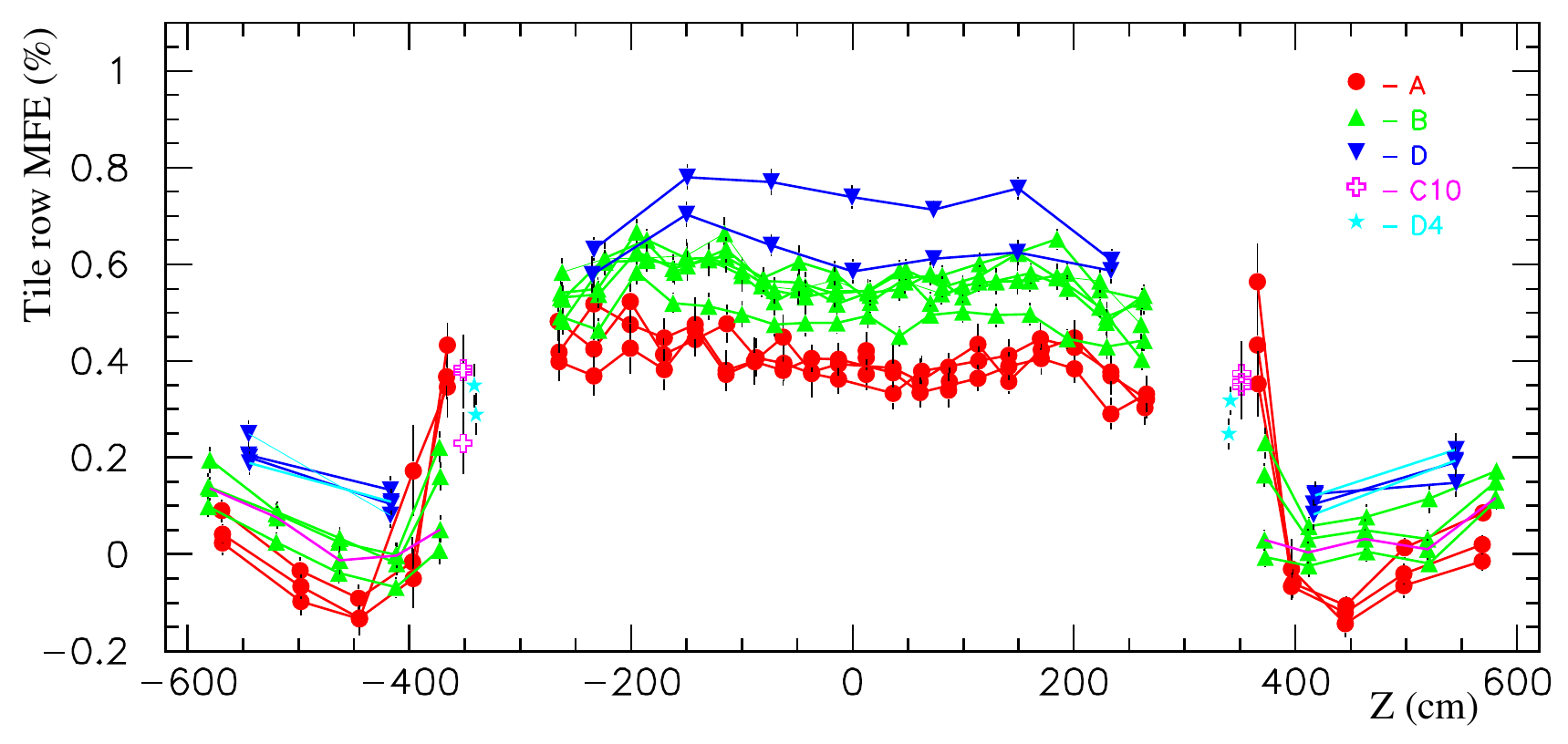}}
  \caption{The effect of the ATLAS magnetic field. (a) The two sets of measurements of A-layer cells with (blue squares) and without (green triangles) magnetic field. The sets are fitted with a curve taking into account combined effect of the Cs isotope decay and PMT gain updrift. (b) Distribution amongst cells at positions along the colliding beam axis Z, of the magnetic field on-off differences for different layers of the calorimeter, with separately shown cells C10 and D4, which are outside of the calorimeter end-plate.}
  \label{fig:mfe}
\end{figure}

As an example, two sets of response measurements for a tile row of an A-cell are shown in figure~\ref{fig:mfe}a. Typically, as in this example, the response to the \Cs source signal is roughly 1\% higher when the magnetic field is on. The effect is more apparent during the first 1200 days before the higher luminosity runs that temporarily decreased the response of the PMTs, however, it can be seen again after day 1400.

The effect of the magnetic field on cells in different layers of the LB, EBA and EBC sections, as a function of the position of the cells along the colliding beams axis, is shown in the  figure~\ref{fig:mfe}b. The effect is seen to be largest in the most external layer D, that is closest to the toroidal magnet's coils. Such maps can be done module by module and cell by cell, providing a very detailed 3D view of the effect of the magnetic field at any point of TileCal. 

\section{Summary and conclusions}

This article gives a comprehensive description of the calibration and monitoring system for the ATLAS Tile Calorimeter, based on movable radioactive \Cs sources. The sources are driven by the flow of a liquid within a system of tubes which traverses every single tile of the detector, supervised by custom-developed hardware and software that allows to control at any time the position and movement of the radioactive sources. 
The TileCal signal readout system includes a parallel branch devoted to reading out the signals produced by the \Cs sources. The acquisition of these signals is coordinated with the source movement controls; the data are stored in a database for later analysis.

The system measures the response of every single tile over the entire path of the optical signal, up to and including the PMTs, thereby providing information about the inner structure of every one of the about 5000 calorimeter cells. The response accuracy is reproducible with a precision of 0.2--0.3\% for a standard cell and about 2\% for individual tiles, fully adequate for the calorimeter calibration and monitoring goals.

The system was developed during the R\&D phase of the Tile Calorimeter and used to investigate the performance of the module prototypes.
During the production phase of modules, it was used to supervise the optical instrumentation of LB modules, at CERN, and to check similar construction steps, performed in collaborating laboratories, of EBA and EBC modules. Throughout this process, the Cs source scans allowed to detect a number of local faults and to certify their repair. The recorded data show that the uniformity of response goals across the calorimeter was met. 
Using the Cs source data as a reference, a number of global settings were made: the response of all TileCal cells was equalised, the dynamic range of the module readout signals was set, the ratio of charge-to-energy deposited in the calorimeter was determined for a sample of modules exposed to test beams, and extended to all modules. 

Periodic source scans of the entire calorimeter installed in the ATLAS cavern are performed, with an approximately monthly frequency. These scans allow to monitor at global and detailed levels the stability of the calorimeter and its components, and most importantly, to maintain the overall energy calibration of the detector. When compared to data taken after completion of module production, the uniformity of response is seen not to have deteriorated.

Furthermore, because the source signal shows the response of the entire TileCal optical system, Cs source measurements are a very useful tool to investigate several instrumental effects, such as the stability of response of PMTs themselves and the impact of the ATLAS magnetic field on the signals from particles. 

Two overall conclusions can be drawn from the two-decade experience with this system:
\begin{enumerate} 
\item The chosen liquid drive method has been proved to be extremely robust, and fully adequate to the requirements of the calibration and monitoring task. 
\item The Cs source system --- hardware, controls, online software and offline analysis tools --- has been of paramount importance in achieving the goals for which the Tile Calorimeter was built and is operated.
\end{enumerate} 


Finally, in the next years, the LHC and ATLAS, will undergo a series of upgrades, necessitating the upgrade of TileCal calibration systems, including Cesium calibration system, due to the new requirements on the radiation hardness, ageing of existing electronics, and changes in the front-end electronics. The upgrades will have to increase the reliability and integration with the calorimeter readout. The new front-end electronics will allow for shorter integration times and faster readout. The hydraulic system will have to undergo the improvements as well. The design of the new control boards and a path to the hydraulic system development are outlined in the TileCal upgrade technical design report~\cite{UTDR}.

\acknowledgments
The authors are very grateful to all members of the ATLAS TileCal community who participated in all the discussions, talks, tests and in the intensive use of the Cs monitoring system during several years. They crucially helped in creating a really useful tool for the TileCal optical instrumentation, inter-calibration and monitoring, and for the setting and maintenance of the calorimeter's electromagnetic energy scale.

\end{document}